\documentclass[iop, twocolappendix, numberedappendix]{emulateapj_mod}

\usepackage{natbib}
\usepackage{graphicx}
\usepackage{txfonts}
\usepackage{url}
\usepackage{color}
\usepackage{rotating}

\makeatletter

\newcommand{\smi}{$\sigma_{\rm{minor}}$}
\newcommand{\smj}{$\sigma_{\rm{major}}$}
\newcommand{\rmaxs}{$R_{\rm{\sigma}}^{\rm{max}}$}
\newcommand{\rmaxh}{$R_{\rm{h3}}^{\rm{max}}$}
\newcommand{\recirc}{$R_{\rm{e,c}}$}
\newcommand{\re}{$R_{\rm{e}}$}
\newcommand{\ret}{$R_{\rm{e/2}}$}

\newcommand{\kpa}{$PA_{\rm{kin}}$}

\newcommand{\lr}{\ifmmode{\lambda_R}\else{$\lambda_{R}$}\fi}
\newcommand{\lre}{\ifmmode{\lambda_{R_{\rm{e}}}}\else{$\lambda_{R_{\rm{e}}}$}\fi}
\newcommand{\lret}{$\lambda_{R_{\rm{e/2}}}$}
\newcommand{\lretwo}{$\lambda_{2R_{\rm{e}}}$}
\newcommand{\vmf}{$V_{\rm{m4}}$}
\newcommand{\smf}{$\sigma_{\rm{m4}}$}
\newcommand{\vmt}{$V_{\rm{m2}}$}
\newcommand{\smt}{$\sigma_{\rm{m2}}$}
\newcommand{\vs}{$V / \sigma$}
\newcommand{\vse}{$(V / \sigma)_{\rm{e}}$}
\newcommand{\at}{ATLAS$^{\rm{3D}}$}
\newcommand{\mk}{\ifmmode{\overline{k_5 / k_1}}\else{$\overline{k_5 / k_1}$}\fi}
\newcommand{\mkt}{${(k_3+k_5) / 2k_1}$}
\newcommand{\kms}{$\,$km$\,$s$^{-1}$}
\newcommand{\msun}{{$M_{\odot}$}}

\newcommand{\kinemetry}{\textsc{kinemetry}}

\slugcomment{}
\shorttitle{Revisiting Galaxy Classification Through High-Order Stellar Kinematics}
\shortauthors{van de Sande et al.}

\begin{document}

\title{The SAMI Galaxy Survey: Revisiting Galaxy Classification Through High-Order Stellar Kinematics}

\author{Jesse van de Sande\altaffilmark{1},
Joss Bland-Hawthorn\altaffilmark{1}, 
Lisa ~M.~R. Fogarty\altaffilmark{1,2},
Luca Cortese\altaffilmark{3}, 
Francesco d'Eugenio\altaffilmark{4},
Scott ~M. Croom\altaffilmark{1,2},
Nicholas Scott\altaffilmark{1,2}, 
James ~T. Allen\altaffilmark{1,2}, 
Sarah Brough\altaffilmark{5},
Julia ~J. Bryant\altaffilmark{1,2,5}, 
Gerald Cecil\altaffilmark{6},
Matthew Colless\altaffilmark{2,4},
Warrick ~J. Couch\altaffilmark{5}, 
Roger Davies\altaffilmark{7},
Pascal ~J. Elahi\altaffilmark{3},
Caroline Foster\altaffilmark{5},
Gregory Goldstein\altaffilmark{8},
Michael Goodwin\altaffilmark{5},
Brent Groves\altaffilmark{4},
I-Ting Ho\altaffilmark{4,9},
Hyunjin Jeong\altaffilmark{10},
~D. Heath Jones \altaffilmark{11},
Iraklis ~S. Konstantopoulos\altaffilmark{5,12},
Jon ~S. Lawrence\altaffilmark{5},
Sarah ~K. Leslie\altaffilmark{2,4,13},
\'Angel ~R. L\'opez-S\'anchez\altaffilmark{5,8},
Richard ~M. McDermid\altaffilmark{5,8}, 
Rebecca McElroy\altaffilmark{1,2}, 
Anne ~M. Medling\altaffilmark{4,14,15}, 
Sree Oh\altaffilmark{16},
Matt ~S. Owers\altaffilmark{5,8}, 
Samuel ~N. Richards\altaffilmark{1,2,5},  
Adam ~L. Schaefer\altaffilmark{1,2,5},
Rob Sharp\altaffilmark{2,4}, 
Sarah ~M. Sweet\altaffilmark{4},
Dan Taranu\altaffilmark{2,3},
Chiara Tonini\altaffilmark{17},
~C. Jakob Walcher\altaffilmark{18},
Sukyoung ~K. Yi\altaffilmark{16}
}

\altaffiltext{1}{Sydney Institute for Astronomy, School of Physics, A28, The University of Sydney, NSW, 2006, Australia}
\altaffiltext{2}{ARC Centre of Excellence for All-Sky Astrophysics (CAASTRO)}
\altaffiltext{3}{International Centre for Radio Astronomy Research, The University of Western Australia, 35 Stirling Highway, Crawley WA 6009, Australia}
\altaffiltext{4}{Research School of Astronomy and Astrophysics, Australian National University, Canberra ACT 2611, Australia}
\altaffiltext{5}{Australian Astronomical Observatory, PO Box 915, North Ryde NSW 1670, Australia}
\altaffiltext{6}{Dept. Physics and Astronomy, University of North Carolina, Chapel Hill, NC 27599-3255, USA.}
\altaffiltext{7}{Astrophysics, Department of Physics, University of Oxford, Denys Wilkinson Building, Keble Rd., Oxford, OX1 3RH, UK.}
\altaffiltext{8}{Department of Physics and Astronomy, Macquarie University, NSW 2109, Australia}
\altaffiltext{9}{Institute for Astronomy, University of Hawaii, 2680 Woodlawn Drive, Honolulu, HI 96822, USA}
\altaffiltext{10}{Korea Astronomy and Space Science Institute, Daejeon 305-348, Korea}
\altaffiltext{11}{English Language and Foundation Studies Centre, University of Newcastle, Callaghan, NSW 2308, Australia}
\altaffiltext{12}{Envizi Suite 213, National Innovation Centre, Australian Technology Park, 4 Cornwallis Street, Eveleigh NSW 2015, Australia}
\altaffiltext{13}{Max-Planck-Institut f\"{u}r Astronomie, K\"{o}nigstuhl 17, 69117 Heidelberg, Germany}
\altaffiltext{14}{Cahill Center for Astronomy and Astrophysics California Institute of Technology, MS 249-17 Pasadena, CA 91125, USA}
\altaffiltext{15}{Hubble Fellow}
\altaffiltext{16}{Department of Astronomy and Yonsei University Observatory, Yonsei University, Seoul 120-749, Republic of Korea}
\altaffiltext{17}{School of Physics, the University of Melbourne, Parkville, VIC 3010, Australia}
\altaffiltext{18}{Leibniz-Institut f\"{u}r Astrophysik Potsdam (AIP), An der Sternwarte 16, D-14482, Potsdam, Germany}


\begin{abstract}

Recent cosmological hydrodynamical simulations suggest that integral field spectroscopy can connect the high-order stellar kinematic moments $h_3$ ($\sim$skewness) and $h_4$ ($\sim$kurtosis) in galaxies to their cosmological assembly history. Here, we assess these results by measuring the stellar kinematics on a sample of 315 galaxies, without a morphological selection, using two-dimensional integral field data from the SAMI Galaxy Survey. A proxy for the spin parameter ($\lambda_{R_{\rm{e}}}$) and ellipticity ($\epsilon_{\rm{e}}$) are used to separate fast and slow rotators; there exists a good correspondence to regular and non-regular rotators, respectively, as also seen in earlier studies. We confirm that regular rotators show a strong $h_3$ versus $V/\sigma$ anti-correlation, whereas quasi-regular and non-regular rotators show a more vertical relation in $h_3$ and $V/\sigma$. Motivated by recent cosmological simulations, we develop an alternative approach to kinematically classify galaxies from their individual $h_3$ versus $V/\sigma$ signatures. Within the SAMI Galaxy Survey, we identify five classes of high-order stellar kinematic signatures using Gaussian mixture models. Class 1 corresponds to slow rotators, whereas Classes 2-5 correspond to fast rotators. We find that galaxies with similar $\lambda_{R_{\rm{e}}}-\epsilon_{\rm{e}}$ values can show distinctly different $h_3-V/\sigma$ signatures. Class 5 objects are previously unidentified fast rotators that show a weak $h_3$ versus $V/\sigma$ anti-correlation. From simulations, these objects are predicted to be disk-less galaxies formed by gas-poor mergers. From morphological examination, however, there is evidence for large stellar disks. Instead, Class 5 objects are more likely disturbed galaxies, have counter-rotating bulges, or bars in edge-on galaxies. Finally, we interpret the strong anti-correlation in $h_3$ versus $V/\sigma$ as evidence for disks in most fast rotators, suggesting a dearth of gas-poor mergers among fast rotators.

\end{abstract}



%

\section{Introduction}
\label{sec:introduction}

\noindent Studying the build-up of mass and angular momentum in galaxies is fundamental to understanding the large variations in morphology and star formation that we see in present-day galaxies. Numerous methods and techniques have been employed, but two are most often compared to simulations. 1) The evolution of the galaxy stellar mass function \citep[e.g.,][]{bundy2006,marchesini2009,baldry2012,muzzin2013,ilbert2013,tomczak2014} reveals the stellar mass density in the universe over time and provides strong constraints on galaxy formation models \citep[see e.g.,][]{delucia2007} but is limited to galaxy samples as a whole. 2) Detailed dynamical studies of stars in present-day galaxies, provide a fossil-record of their individual assembly history \citep[e.g.,][]{dezeeuw1991, bender1994, cappellari2016}.

A major development for measuring the stellar dynamics in galaxies came with the introduction of visible-light integral field spectrographs \citep[e.g., SAURON;][]{bacon2001}. The projected angular momentum and spin parameter could now be estimated rigorously in two dimensions, weighted by the surface brightness, in each galaxy. Two-dimensional (2D) stellar kinematics measurements led to a new technique for classifying galaxies using a proxy for the spin parameter (\lr) within one effective radius (\re) to define slow ($\lre < 0.1$) and fast ($\lre > 0.1$) rotating galaxies \citep[SAURON survey;][]{bacon2001,dezeeuw2002,emsellem2004}. The \at\, survey \citep{cappellari2011a} refined the slow versus fast criterion \citep{emsellem2011} by studying a larger sample of 260 local early-type galaxies. They showed that only 12\% of the galaxies in their sample are slow-rotating galaxies with no disk component, whereas the majority (86\%) are fast-rotating galaxies with ordered rotation and regular velocity fields and disks \citep{krajnovic2011, emsellem2011}. Slow and fast rotating galaxies have very different dynamical properties, which suggests that there are at least two formation paths for creating the two kinematic classes in early-type galaxies.

From a theoretical perspective, many studies are aimed at explaining these different kinematic classes and linking this to the build-up of mass and angular momentum \citep[for a recent review see ][]{naab2014}. It was recognized early on that it is difficult to create slowly rotating early-type galaxies through major mergers of spheroids \citep{white1979a}. Simulations of merging cold disks were successful in creating dispersion dominated spheroids \citep{gerhard1981,farouki1982,negroponte1983,barnes1988,barnes1992a,hernquist1992,barnes1992b,hernquist1993,heyl1994} 
and modeled collisions of unequal mass mergers turned out to have significant impact for creating flattened systems with faster rotation \citep{bekki1998,naab1999,bendo2000,naab2003,bournaud2004,bournaud2005,jesseit2005,gonzalezgarcia2005,bournaud2005,naab2006a,bournaud2007,jesseit2007}. Many early-type formation models, however, have difficulty in reproducing the observed population of slow rotators in the nearby Universe \citep[e.g.,][]{bendo2000,jesseit2009,bois2011}. Using binary-disk merger simulations, most merger remnants are consistent with fast rotators \citep{bois2010,bois2011}. While the mass ratio of the progenitors in binary-disk mergers seem to be the most-important parameter for creating slow rotators, the binary-disk mergers also require specific spin-orbit alignments \citep{jesseit2009,bois2010,bois2011}. The slow rotators that are created, however, are relatively flat systems \citep[$0.45 < \epsilon_{\rm{e}} < 0.65$;][]{bois2011} instead of the observed round galaxies \citep[$0 < \epsilon_{\rm{e}} < 0.45$;][]{emsellem2011}, and hold a kinematically distinct core \citep[e.g.,][]{jesseit2009,bois2010}. Furthermore, \citet{bois2010} study re-mergers, and find relatively round fast rotators or galaxies close to the selection criterion for slow rotators, but no true slow rotators as identified by the \at\, survey. 

This is in contrast with \citet{cox2006} who find that collisions of equal-mass disk galaxies with 40\% gas can produce slow-rotating elliptical galaxies. Their results suggest that remnants formed from dissipational mergers of equal-mass disk galaxies better match the observed data than dissipationless merger remnants. \citet{taranu2013} study collisionless simulations of dry mergers and also find that group-central galaxy remnants have properties similar to ellipticals. Yet, their results suggest that dissipation is not necessary to produce slow-rotating galaxies; instead, multiple, mostly dry minor mergers are sufficient.

Using cosmological hydrodynamical zoom-in simulations of 44 individual central galaxies, \citet{naab2014} link the assembly history of these galaxies to their stellar dynamical features. These simulations follow the growth and evolution of the galaxy from $z=43$ to $z=0$ and give a more realistic insight into the formation paths for slow and fast rotators when compared with previous idealized, binary merger simulations. Their analysis of the stellar kinematic data follows the \at\, approach. They find that the 2D velocity and velocity dispersion fields are in good qualitative agreement with the \at\, kinematics. The simulated galaxies show a similar diversity in kinematical classifications when compared with observed galaxies, producing fast and slow rotators as well as galaxies with counter-rotating cores. The striking result from \citet{naab2014} is, however, that there are not two unique formation histories for fast and slow rotators, and that the formation mechanism for massive galaxies is complex.

Although \citet{naab2014} showed that the detailed formation history cannot be constrained from the spin parameter alone, when combined with the high-order kinematic signatures, different merger scenarios can be distinguished. High-order kinematic signatures are defined as the deviations from a Gaussian line of sight velocity distribution  (LOSVD). The skewness and kurtosis are parameterized with Gauss-Hermite polynomials $h_3$ and $h_4$, respectively 
\citep{vandermarel1993,gerhard1993}. 

The classical interpretation is that in early-type galaxies the presence of a stellar disk leads to asymmetric line profiles \citep[$h_3$;][]{gerhard1993,vandermarel1993}. For fast-rotating galaxies a strong anti-correlation has been observed between the high-order Gauss-Hermite moment $h_3$ and \vs\, \citep{bender1994}, which originates mostly from stars on z-tube orbits \citep{rottgers2014}. Non-rotating galaxies often show flat-topped or peaked line profiles ($h_4$), which is associated with radial anisotropy if $h_4$ is positive, or tangential anisotropy if $h_4$ is decreased \citep{gerhard1993,vandermarel1993,gerhard1998,thomas2007}. Furthermore, positive values for $h_4$ are also found when the LOSVD traces regions with significantly different circular velocities \citep{gerhard1993}. More complex $h_3$ and $h_4$ signatures can also arise from the presence of a bar \citep{bureau2005} or from disk regrowth in bulges \citep{naab2001}.
 
Subsequently, the strong anti-correlation between $h_3$ and \vs\, was also seen in simulations 
\citep{bendo2000,jesseit2005,naab2006c,gonzalezgarcia2006,jesseit2007,hoffman2009,hoffman2010}, revealing that the presence of a dissipational component changes the asymmetry of the LOSVD towards steep leading wings. \citet{naab2014} find, however, that fast rotators with a gas-rich merger history show the anti-correlation between $h_3$ and \vs, whereas fast rotators with a gas-poor merger history do not \citep[see also ][]{naab2001}.

To better understand the assembly and merger history of galaxies, we thus have to study the high-order kinematic features. \citet{krajnovic2006, krajnovic2011} explore the high-order kinematic features in SAURON and \at\, galaxies. By first separating galaxies into having regular and non-regular rotational velocity fields, as based on the kinematic asymmetry, they find distinct high-order kinematic features between the two groups of galaxies. Regular rotators and barred galaxies have a degree of correlation (from the bar) combined with anti-correlation (from the disk) of $h_3$ and \vs, whereas non-regular rotators show no correlation between $h_3$ or $h_4$ and \vs. No fast-rotating galaxies without an $h_3$-\vs\, anti-correlation were identified, either because the high-order signatures were stacked and analyzed together, or because no such galaxies were present in the \at\, sample. Therefore, in order to test the predictions by \citet{naab2014}, we still need to analyze a larger sample of galaxies, and classify the kinematic signatures for each galaxy individually.

The introduction of new multi-object integral field spectrographs (IFS) such as SAMI \citep[Sydney-AAO Multi-object Integral field spectrograph; ][]{croom2012} now makes it possible to analyze the high-order kinematic features for a large ($N>1000$) number of galaxies with a broad range in mass and environment. The SAMI Galaxy Survey \citep{bryant2015} will observe $\sim3600$ galaxies by employing the revolutionary new imaging fibre bundles, or \textit{hexabundles} \citep{blandhawthorn2011,bryant2011,bryant2012a,bryant2014}. The survey is set up to have a spectral resolution of $R\sim1700$ in the blue (3700-5700\AA), and $R\sim4500$ in the red (6300-7400\AA), which we show is sufficient to measure $h_3$ and $h_4$ down to $\sigma> 75$\kms. Other large IFS surveys, such as the CALIFA Survey \citep[N $\sim600$; ][]{sanchez2012}, and the SDSS-IV MaNGA Survey \citep[Sloan Digital Sky Survey Data; Mapping Nearby Galaxies at APO; N $\sim10000$; ][]{bundy2015}, also have the capability to measure high-order stellar kinematics.

In this paper we present our methods for measuring the stellar kinematic parameters in the SAMI Galaxy Survey, with the main goal to explore the different classes of high-order ($h_3$, $h_4$) kinematic signatures that galaxies exhibit. We also investigate which uncertainties arise due to our data quality and from the different assumptions that are made. Our second goal is to link the observed high-order stellar kinematic moments to those predicted by the cosmological hydrodynamical simulation to get an insight on the assembly history of galaxies.

The paper is organized as follows. Section \ref{sec:data} describes the SAMI Galaxy Survey data in more detail. In Section \ref{sec:msk}, we describe our method for extracting the stellar kinematics. The stellar kinematic measurements are used for measuring the kinematic asymmetry and $\lambda_{r_e}$, and we define a sample of regular, non-regular, slow and fast rotators in Section \ref{sec:dp}. The high-order kinematic features are explored for the sample as a whole and for individual galaxies in Section \ref{sec:hok}. We discuss the implications of this work in Section \ref{sec:discussion}, and summarize and conclude in Section \ref{sec:conclusions}. Finally, the optimization of our method is described in Appendix \ref{sec:app}. Throughout the paper we assume a $\Lambda$CDM cosmology with $\Omega_\mathrm{m}$=0.3, $\Omega_{\Lambda}=0.7$, and $H_{0}=70$ km s$^{-1}$ Mpc$^{-1}$. All broadband data are given in the AB-based photometric system.

\section{Data}
\label{sec:data}


\begin{figure}
\epsscale{1.15}
\plotone{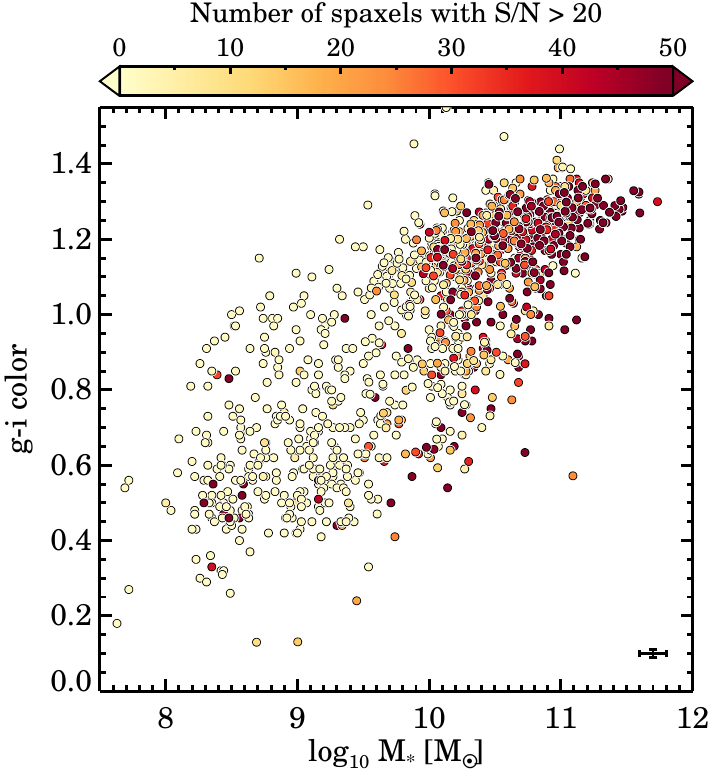}
\caption{Color ($g-i$) versus stellar mass for the full v0.9.1 SAMI sample. The data are color-coded by the number of spaxels in the galaxy with S/N $>$ 20 \AA$^{-1}$\, which is the minimum required for an accurate measurement of $h_3$ and $h_4$ (see Section \ref{subsec:qc}). At low stellar masses ($M_* < 10^{10}$\msun), we find that almost all galaxies have fewer than 20 high-S/N spaxels, whereas at high stellar mass ($M_* > 10^{11}$\msun) almost all galaxies have more than 40 high-S/N spaxels. The median uncertainty is shown in the bottom-right corner.
}
\label{fig:fig1}
\end{figure}
%

\subsection{SAMI Galaxy Survey}
\label{sec:sgs}

The SAMI instrument and Galaxy Survey is described in detail in \citet{croom2012} and \citet{bryant2015}. Here, we briefly outline the main characteristics of the instrument, sample selection, and global galaxy parameters.

SAMI is a multi-object integral field spectrograph with 13 IFUs deployable over a 1$^\circ$ diameter field of view, mounted at the prime focus of the 3.9m Anglo Australian Telescope (AAT). Each IFU, or hexabundle, is made out of 61 individual fibers. Even though the hexabundles have a high filling factor of 75\%, observations were carried out using a dither pattern to create data cubes with 0\farcs5 spaxel size \citep{sharp2015,allen2015}. The fibers are 1\farcs6 in size, and combine into a hexabundle which covers a $\sim15^{\prime\prime}$ diameter on the sky. All 819 fibers, including 26 individual sky fibers, are fed into the AAOmega dual-beamed spectrograph \citep{saunders2004, smith2004, sharp2006}. For the SAMI Galaxy Survey, the 580V grating is used in the blue arm of the spectrograph, which results in a resolution of R $\sim1700$ with wavelength coverage of 3700-5700\AA. In the red arm, the higher resolution grating 1000R is used, which gives an R $\sim4500$ over the range 6300-7400\AA.

The SAMI Galaxy Survey aims to observe $\sim3600$ galaxies. The redshift range of the survey, $0.004<z<0.095$, was chosen such that Mgb $\lambda$5177 and [SII] $\lambda 6716,6731$ fall within the wavelength range of the blue and red arm, respectively. This limited redshift range results in spatial resolutions of 0.1 kpc per fiber at $z=0.004$ to 2.7 kpc at $z=0.095$. The survey has four volume-limited galaxy samples derived from cuts in stellar mass in the Galaxy and Mass Assembly (GAMA) G09, G12 and G15 regions \citep{driver2011}. GAMA is a large campaign that combines large multi-wavelength photometric data with a spectroscopic survey of $\sim$300,000 galaxies carried out using the AAOmega multi-object spectrograph on the AAT. Furthermore, targets were selected from eight high-density cluster regions sampled within radius $R_{200}$ with the same stellar mass limit as for the GAMA fields \citep[Owers et al. in prep; ][]{bryant2015}.
The aim of the SAMI galaxy survey selection is to cover a broad range in galaxy stellar mass (M$_* = 10^{8}-10^{12}$\msun) and galaxy environment (field, group, and clusters).

\subsection{Ancillary Data}

For galaxies in the GAMA fields, the aperture matched $g$ and $i$ photometry from the GAMA catalog  \citep{hill2011,liske2015} are used to derive $g-i$ colors, which were measured from reprocessed SDSS Data Release Seven \citep{york2000, kelvin2012}. For the cluster environment, we use photometry from the SDSS \citep{york2000} and VLT Survey Telescope ATLAS imaging data \citep[Owers et al. in prep; ][]{shanks2013}. Stellar masses are derived from the rest-frame i-band absolute magnitude and $g-i$ color by using the color-mass relation following the method of \citet{taylor2011}. For the stellar mass estimates, a \citet{chabrier2003} stellar initial mass function (IMF) and exponentially declining star formation histories are assumed. For more details see \citet{bryant2015}. 

In Figure \ref{fig:fig1}, we show the $g-i$ color versus stellar mass, color-coded by the number of \textit{spaxels} (spatial pixels) with stellar continuum signal-to-noise ratio (S/N) $ > 20$ \AA$^{-1}$. SAMI galaxies with low stellar mass ($M_* < 10^{10}$\msun) rarely have more than 10 individual un-binned spaxels with relatively high-S/N. Galaxies around $M_* \sim 10^{10}$\msun\, have on average 10-40 good quality spaxels, whereas at higher stellar masses ($M_* > 10^{11}$\msun) almost all galaxies have more than 40 good quality spaxels.

Galaxy sizes are derived from GAMA-SDSS \citep{driver2011}, SDSS \citep{york2000}, and VST \citep[Owers et al. in prep;][]{shanks2013} imaging. The  Multi-Gaussian Expansion \citep[MGE;][]{emsellem1994} technique and the code from \citet{cappellari2002} is used for measuring effective radii, ellipticity, and positions angles. For more details, we refer to D'Eugenio et al. (2016; in prep). 
Throughout the paper, \re\, is defined as the semi-major axis effective radius, and \recirc\, as the circularized effective radius, where \recirc\, = \re $\sqrt{q}$, where $q$ is the axis ratio $b/a = 1-\epsilon$. The ellipticity used in this paper, $\epsilon_{\rm{e}}$, is the average ellipticity of the galaxy within one effective radius measured from the best-fitting MGE model, not the global luminosity-weighted ellipticity from the MGE fit.
%

%

\begin{deluxetable*}{c c c c c c c c }[!th]
\tabletypesize{}
\tabletypesize{\scriptsize} 
\tablecolumns{9} 
\tablewidth{0pt} 
\tablecaption{SAMI Spectral Resolution in Blue and Red}
\tablehead{
\colhead{Arm} & \colhead{$\lambda$-range [\AA]} & \colhead{$\lambda$-central [\AA]} & \colhead{FWHM [\AA]} & \colhead{$\sigma$ [\AA]} & 
\colhead{$R_{\lambda-\rm{central}}$} & \colhead{$\Delta v$ [\kms]} & \colhead{$\Delta \sigma$ [\kms]}} \\
\startdata
Blue & 3750-5750 & 4800 & $2.65_{-0.09}^{+0.12}$ & 1.13 & 1812 & 165.5 & 70.3 \\
Red & 6300-7400 & 6850 & $1.61_{-0.05}^{+0.07}$ & 0.68 & 4263 & 70.3 & 29.9 \\
\enddata
\tablecomments{Line-spread-function parameters derived from unblended CuAr arc lines. For both the blue and red arm we give the wavelength range ($\lambda$-range), the central wavelength ($\lambda$-central), the median FWHM of the best-fit Gaussian to the spectral instrumental LSF in \AA, the standard deviation of this Gauss in \AA, the spectral resolution at $\lambda$-central ($R_{\lambda-\rm{central}}$), the velocity resolution (FWHM) in \kms\, ($\Delta v$), and the dispersion resolution ($1\sigma$) in \kms\, ($\Delta \sigma$).}
\label{tbl:tbl1}
\end{deluxetable*}


\section{Stellar Kinematics and Sample Selection}
\label{sec:msk}

In this section we describe how the stellar kinematic measurements are derived from the SAMI data by using the penalized pixel fitting code (\textsc{pPXF}; Cappellari \& Emsellem 2004). The SAMI kinematic pipeline was run on the 1404 galaxy cubes that make up the SAMI Galaxy Survey internal v0.9.1 data release. This number includes 24 repeat galaxy observations (see Appendix \ref{subsec:app_ro}). In total, the stellar kinematic parameters from approximately \mbox{$400,000$} spectra are extracted. The stellar kinematic measurements will be released in 2017 as part of the second SAMI Galaxy Survey Data Release.

\subsection{Spectral Resolution}
\label{subsec:sr}

SAMI is setup to have a resolution of R$\sim$1700 in the blue 3700-5700\AA, and R$\sim$4500 in the red 6300-7400 \AA\, \citep{croom2012,bryant2015}. Given the importance of the instrumental resolution and spectral profile for the stellar kinematic measurements, here we re-derive the full-width-at-half-maximum (FWHM) of the spectral instrumental line-spread-function (LSF) of the extracted spectra, and test whether the instrumental profile is well approximated by a Gaussian function. We use 24 unsaturated, unblended CuAr arc lines in the blue arm, and 12 lines in the red arm, from 16 frames between $05/03/2013$ and $17/08/2015$, for all 819 fibers.

Two functions are used for fitting the arc lines: a Gaussian and a Gaussian with skewness and kurtosis, as parameterized with Gauss-Hermite polynomials $h_3$ and $h_4$, respectively \citep{vandermarel1993,gerhard1993}.
There is an excellent agreement between the median resolution from the Gaussian and high-order moment fit. The median value for $h_3$ and $h_4$ is -0.01 in the blue arm and 0.00 in the red arm, with a 1$\sigma$ spread of 0.016. These results indicate that SAMI's instrumental profile is well approximated by a single Gaussian function. 

Key resolution quantities for SAMI are given in Table \ref{tbl:tbl1}. In the blue arm, we find a median resolution of: FWHM$_{\rm{blue}}= 2.65$ \AA, and in the red arm of: FWHM$_{\rm{red}} = 1.61$ \AA. The fiber-to-fiber FWHM variation is 0.05\AA\, (1-$\sigma$ scatter) in the blue, and 0.03\AA\, in the red. Over a period of two years, we find FWHM variations of 0.04\AA\, in the blue arm, and 0.03\AA\, in the red arm. The FWHM decreases with increasing wavelength in the red arm by 0.15\AA, but we do not find a wavelength dependence in the blue. 

\subsubsection{Combining Blue and Red Arm}
\label{subsec:cbr}

Both the blue and red spectra are used for fitting the stellar kinematics. While there a fewer strong features in the red spectra than in the blue, and H$\alpha$ is masked, adding the red arm helps to constrain the templates (see Section \ref{subsec:otc} and \ref{subsec:app_tc}). Before the blue and red spectra are combined, we first convolve the red spectra to the instrumental resolution in the blue. For the convolution a Gaussian kernel is used, with an FWHM set by the square root of the quadratic difference of the red and blue FWHM. We assume a constant resolution as a function of wavelength as given by the median value found in Section \ref{subsec:sr}. The red spectrum is interpolated onto a grid with the same wavelength spacing as in the blue, and then combined with the blue spectrum with a gap in between. We note that the resolution-degraded red spectra are only used for the stellar kinematic measurement; emission line studies with SAMI use the native red resolution \citep[see e.g.,][]{ho2014}. We de-redshift the spectra to a rest-frame wavelength grid by dividing the observed wavelengths by  $(1+z_{\rm{spec}})$ of the galaxy. All galaxies are fitted at their native redshift-corrected SAMI resolution, i.e., the spectra are not homogenized to a common resolution after de-redshifting. The spectrum is then rebinned onto a logarithmic wavelength scale with constant velocity spacing, using the code \textsc{log\_rebin} provided with the \textsc{pPXF} package. The adopted velocity scale is 57.9 \kms.

\subsubsection{Annular Spectra Extraction}
\label{subsec:ase}

Annular binned spectra are used for deriving local optimal templates as opposed to deriving an optimal template for each individual spaxel (see Appendix \ref{subsec:app_tc}). Individual spaxels generally do not meet our S/N requirement of 25 \AA$^{-1}$, which is needed to derive a reliable optimal template. Annular binned spectra reach our S/N requirement more easily, while also accounting for strong stellar population gradients in late-type galaxies.

For each galaxy, we define five equally-spaced, elliptical annuli, that follow the light distribution of the galaxy (see for example Figure \ref{fig:fig24}). In each annulus, we derive the mean flux with an optimal inverse-variance weighting scheme to increase the S/N. In some cases the individual annular spectra in the five bins do not meet our S/N requirement. When this is the case, the annular bins are combined from the outside inwards until the target S/N of 25 \AA$^{-1}$\, is obtained. For each galaxy we obtain five annular binned spectra if the average S/N of the galaxy is relatively high, and only one annular binned spectrum if the average S/N is relatively low.

\subsection{Running \textsc{pPXF}}
\label{sec:rp}

We run \textsc{pPXF} in two different modes, producing two final data products. The first data product consists of fits using a Gaussian LOSVD, i.e., fitting only the stellar velocity and stellar velocity dispersion (hereafter \vmt\, and \smt). The velocity and dispersion maps from the Gaussian LOSVD fits are used for measuring $\lambda_r$. 

In the second mode, a truncated Gauss-Hermite series \citep{vandermarel1993,gerhard1993} is used to parameterize the LOSVD. We fit four kinematic moments: \vmf, \smf, $h_3$ and $h_4$, where $h_3$ and $h_4$ are related to the skewness and kurtosis of the LOSVD.


\subsubsection{Additive Polynomials}
\label{subsec:ap}

We use an additive Legendre polynomial to correct for possible mismatches between the stellar continuum emission from the observed galaxy spectrum and the template due to small errors in the flux calibration and minor template mismatch effects. If no such correction was applied, \textsc{pPXF} could try to correct for this discrepancy by changing the optimal template and/or fitted LOSVD parameters. After experimenting with different order polynomials, we find that for the blue and red spectrum combined, a 12th order additive Legendre polynomial is required to remove residuals from small errors in the flux calibration (see Appendix \ref{subsec:app_ap}).

\subsubsection{Noise Estimate}
\label{subsec:ne}

A good estimate of the noise is crucial for \textsc{pPXF} to accurately measure the LOSVD if there is a significant spread in S/N along the spectrum, as is the case for SAMI. Whereas the original noise spectrum, as derived from the variance cubes, is a good measure for the noise of individual spaxels, we found small amplitude offsets of the noise spectrum for the annular bins as compared to fitting residuals. In order to get an accurate scaling measure of the noise spectrum, we therefore use the residual of the galaxy spectrum minus the best-fitting template. This involves running \textsc{pPXF} multiple times.

First, \textsc{pPXF} is run with equal weights at every wavelength. From the residual of the fit we calculate the standard deviation, that is then compared to the mean of the original noise spectrum. The difference between the two values is used to scale the original noise spectrum. 

There is a significant improvement in the stellar kinematic maps when we use the scaled noise spectrum for weighting, compared to simply using equal weights at every wavelength. When using equal weights in the fits, we find on average significantly higher values for $h_4$, which disappear when the scaled noise spectrum is used. This is likely due to the strongly varying response of the detector and spectrograph at the wavelength edges of the blue and red arm, which translates into varying noise as a function of wavelength.

\subsubsection{Removing Emission Lines and Outliers}

We remove emission lines and outliers by using a combination of initial masking and the CLEAN parameter in \textsc{pPXF}. Masking is always done around the following lines: [OII], H$\delta$, H$\gamma$, H$\beta$, [OIII], [OI], H$\alpha$, [NII], and [SII], even if no emission lines are present after a cursory inspection of the data. While the H$\beta$ and H$\alpha$ absorption lines could potentially be used for measuring the stellar kinematics, weak emission is often present and could bias the LOSVD measurement if not properly masked.

With the new noise spectrum from the first \textsc{pPXF} run (Section \ref{subsec:ne}), \textsc{pPXF} is run a second time with the CLEAN parameter set. \textsc{pPXF}'s CLEAN function performs a three-sigma outlier clipping based on the residual between the best-fit template and the observed galaxy spectrum. For the annular binned spectra in our test sample (Appendix \ref{sec:app}), we visually confirmed that CLEAN removes all emission lines and outliers that could be visibly classified, while keeping the pixels in the spectrum that are not affected. 

\subsubsection{Optimal Template Construction}
\label{subsec:otc}

We derive optimal templates for each annular bin as described in Section \ref{subsec:ase}. For deriving each optimal template, \textsc{pPXF} is run three times as described above. The first run is for getting a precise measure of the noise scaling from the residual from the fit. The second run, with the new estimate for the noise spectrum, is for the masking of emission lines and outliers with the optimal template from the first run. \textsc{pPXF} is used a third time to derive the optimal template which will be used for individual spaxel fitting. Our default library for deriving optimal templates is the MILES stellar library \citep{sanchezblazquez2006}. This library consists of 985 stars spanning a large range in atmospheric parameters. We convolve the template spectra from their original resolution of 2.50 \AA\, \citep{falconbarroso2011}, to the resolution of the de-redshifted SAMI spectra used for fitting. Thus, all galaxies are fitted at their native redshift-corrected SAMI resolution, and not homogenized to a common resolution after de-redshifting. For the convolution a Gaussian kernel is used, where the FWHM of the Gaussian is set by the square root of the quadratic difference of the SAMI and MILES FWHM. In Appendix \ref{subsec:app_tc} we tested the impact of using the MILES stellar library versus templates constructed from stellar population synthesis (SPS) models. The MILES stellar library was found to have the best overall fit quality, from the residuals of spectrum minus the best-fitting template.

\begin{figure*}[!ht]
\epsscale{1.15}
\plotone{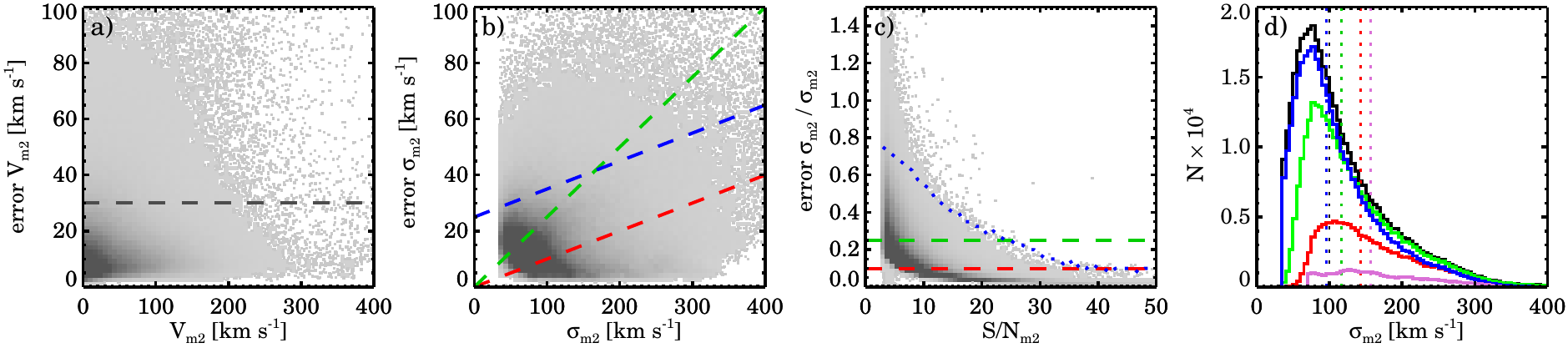}
\caption{Quality cuts on the stellar kinematic data from the stellar velocity and velocity dispersion from second-order moment fits. Panel a): uncertainty on the velocity versus velocity. In gray, we show the density of all spaxels in our sample with S/N $>3$ and $\sigma>35$\kms. We show Q$_1$, i.e., where $V_{\rm{error}} < 30$ \kms, as the dashed grey line.
Panel b): uncertainty on the velocity dispersion versus velocity dispersion. Different lines show three different quality cuts. Q$_{\rm{red}}$: 10\% uncertainty,  Q$_{\rm{green}}$: 25\% uncertainty, and Q$_{\rm{blue}}$: $\sigma_{\rm{error}} < \sigma *0.1 + 25$\kms. At low velocity dispersion ($<100$\kms) Q$_{\rm{red}}$ and Q$_{\rm{green}}$ remove a relatively large fraction of the spaxels, which would bias our sample towards higher median velocity dispersions.
Panel c): ratio of the uncertainty on the velocity dispersion and the velocity dispersion versus the signal-to-noise. The blue dotted line shows Q$_{\rm{blue}}$. With this quality cut, the uncertainty on the velocity dispersion at S/N $<$ 20 \AA$^{-1}$\, is always less than 75\% with a median of 12.6\%. For S/N $>$ 20 \AA$^{-1}$\, the median uncertainty is 2.6\%. 
Panel d): distribution of the measured velocity dispersions in all spaxels after different quality cuts. The black line shows the distribution for all spaxels with S/N $>3$ and $\sigma>35$\kms, whereas the red, green, and blue line show the samples with the quality cuts from the left panels. We adopt Q$_{\rm{blue}}$ as the quality cut for the velocity dispersion measurements, hereafter referred to as Q$_2$. In purple, the stricter quality cut Q$_3$ is shown ($\sigma>70$\kms; S/N $>$ 20 \AA$^{-1}$), which is required for reliable measurements of $h_3$ and $h_4$ (see Section \ref{subsec:qc}). The dotted vertical lines show the median of each distribution. The median of the sample with the blue quality cut is closest to the median of all spaxels.
}
\label{fig:fig2}
\end{figure*}
%


\subsubsection{Full Spaxel Fitting}
\label{subsec:fsf}

After the optimal template is constructed for each annular bin, we run \textsc{pPXF} three times on each galaxy spaxel. \textsc{pPXF} is allowed to use the optimal templates from the annular bin in which the spaxel lives, as well as the optimal templates from neighboring annular bins. This removes any possible remaining template mismatch for spaxels close to the edges of the annular bins, that could arise from radial stellar population gradients. For a high S/N galaxy that has the maximum of five annular optimal templates, we provide \textsc{pPXF} with two annular-optimal templates for the spaxels within the central annular bin, i.e., the templates derived from the central and second annular bin. For spaxels in the second annular bin, \textsc{pPXF} is provided with three annular-optimal templates, as derived from the central, the second, and the third annular bin, and so on. If a galaxy has very low S/N overall, and had only one annular bin from which the optimal template was derived, this template is fit to all spaxels. \textsc{pPXF} is allowed to weight and combine the different annular-optimal templates.

For each spaxel, we estimate the uncertainties on the LOSVD parameters from 150 simulated spectra. We construct these spectra in the following way. The best-fit template is first subtracted from the spectrum. The residuals are then randomly rearranged in wavelength space within eight wavelength sectors. We use eight sectors to ensure that residuals from noisier regions in the spectrum (e.g., blue arm) are not mixed with residuals from less noisy regions (e.g., red arm). The re-shuffled residuals are added to the best-fit template. We refit this simulated spectrum with \textsc{pPXF}, and the process is then repeated 150 times using the same number of templates as for the actual spaxel fit. Our quoted uncertainties are the standard deviations of the resulting simulated distributions.

\subsubsection{Quality Cuts}
\label{subsec:qc}

The quality of each stellar kinematic fit depends on a number of factors: most importantly the S/N of the spectra, but also on the age of the stellar population, if the velocity dispersion is close to, or lower than, the instrumental resolution, and the presence of strong emission lines. If we were to apply a strong cut in mean S/N = 40 \AA$^{-1}$ as in, for example, the \at\, survey, a large fraction of the galaxies in the SAMI sample would be excluded from the sample and the spatial coverage for the remaining individual galaxies would decrease as well, or, if we Voronoi-bin the data, the spatial resolution would decrease.

\citet{fogarty2015} used Monte Carlo (MC) simulations to show that the \smt, for SAMI spectra with S/N=5, can be recovered with an accuracy of $\pm20$\kms\, at \smt=50\kms. Therefore, instead of setting a strict limit on the minimal S/N, we explore a quality cut based on the velocity dispersion and its uncertainty that keeps the maximum number of spaxels without including unreliable measurements.

Figure \ref{fig:fig2}a and \ref{fig:fig2}b show the uncertainty on the stellar velocity and velocity dispersion for all \mbox{$\sim400,000$} spaxels with S/N $>3$ and \smt $>$ FWHM$_{\rm{instr}}/2 \sim 35$\kms. We exclude all spaxels with \smt $<35$\kms, where the systematic uncertainties, due to the instrumental spectral resolution, start to dominate over the random uncertainties (Appendix \ref{subsec:app_pb}, see also \citeauthor{fogarty2015} \citeyear{fogarty2015}). For the SAMI velocity measurements, most spaxels have uncertainties less than 20\kms\, as seen from the higher density of spaxels in the bottom left corner of Figure \ref{fig:fig2}a. In order to keep the majority of spaxels without sacrificing the quality of our results, we exclude all spaxels where the maximum velocity uncertainty $>30$\kms\,  (hereafter Q$_1$), as indicated by the gray dashed line in Figure \ref{fig:fig2}a. 

For the velocity dispersion, three different quality cuts are tested (Figure \ref{fig:fig2}b). Q$_{\rm{red}}$: a conservative selection in which the uncertainty on the velocity dispersion has to be less than 10\% (red line), Q$_{\rm{green}}$: a less strict quality cut of 25\% (green line), and Q$_{\rm{blue}}$: a relative quality cut where the uncertainty on the velocity dispersion has to be less than 10\% plus an additional 25\kms\, (blue line). We find that Q$_{\rm{red}}$ and Q$_{\rm{green}}$ remove a relatively large number of spaxels when the dispersion is low, which is not surprising given that 10\% (25\%) of 50\kms\, is only 5\kms\, (12.5\kms). A fractional quality cut based on the velocity dispersion therefore biases our sample towards higher velocity dispersions. The Q$_{\rm{blue}}$ cut softens the limit on the relative uncertainty for low velocity dispersion. This way, we keep a large fraction of the low velocity dispersion spaxels, while keeping a strict quality cut for the higher velocity dispersions. A total of \mbox{347,538} spaxels meet our selection criteria Q$_{\rm{blue}}$ out of the initial \mbox{408,666} SAMI spaxels with S/N$ > 3$ \AA$^{-1}$ and \smt\,$>$35\kms

With Q$_{\rm{blue}}$, the ratio of the velocity dispersion uncertainty and velocity dispersion is always less than 75\% (Figure \ref{fig:fig2}c). Furthermore, in Figure \ref{fig:fig2}d, the median velocity dispersion of the sample with and without the Q$_{\rm{blue}}$ cut are in good agreement (black versus blue distribution), whereas Q$_{\rm{red}}$ and Q$_{\rm{green}}$ bias the sample towards higher velocity dispersions. Therefore, we adopt Q$_{\rm{blue}}$ as the quality cut for the velocity dispersion measurements, which we hereafter refer to as Q$_2:\sigma_{\rm{error}} < \sigma *0.1 + 25$\kms.

Finally, in Appendix \ref{subsec:app_pb} we show that a reliable estimate of $h_3$ and $h_4$ requires an additional quality cut of S/N $>20$ \AA$^{-1}$ and $\sigma>70$\kms, which we define as Q$_3$. This stricter quality cut Q$_3$ is shown in Figure \ref{fig:fig2}d as the purple line. From this figure it is clear that only a relatively small number \mbox{($\sim10\%$)} of spaxels pass Q$_3$: \mbox{35,798} versus \mbox{347,538} (Q$_2$).

\subsubsection{High-Order Moments}
\label{subsec:ihom}

We parameterize the skewness and kurtosis of the LOSVD, i.e., deviations from Gaussian line profiles, by using Gauss-Hermite Polynomials, with $h_3$ being related to the skewness and $h_4$ related to the kurtosis 
\citep{vandermarel1993,gerhard1993}. Gauss-Hermite polynomials are used as opposed to using, for example, a decomposition into a double Gaussian for two reasons. First, the uncertainties in the six parameters in a two-Gaussian decomposition are highly correlated, whereas Gauss-Hermite polynomials are orthogonal, reducing the degeneracy in the LOSVD fit \citep{vandermarel1993}. Secondly, a two-Gaussian decomposition \textit{a priori} assumes that two kinematic distinct components are present, which is not always the case.

\textsc{pPXF} was designed to employ a maximum penalizing likelihood, i.e., forcing a solution to a Gaussian LOSVD, if the high-order moments are unconstrained by the data \citep{cappellari2004}. Following the code documentation, we derive an optimal penalizing bias value for SAMI spectra \citep[see also][]{cappellari2011a}, as the automatic penalizing bias in \textsc{pPXF} is often too strong. We define the \textit{ideal} bias as one that reduces the scatter in the velocity dispersion, $h_3$ and $h_4$, without creating a systematic offset in the velocity and velocity dispersion. By running a large ensemble of Monte Carlo simulations, a simple analytic expression was obtained for the ideal penalizing bias for SAMI spectra as a function of S/N (see Appendix \ref{subsec:app_pb}):
\begin{equation}
\label{eq:eq1}
Bias = 0.0136 + 0.0023 \rm{(S/N)} - 0.000009 \rm{(S/N)}^2.
\end{equation}
For every spaxel, this optimal bias setting is then fed into \textsc{pPXF}.

\begin{figure}
\epsscale{1.15}
\plotone{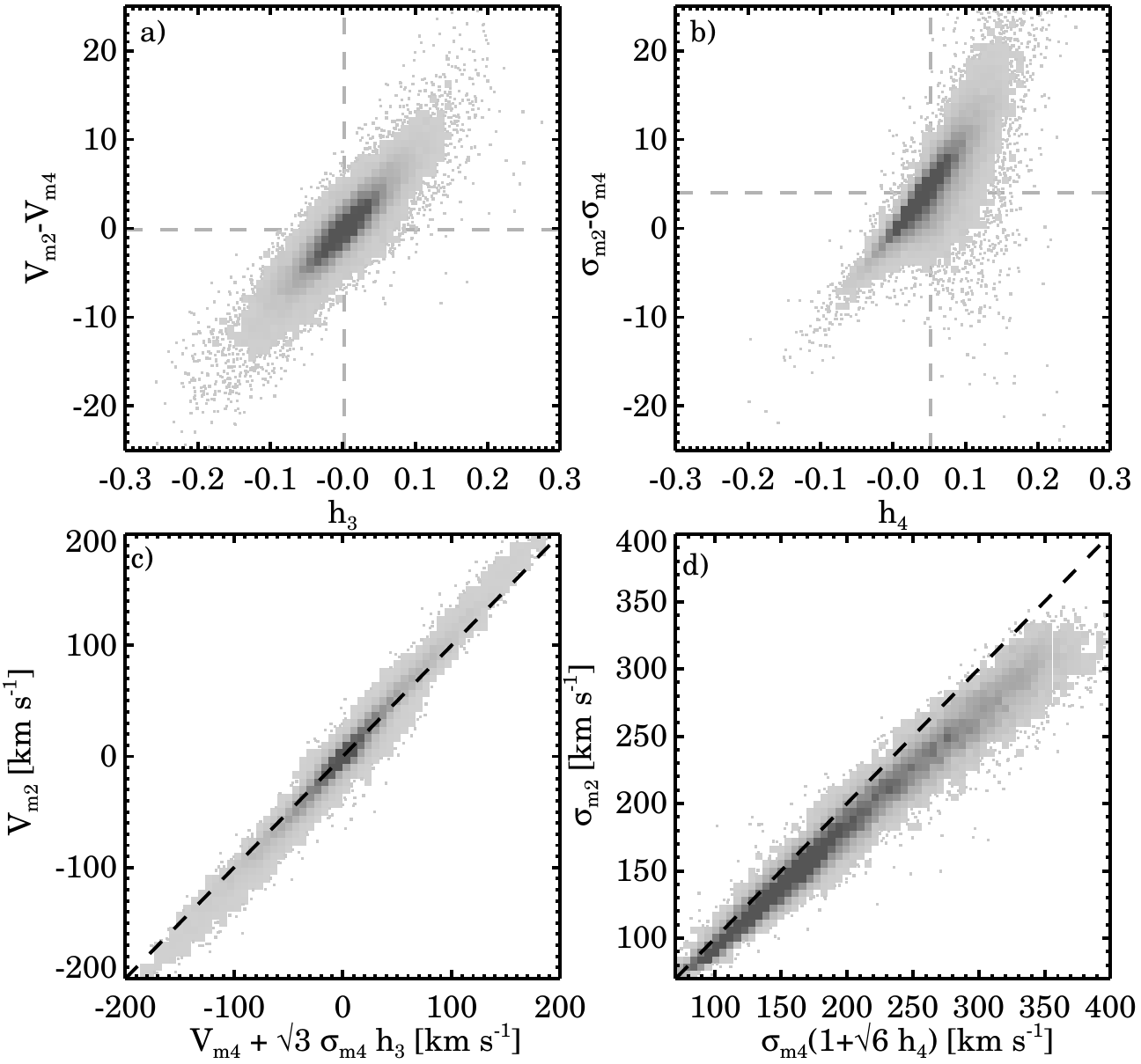}
\caption{Relation between the measured LOSVD parameters of second and fourth order moment fits. The gray squares show the density of all spaxels with reliable measurements (Q$_3$), and the gray dashed lines show the median along each axes. In panel a) we find a strong correlation between \vmt~-~\vmf\, and $h_3$ as expected, whereas panel b) shows a strong correlation between \smt~-~\smf\, and $h_4$. We find more scatter in the \smt~-~\smf\, and $h_4$ relation towards the positive $h_4$ side. In panel c) \& d) we show the best-estimates of the true LOSVD moments  versus the best-fitted Gaussian moments (Equation \ref{eq:eq2}-\ref{eq:eq3}). The \vmt\, agrees well with $\tilde{V}$, with a small offset of 10\kms\, for $|$\vmt$|$\, $>$ 100\kms. We find an offset in \smt\, versus $\tilde{\sigma}$, where the best-estimate of the true $\sigma$ is larger than \smt\, ($\sim10 $\kms\, at \smt\, = 200 \kms, $\sim25 $\kms\, at \smt\, = 300 \kms).
}
\label{fig:fig3}
\end{figure}

If the LOSVD is a Gaussian, the $m=2$ and $m=4$ parameters are the same. In the case of a non-Gaussian LOSVD, due to the Gauss-Hermite polynomial parameterization of the skewness and kurtosis of the LOSVD fit, the velocity \vmf\, and velocity dispersion \smf\, deviate from the velocity \vmt\, and velocity dispersion \smt\, of a pure Gaussian LOSVD. Furthermore, the best-estimates of the true LOSVD moments can be calculated by \citep[Eq. 18,][]{vandermarel1993}:
\begin{equation}
\label{eq:eq2}
\tilde{V} \approx V_{\rm{m4}} + \sqrt{3}\,\sigma_{\rm{m4}}\,h_3
\end{equation}
\begin{equation}
\label{eq:eq3}
\tilde{\sigma} \approx \sigma_{\rm{m4}}\,(1+\sqrt{6}\,h_4).
\end{equation}
In Figure \ref{fig:fig3}, we show the difference between \vmt\, and \vmf, \smt\, and \smf\, versus $h_3$ and $h_4$, and compare these values to the best-estimates of the true moments $\tilde{V}$ and $\tilde{\sigma}$. Only spaxels that meet selection criteria Q$_3$ are shown. As expected, there is a strong correlation between \vmt~-~\vmf\, and $h_3$: if the LOSVD is highly skewed, the velocity difference between the second and fourth order moments fits will be larger ($h_3=0.1$, $\Delta $(\vmt~-~\vmf)$ \simeq 10$\kms). For the velocity dispersion we find a strong correlation between \smt~-~\smf\, and $h_4$. If the line has strong kurtosis, then the velocity dispersion difference between the second and fourth order moments fits will be larger ($h_4=0.1$, $\Delta $(\smt~-~\smf)$ \simeq 15$\kms).

%

\begin{figure}
\epsscale{1.15}
\plotone{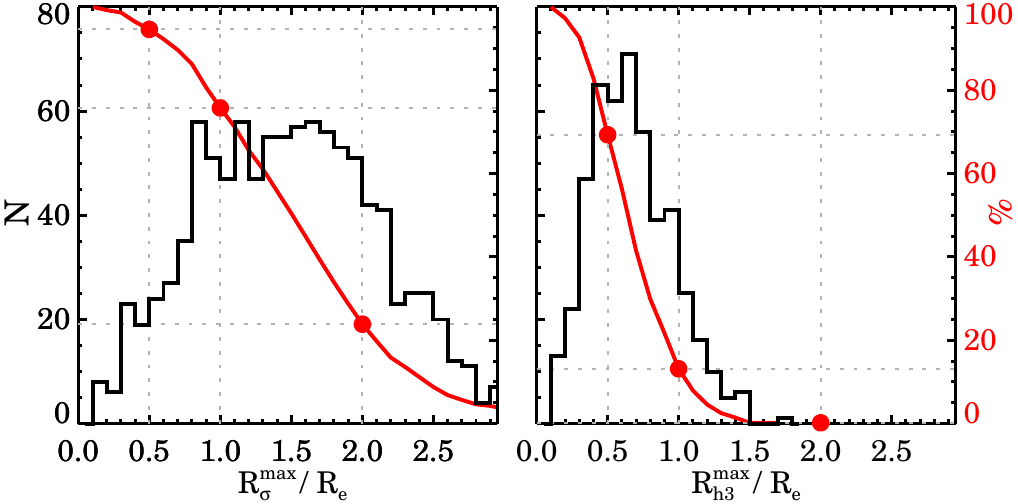}
\caption{Comparison of the ratio between the maximum kinematic aperture radius and the effective semi-major axis for SAMI galaxies. We show the distribution of \rmaxs/\re\, in the left panel, and \rmaxh/\re\, in the right panel. The red lines show the cumulative fraction of galaxies with  \rmaxs/\re\, $>$ value (scale on right-axis). In total, 1035 galaxies meet our selection criteria for the second-order moment fits (Q$_2$ and \re\, and \rmaxs\, $>$ HWHM$_{\rm{PSF}}$), and 479 galaxies for the high-order quality cuts (Q$_3$ and \re\, and \rmaxs\, $>$ HWHM$_{\rm{PSF}}$). For the stellar velocity distribution maps, we cover \rmaxs $>$ \re\, for 76\% of the galaxies, and 24\% meet \rmaxs $>$ 2\re. For the high-order moment fits, \rmaxh\, is greater than \ret\, for 69\% of the galaxies, and 13\% meet \rmaxh $>$ \re.\\}
\label{fig:fig4}
\end{figure}
%

The dashed lines in Figure \ref{fig:fig3} show the median values of the data. There is no systematic offset from zero in the relation between \vmt~-~\vmf\, and $h_3$. For \smt~-~\smf\, and $h_4$, however, we find a systematic offset towards lower values of \smf, and higher values of $h_4$. Moreover, there is more scatter in the \smt~-~\smf\, and $h_4$ relation towards the positive $h_4$ side. We investigated whether the positive $h_4$ values are the result of instrumental resolution, template mismatch, or different seeing conditions. However, no correlation of $h_4$ with $\sigma_{\rm{obs}}$, S/N, FWHM of the point-spread-function (PSF), host galaxy's stellar mass or color were detected, which excludes the aforementioned possible issues.

Figure \ref{fig:fig3}c-d shows the best-estimates of the true LOSVD moments $\tilde{V}$ and $\tilde{\sigma}$ versus the best-fitted Gaussian moments. We find that \vmt\, agrees well with $\tilde{V}$, albeit with a small offset of 10\kms\, for $|$\vmt$|$\, $>$ 100\kms. For \smt\, versus $\tilde{\sigma}$, there is an offset in the one-to-one relation, such that the best-estimate of the true $\sigma$ is larger than \smt. At \smt\, = 200 \kms\, there is an offset of $\sim10 $\kms\, that goes up to $\sim25 $\kms\, at \smt\, = 300 \kms). The offset can be largely explained by the overall positive $h_4$ that we measure. 

In \citet{cappellari2006} a similar test was performed, and they find that their results are consistent, within the errors, if Equation \ref{eq:eq3} is used. \citet{veale2017} also find positive $h_4$ values for 41 massive early-type galaxies in the MASSIVE survey, similar to the results presented here. This suggest that some physical effect could be responsible for our positive $h_4$ values, rather than residual template mismatch. However, given the current unknown nature of the positive $h_4$ values in our data we do not explore this matter further. Furthermore, we caveat that $\sigma_{\rm{obs}}$ is highly sensitive to the wings of the LOSVD, which are generally poorly constrained by the data. For this reason, it is unclear whether Equation \ref{eq:eq3} can provide a better estimate of the true velocity dispersion rather than $\sigma_{\rm{obs}}$ alone. Detailed N-Body simulations, with realistic LOSVDs, and accounting for template mismatch, are needed to demonstrate this.

\subsection{Sample Selection and Morphology}
\label{subsec:as}

We visually checked all 1380 unique SAMI kinematic maps and flagged 75 galaxies with irregular kinematic maps due to mergers or nearby objects that influence the stellar kinematics of the main object. These objects are excluded from further analysis in this paper.

For each galaxy, we calculate the maximum aperture out to which there are reliable data. \rmaxs\, and \rmaxh\, are defined as the semi-major axis of an ellipse where at least 75\% of the spaxels meet our velocity dispersion (Q$_2$) or $h_3$ quality criteria (Q$_3$) respectively. The axis ratio and position angle of the ellipse are obtained from the 2D MGE fits to the imaging data. A total of 270 galaxies have \re\, or \rmaxs\, less than the Half Width at Half Maximum of the PSF (HWHM$_{\rm{PSF}}$), and are therefore excluded from the sample. This brings the number of galaxies with usable stellar velocity and stellar velocity dispersion maps to 1035. 

Figure \ref{fig:fig4} shows the ratio of the maximum aperture radius and the effective semi-major radius for galaxies in the SAMI Galaxy Survey. The left panel shows the results for \rmaxs, the right panel for \rmaxh. In red we show the cumulative fraction of galaxies with \rmaxs/\re\, $>$ value (scale on right-axis). Out of the 1035 galaxies with usable $V$ and $\sigma$ data, 76\% ($N=784$) have \rmaxs $>$ \re, and 24\% ($N=247$) have \rmaxs $>$ 2\re. With the stricter quality cut Q$_3$ for the high-order moment fits, a total of 479 SAMI galaxies have \re\, and \rmaxh\, $>$ HWHM$_{\rm{PSF}}$. Of these 479 galaxies, 69\% ($N=332$) have \rmaxh $>$ \ret, and 13\% (N=63) have \rmaxh $>$ \re.

In order to measure reliable high-order signatures of galaxies within the SAMI galaxy survey, we apply another quality cut to our sample. We only consider galaxies that have enough high-quality spaxels (Q$_3$) to fill an area greater than the maximum seeing aperture. The maximum allowed seeing in our data is $3^{\prime\prime}$ (FWHM), which corresponds to an aperture containing thirty spaxels. In our sample, we find that 321 galaxies have a minimum of thirty spaxels that meet Q$_3$ and the other quality flags as mentioned before. Finally, given the low number (six) of low-mass galaxies with thirty high-quality spaxels or more (see also Figure \ref{fig:fig1}), the sample is further restricted to galaxies with stellar mass $M_* > 10^{10}$\msun. Our final sample contains a total of 315 galaxies that meet all these selection criteria.

Our sample has no selection on morphology, age, or galaxy type. However, due to the quality cuts in S/N and requiring that $\sigma_{\rm{obs}}>70$\kms, our sample might be biased towards early-type galaxies. We therefore perform a basic visual classification on our sample using the available GAMA-SDSS, SDSS, and VST imaging. Within our sample of 315 galaxies, 82\% are early-type and 18\% are late-type galaxies. Note that with the relatively poor imaging-quality, and the fact that visual-classifications can vary from observer to observer, this number if a rough approximation only.


\begin{figure*}
\epsscale{1.0}
\plotone{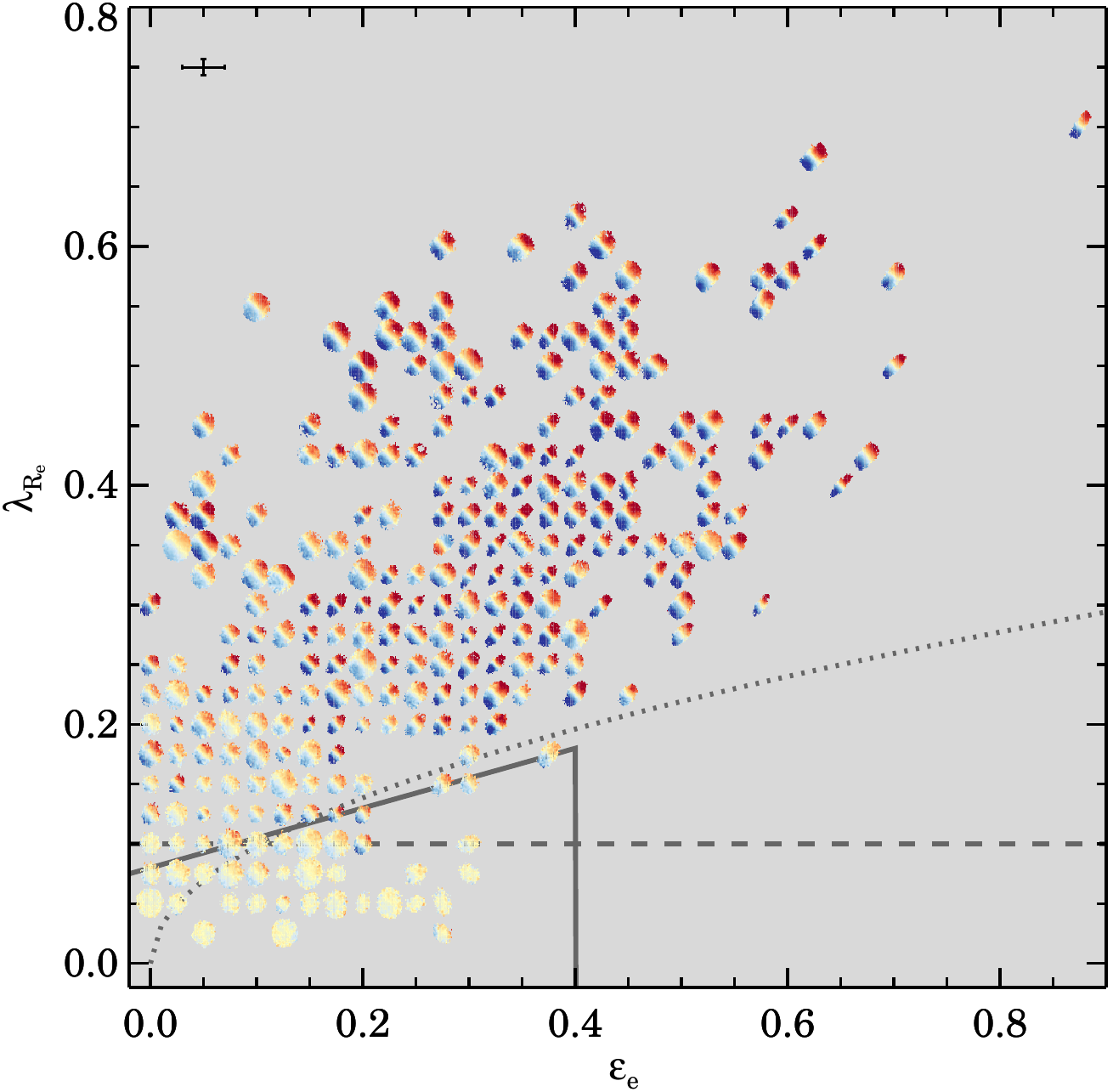}
\caption{Proxy for the spin parameter \lre\, versus ellipticity $\epsilon_{\rm{e}}$ for galaxies with $M_* > 10^{10}$\msun. For each galaxy we show its velocity map aligned to 45$^{\circ}$ using the kinematic position angle, with the velocity range set by the stellar mass Tully-Fisher relation \citep{dutton2011}. A regularization algorithm is applied to avoid overlap of the velocity maps. The median uncertainty is shown in the top-left corner. Different lines show suggested separations between slow and fast rotating galaxies from \citet[][dashed line]{emsellem2007}, \citet[][dotted line]{emsellem2011}, and \citet[][solid line]{cappellari2016}. Above the separation lines, we predominantly find galaxies that show clear signs of rotation with regular velocity fields. However, when using the \citet{emsellem2007} or \citet{cappellari2016} fast/slow classification, we find galaxies with regular rotation that would be classified as slow rotators, and vice versa.
}
\label{fig:fig5}
\end{figure*}
%


\section{Classifying galaxies from 2nd-order moment stellar kinematics}
\label{sec:dp}

In this section, we will revisit existing galaxy classifications based on 2D stellar velocity and dispersion profiles. Our aim is to find a clean separation for SAMI galaxies into different groups: fast versus slow rotators, and regular versus non-regular rotators. In the next section, these groups will then be used to analyze the stacked $h_3$-\vs\, signatures for galaxies with similar rotational properties.


%
\begin{figure*}
\epsscale{1.15}
\plotone{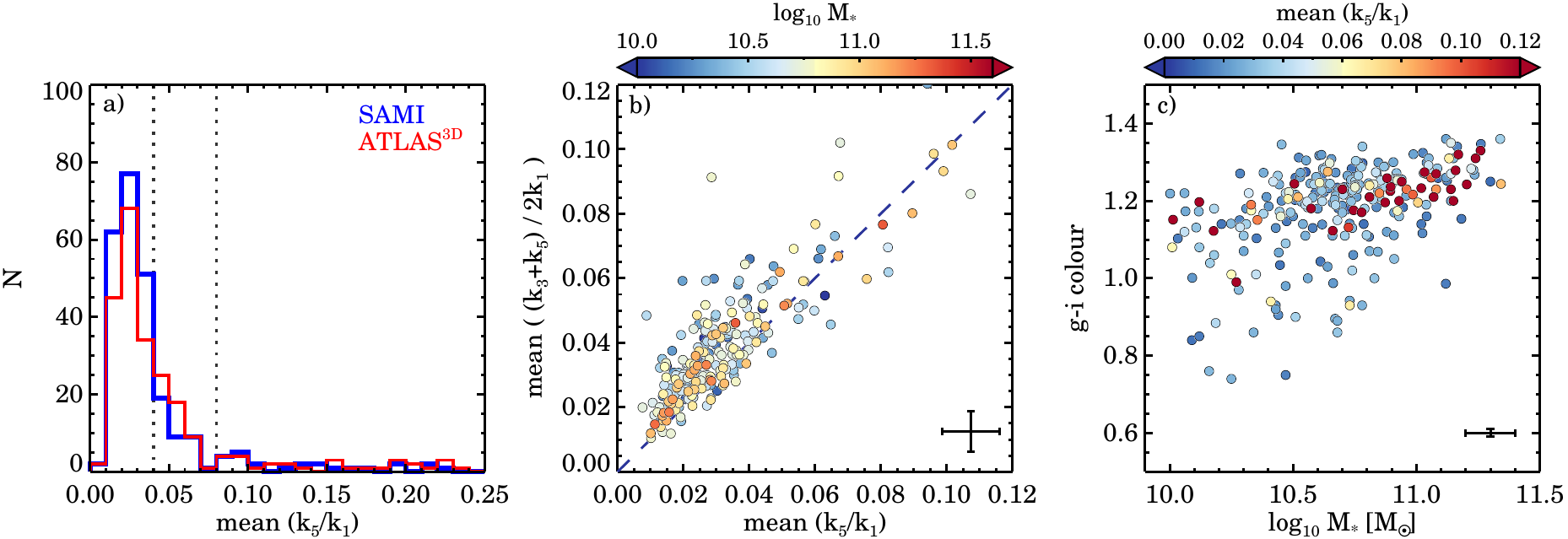}
\caption{Derived \kinemetry\, parameters compared to global galaxy properties. a): Distribution of \mk\, for SAMI galaxies (blue) and \at\, galaxies (red). The blue dashed vertical line shows the medians of the distributions. There is excellent agreement between the kinematic asymmetry distributions of the SAMI Galaxy Survey and the \at\, survey, which is remarkable given the differences in overall morphology between the samples. b): Comparison between different definitions of the kinematic asymmetry color-coded by stellar mass. We find a good agreement between \mk\, and the mean of \mkt, with a scatter of 0.009 (1-$\sigma$). c): $g - i$ color versus stellar mass, color-coded by the kinematic asymmetry. Most non-regular rotators ($\mk>0.08$) reside on the massive end of the red sequence, whereas quasi-regular ($0.04<\mk \leq 0.08$) and regular rotating ($\mk \leq 0.04$) galaxies are evenly distributed in the color-mass plane. The median uncertainty is shown in the bottom-right corner.
}
\label{fig:fig6}
\end{figure*}
%
%

\subsection{Separating Fast and Slow Rotators}
\label{subsec:lambda_r}

Following \citet{emsellem2007, emsellem2011}, we use the spin parameter approximation \lr\, to investigate separating fast-rotating galaxies from slow-rotating galaxies. For each galaxy, \lr\, is derived from the following definition \citep{emsellem2007}:

\begin{flushleft}
\begin{equation}
\label{eq:eq4}
\lambda_{R} = \frac{\langle R |V| \rangle }{\langle R \sqrt{V^2+\sigma^2} \rangle } = \frac{ \sum_{i=0}^{N_{spx}} F_{i}R_{i}|V_{i}|}{ \sum_{i=0}^{N_{spx}} F_{i}R_{i}\sqrt{V_i^2+\sigma_i^2}},
\label{eq:lr}
\end{equation}
\end{flushleft}

\noindent where the subscript $i$ refers to the spaxel position within the ellipse, $V_{i}$ is the stellar velocity in \kms, $\sigma_{i}$ the velocity dispersion in \kms, $R_{i}$ the radius in arcseconds, and $F_{i}$ the flux in units of erg cm$^{2}$ s$^{-1}$ \AA$^{-1}$ of the  $i^{th}$ spaxel. We sum over all spaxels $N_{spx}$ that meet the quality cut for the second-moment fits as described in Section \ref{subsec:qc} within an ellipse with semi-major axis \re~ and axis ratio $b/a$. Note that $R_{i}$ is the semi-major axis of the ellipse on which spaxel $i$ lies, not the circular projected radius to the center as is used by e.g., \at\, \citep{emsellem2007}. A different approach of using the intrinsic radius $R_{i}$ (semi-major axis) over the projected radius (circular) is used here, as the intrinsic radius follows the light profile of the galaxy more accurately. Our current method assigns the same weight $R_{i}$ for all spaxels on the same isophote, and will thus be less dependent on inclination; a spaxel on the minor axis will be weighted the same whether the galaxies is observed face-on or edge-on. However, by using the intrinsic radius rather than the projected, \lr\, is expected to be lower, as more weight will be given to spaxels on the minor-axis, which typically have low velocity values. We quantify the effect by measuring \lre\, using both methods. For round objects ($\epsilon<0.4$), the effect is small, i.e., we find a median $\lambda_{R_{\rm{proj}}} - \lambda_{R_{\rm{intr}}} = 0.01$. The effect becomes more pronounced for flattened objects ($\epsilon>0.4$), for which we find a median $\lambda_{R_{\rm{proj}}} - \lambda_{R_{\rm{intr}}} = 0.04$, with a maximum difference of 0.09. \lr\, within one effective radius is only considered reliable and used in our analysis when the fill factor of good spaxels ($Q_3$; Section \ref{subsec:qc}) within \re\, is greater than $75\%$
\footnote{Note that in Table B1 from \citet{emsellem2011}, \lret\, and \lre\, are quoted regardless of the \re\, coverage factor. Galaxies with $R_{\rm{max}}/R_{\rm{e}}<0.5$ therefore have identical \lret\, and \lre\, values. Only 43\% of the \at\, kinematic maps extend beyond one \re, so we caution using these values without selecting on $R_{\rm{max}}$ first.}.
Out of our 315 galaxies, 269 ($85\%$) have \lr\, measurements out to one \re. For more details on \lr\, from SAMI data, see also \citet{cortese2016}. All derived \lre\, values are given in Table \ref{tbl:tbl2}.

Figure \ref{fig:fig5} shows \lre\, versus ellipticity $\epsilon_{\rm{e}}$. For each galaxy, we show the velocity map to highlight the rotational properties. To avoid overlap between the galaxy velocity maps, the data are first put on a regular grid with spacing 0.02 in \lre\, and $\epsilon_{\rm{e}}$. We position every galaxy to a closest grid point, or its neighbor, if its closest grid point is already filled by another galaxy. The size of the grid and velocity maps are chosen such that no galaxy is offset by more than one grid point from its original position. The stellar mass Tully-Fisher \citep{dutton2011} relation is used for the velocity scale: for a galaxy with stellar mass $M_* > 10^{10}$\msun\, the velocity scale of the velocity map ranges from $-95 < V\,[$\kms$] < 95$, whereas a galaxy with stellar mass $M_* > 10^{11}$\msun\, is assigned a velocity scale from $-169 < V\,[$\kms$] < 169$. The kinematic position angle \kpa\, is used to align the major axis of all galaxies to 45$^{\circ}$. 

In the SAURON and \at\, survey, fast and slow rotators are separated as based on their position in \lre-$\epsilon_{\rm{e}}$ space. In the SAURON survey, galaxies above and below \lre=0.1 were defined as fast and slow rotator respectively \citep{emsellem2007}, whereas in the \at\, survey slow rotators are defined to have \lre$ < 0.31\sqrt{\epsilon_{\rm{e}}}$, and fast rotators are selected by \lre$\geqslant 0.31\sqrt{\epsilon_{\rm{e}}}$ \citep{emsellem2011}. We show the SAURON and \at\, relation in Figure \ref{fig:fig5} as the dashed and dotted grey line. Recent results from the SAMI Pilot Survey \citep{fogarty2014} and the CALIFA survey \citep{sanchez2012} motivated \citet{cappellari2016} to propose a refinement of the fast-slow rotators division, presented here as the solid line.

For SAMI galaxies, the majority of the galaxies with clear rotation reside above the \at\, and \citet{cappellari2016}  relations. In the bottom left region, however, where $\epsilon_{\rm{e}}\lesssim 0.15$ and \lre$\lesssim 0.15$, we find a number of galaxies with no clear sign of rotation that would be fast rotators according to the \at\, relation. In addition, there are several galaxies with regular velocity fields that are below the \at\, and \citet{cappellari2016} relations. The SAURON relation of \lre=0.1 appears to be most effective in separating galaxies with and without regular velocity fields. We will return to this issue in Section \ref{sec:discussion}.

\vspace*{0.5cm}

\subsection{Kinemetry: Regular and Non Regular Rotators}
\label{subsec:kinemetry}

We use \kinemetry\, \citep{krajnovic2006,krajnovic2008} to estimate the kinematic asymmetry of the galaxies in our sample. Our aim is to separate galaxies with regular rotation from galaxies with non-regular rotation following the method by \citet{krajnovic2006,krajnovic2011}. In \kinemetry, the assumption is that the velocity field of a galaxy can be described with a simple cosine law along ellipses: $V(\theta) = V_{\rm{rot}}\cos{\theta}$, with $V_{\rm{rot}}$ the amplitude of the rotation and $\theta$ is the azimuthal angle. Kinematic deviations from the cosine law can be modeled by using Fourier harmonics. The first order decomposition $k_1$ is equivalent to the rotational velocity, whereas the high-order terms ($k_3$, $k_5$) describe the kinematic anomalies. The kinematic asymmetry can be quantified by using the amplitudes of the Fourier harmonics. Following \citet{krajnovic2011} the kinematic asymmetry is defined as the mean (dimensionless) ratio $k_5/k_1$.

Our method for measuring the kinematic asymmetries is as follows: for each galaxy in our sample, we first mask all spaxels that do not pass the velocity quality cut Q$_{1}$ (see Section \ref{subsec:qc}). For determining the amplitude of the Fourier harmonics the \kinemetry\, routine \citep{krajnovic2006} is used. In the fit, the position angle is a free parameter, whereas the ellipticity is restricted to vary between $\pm0.1$ of the photometric ellipticity. This approach was chosen as opposed to leaving both parameters completely free for \kinemetry\, to determine, because ellipticity is not well constrained from the velocity field alone. An average separation of 1.75 spaxels between the semi-major axis of the \kinemetry\, ellipses is used, because of the covariance of the spaxels in the SAMI data. 

For each ellipse, the \kinemetry\, routine determines a best-fitting amplitude for $k_1$, $k_3$, and $k_5$. The \kinemetry\, routine is also used to determine the mean surface brightness in each ellipse from the SAMI flux images, with the same input radii, ellipticity and \kpa. For each galaxy we then determine the luminosity-weighted average ratio \mk\, within one effective radius. The uncertainty on \mk\, is estimated from Monte Carlo simulations. The radial $k_5/k_1$ values are perturbed randomly within their measurement uncertainty range, and the mean \mk\, value is re-derived. The process is repeated 1000 times, and the uncertainty on \mk\, is then estimated from the standard deviation of the distribution of simulated \mk\, values. The derived \mk\, values are given in Table \ref{tbl:tbl2}.

%

\begin{deluxetable*}{ c c c c c c c c c c}[!th]
\tabletypesize{\scriptsize} 
\tablecolumns{10} 
\tablewidth{0pt} 
\tablecaption{Compilation of all Measured Quantities}
\tablehead{
\colhead{CATID} & \colhead{$z_{\rm{spec}}$} & \colhead{$\log M_{*}$/\msun} & \colhead{$g - i$}  & \colhead{\re [kpc]}   & \colhead{${\epsilon_{\rm{e}}}$}   & \colhead{\rmaxs/\re} & \colhead{\rmaxh/\re} & \colhead{\lre} & \colhead{\mk} } \\
\startdata
       15165  &     0.0775  &    11.15  &     1.31  &    4.141  &    0.072  &    1.562  &    0.464  &    0.429  $\pm$   0.008  &   0.0281  $\pm$  0.0090  \\
       15481  &     0.0541  &    11.08  &     1.27  &    4.668  &    0.014  &    1.357  &    0.557  &    0.057  $\pm$   0.006  &   0.2123  $\pm$  0.0685  \\
       22582  &     0.0778  &    11.06  &     1.28  &    3.072  &    0.142  &    2.040  &    0.781  &    0.141  $\pm$   0.007  &   0.0299  $\pm$  0.0150  \\
       22595  &     0.0790  &    11.12  &     1.36  &    3.606  &    0.308  &    1.424  &    0.599  &    0.388  $\pm$   0.006  &   0.0237  $\pm$  0.0055  \\
       22887  &     0.0363  &    10.47  &     1.09  &    5.832  &    0.440  &    1.395  &    0.384  &    0.521  $\pm$   0.006  &   0.0334  $\pm$  0.0038  \\
\nodata & \nodata & \nodata & \nodata & \nodata & \nodata & \nodata & \nodata & \nodata & \nodata \\
\enddata
\tablecomments{This Table will be published in its entirety in the electronic edition of ApJ. A portion is shown here for guidance regarding its form and content.}
\label{tbl:tbl2}
\end{deluxetable*}


We compare our kinematic asymmetry values to those of the \at\, survey \citep{krajnovic2011} in Figure \ref{fig:fig6}a. The distribution for SAMI galaxies is shown in blue and for \at\, in red. There is an excellent agreement between the results from the two surveys, but in the SAMI sample there are slightly more galaxies with low \mk\, values. We note that our sample contains both early-type and late-type galaxies, whereas the \at\, sample only consisted of early-type galaxies. For \at\, galaxies, a limit of $\mk<4$\% was chosen for the velocity map to be well-described by the cosine law. Galaxies below that limit were named regular rotators and galaxies above the limit were called non-regular rotators. Here, the same limit of $\mk \leq 0.04$ is adopted for regular rotators, but we use $\mk > 0.08$ for non-regular rotators. We define the class between $0.04 < \mk \leq 0.08$ as quasi-regular rotators, because the distribution of \mk does not show a sharp transition between regular and non-regular rotators. 

Within one effective radius, 71\% of galaxies are classified as regular rotators ($\mk \leq 4$\%) and 29\% are classified as quasi regular or non-regular rotators ($\mk > 0.04$). We also perform a visual classification of the velocity maps into regular versus non-regular rotation. We find that 76\% of galaxies have regular velocity fields, and 24\% have non-regular velocity fields within one effective radius. This ratio agrees well with the automated \kinemetry\, classification. For the 23 galaxies that were visually "mis-classified" as regular rotators, we find a median \mk = 0.049, with a scatter of 0.013, close to the regular/non-regular selection criteria.

Note that \citet{krajnovic2011} used their \kinemetry\, results to come up with a more elaborate classification scheme, by including the kinematic position angle as a function of radius ($\Gamma_{\rm{kin}}$) and visual classification of the stellar velocity and dispersion fields. They split non-regular rotators into four subgroups (low-velocity, counter rotating cores, kinematically decoupled cores, and galaxies with two-sigma peaks), and regular rotators are split into two groups (regular morphology and bar/ring galaxies). We do not extend our \kinemetry\, analysis beyond the use of \mk, as this is beyond the scope of the paper.


\begin{figure*}
\epsscale{1.10}
\plotone{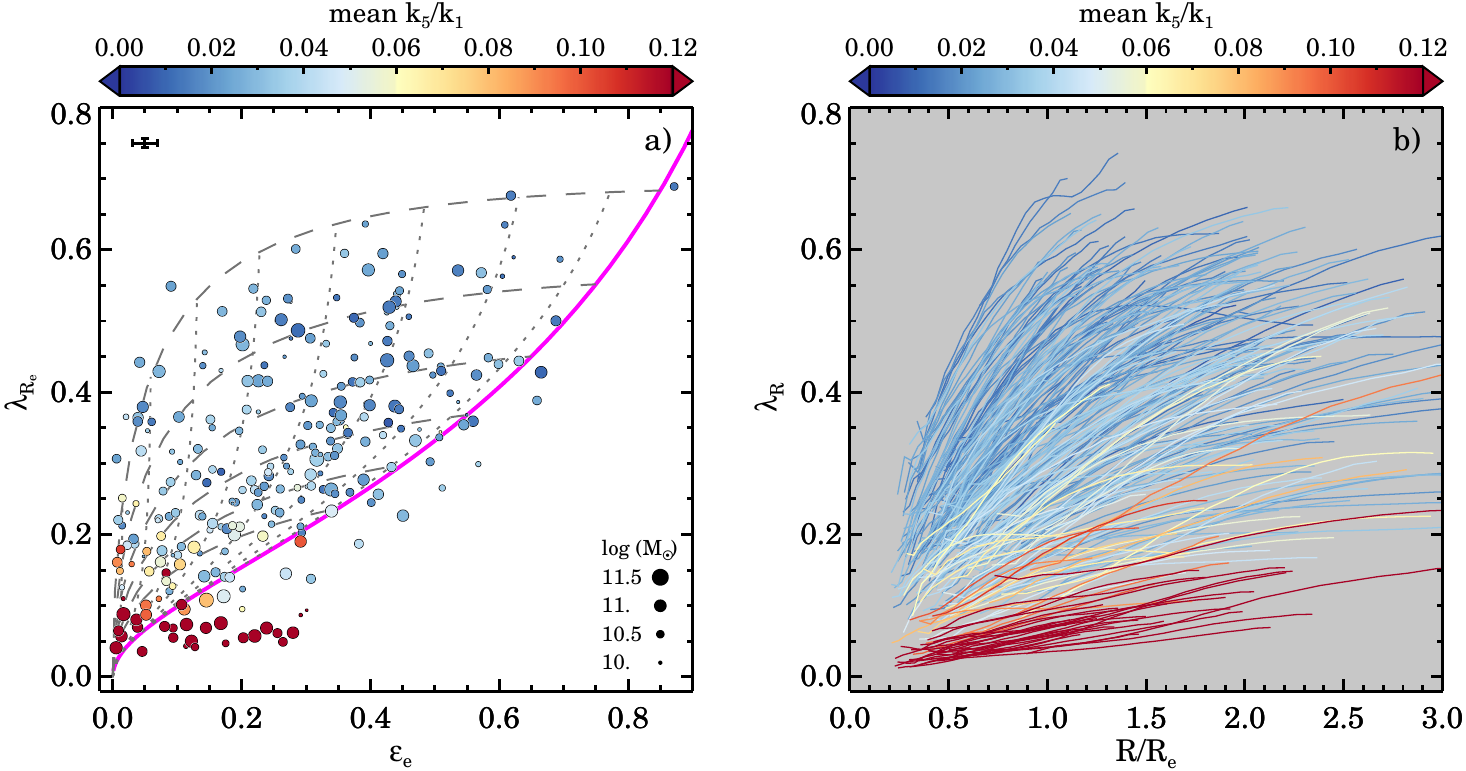}
\caption{Proxy for the spin parameter (\lr) compared to the mean kinematic asymmetry for galaxies with $M_* > 10^{10}$\msun. a): \lre vs. ellipticity. The data are color-coded by the mean asymmetry \mk\, of the galaxy; the size of symbols shows the stellar mass of the galaxy. The median uncertainty is shown in the top-left corner. 
The solid magenta is the theoretical prediction for the edge-on view of axisymmetric galaxies with $\beta_z = 0.70\times \epsilon_{\rm{intr}}$, while the gray dashed lines corresponds to the locations of galaxies with different intrinsic ellipticities $\epsilon_{\rm{intr}}$=0.85-0.35 \citep[see][]{cappellari2007,emsellem2011}. The dotted lines show the model with different viewing angle from edge-on (magenta line) to face-on (towards origin). Our results confirm that the majority (92\%) of galaxies with regular velocity fields ($\mk\leq0.04$) are consistent with being rotating axisymmetric systems with a range in intrinsic ellipticities $\epsilon_{\rm{intr}}$=0.85-0.35. Almost all galaxies with \lre$<0.1$ have non-regular rotation, but we do not find evidence for a sharp transition between regular and non-regular rotators. b): Curves of growth of \lr\, color-coded by the mean asymmetry \mk. The lines have been smoothed with a boxcar average filter with a width of four, for presentation purposes only. Most galaxies with non-regular velocity fields ($\mk > 0.08$) show a linear increase in \lr, whereas galaxies with regular velocity fields ($\mk \leq 0.04$) have a steep increase in \lr\, and then turn-over to flat \lr\, profiles. The growth curves for the quasi-regular rotators overlap with regular and non-regular rotators.
}
\label{fig:fig7}
\end{figure*}
%

Figure \ref{fig:fig6}b compares the \mk\, values to a different definition of the kinematic asymmetry by Bloom et al. (2016, submitted): \mkt\, \citep[see also ][]{shapiro2008}. The second definition uses both $k_3$ and $k_5$ and is slightly more robust when the S/N is low. We find the mean \mkt\, values to be slightly higher when compared to the \mk\, values, with a $1-\sigma$ scatter of 0.009. If we adopt the same selection criteria for both \mk\, and the mean \mkt, more galaxies (119 versus 82, respectively) would be classified as quasi-regular or non-regular if the mean \mkt\, definition were used.

The color ($g-i$) versus stellar mass relation is shown in Figure \ref{fig:fig6}c. The data are color-coded by kinematic asymmetry \mk. Galaxies with regular rotation fields are shown in blue, and non-regular rotators are shown as red, with quasi-regular rotators in between. Most galaxies with high \mk\, values are on the massive end of the red sequence above $\log_{10} M_*/$\msun $> 10.7$. There are also a few non-regular rotating galaxies below $\log M_*/$\msun $<10.5$.

We re-investigate the relation between \lre\, and ellipticity in the left panel of Figure \ref{fig:fig7}, but this time color code the data by kinematic asymmetry. Our observational data are first compared to simple galaxy models with different intrinsic ellipticities and viewing angles as presented in \citet{cappellari2007,emsellem2011}. 
These tracks are derived from models based on $V/\sigma$ and make use of the tight relation between $V/\sigma$ and \lr, with a conversion factor $\kappa$ \citep[e.g., Equation B1 from][]{emsellem2011}. We remeasure $\kappa$ because our definition of \lr\, (Equation \ref{eq:eq4}) is slightly different from \citet{emsellem2011}. We find a lower value for $\kappa$ than \citet{emsellem2011}: 0.94 versus 1.1 respectively. For the models shown here we use $\kappa = 0.94.$ Using the SAURON sample, \citet{cappellari2007} showed that regular rotating galaxies appear to be bounded by the anisotropy parameter $\beta_z = 0.70\times \epsilon_{\rm{intr}}$, where $\beta_z = 1- (\sigma_z/\sigma_R)^2$. This relation is illustrated by the solid magenta line in Figure \ref{fig:fig7} for an asymmetric galaxy viewed edge-on \citep[see e.g., ][]{emsellem2011,deugenio2013}. The same model observed under different viewing angles, from edge-on (magenta line) to face-on (towards origin), is shown by the dotted gray lines. Furthermore, we show models with different intrinsic ellipticities ($\epsilon_{\rm{intr}}$=0.85-0.35) as the gray dashed lines. 

We find that the majority (92\%) of galaxies with regular velocity fields ($\mk \leq 0.04$) are consistent with being rotating axisymmetric systems with a range in intrinsic ellipticities $\epsilon_{\rm{intr}}$=0.85-0.35. This confirms previous results from \citet{emsellem2011}. However, an observational bias may be present in \lre\, as our results differ from \citet{emsellem2011} in two ways: 1) there is lack of galaxies with $\lre<0.05$, and 2) there is dearth of flat objects with $\epsilon_{\rm{e}}>0.4$ and $\lre>0.6$. The first could be explained due to noise, which increases \lre\, \citep[e.g.,][]{emsellem2007}, whereas the second might be due to the effect of seeing, which decreases \lre and because we use elliptical apertures rather circular \citep[as adopted by][]{emsellem2011}, which lower \lre\, on average by 0.05 when $\epsilon_{\rm{e}}>0.4$.

In Appendix \ref{subsec:app_sims} we investigate the effect of seeing and measurement on \lre\, and \mk\, as measured by SAMI. Kinematic maps from the \at\, survey are used as an input, which are rebinned to match the SAMI spatial resolution. The reconstructed LOSVD is then smeared by a Gaussian PSF with varying FWHM, mimicking the different seeing conditions for the SAMI Galaxy Survey. For the typical seeing of FWHM$_{\rm{PSF}}$ = 2\farcs0, we find that the increase due to measurement errors and the decrease due to seeing cancel out for galaxies with $\lre<0.2$. For galaxies with $\lre>0.2$, seeing is the dominant effect and causes a decrease in \lre\, varying between -0.02 and -0.09 with a median of -0.05.

Thus, seeing and the use of elliptical apertures for deriving \lr\, are the likely causes for the dearth of flat objects with $\lre>0.6$. However, we do not believe the lack of galaxies with $\lre<0.05$ as compared to \citet{emsellem2011} to be fully caused by measurement noise. Instead, we note that the galaxies with $\lre<0.05$ in \citet{emsellem2011} are only measured out to 0.25-0.6 \re. If these galaxies had been observed out to one \re\, the minimum \lre\, values in \citet{emsellem2011} would have been higher. This is also clear from Figure \ref{fig:fig7}b, which shows that there would be significantly more galaxies with $\lr<0.05$ if the aperture would be only go out to \ret.

From the \kinemetry\, and spin parameter results combined, we find that galaxies with \lre$ > 0.2$ are predominantly regular rotators, whereas galaxies below $\lre = 0.1$ are almost all non-regular rotators. There is no evidence for a strong dichotomy between regular and non-regular galaxies, but a transition zone at $0.1 < $\lre$ < 0.2$, where galaxies go from slow and non-regular rotation to fast and regular. A clear dichotomy is also missing in Figure \ref{fig:fig7}b, where we show the radial \lr\, profiles color-coded by \mk. Non-regular rotating galaxies ($\mk> 0.08$; red) show a slow linear increase in \lr, whereas for regular rotators ($\mk \leq 0.04$; blue color) we find a steep \lr\, relation with a turnover around $1<R/R_{\rm{e}}<2$. Quasi-regular rotators ($0.04 <\mk< 0.08$; beige) show a variety of \lr\, profiles, but most reside within the transition zone between non-regular and regular rotators. Figure \ref{fig:fig7}b suggests that our results from Figure \ref{fig:fig7}a do not depend on the choice of aperture for \lr. We would find the same results if \lret\, or \lretwo\, instead of \lre\, are used.



\begin{figure*}
\epsscale{0.85}
\plotone{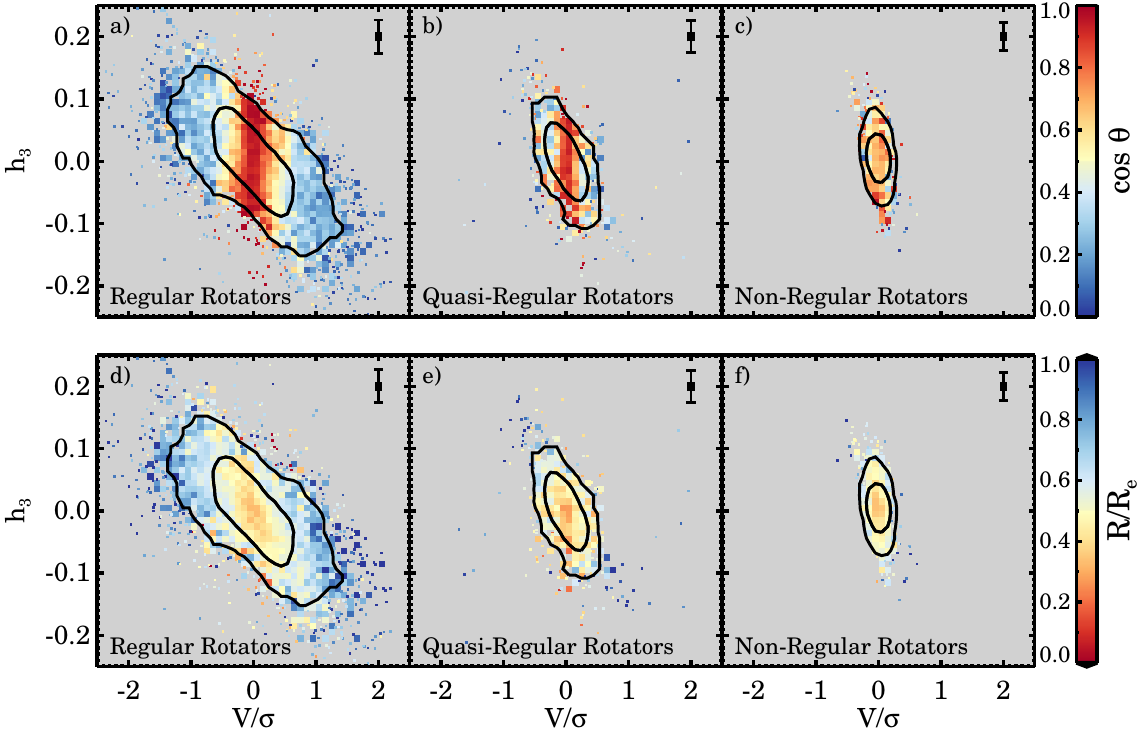}
\caption{Skewness $h_3$ versus \vs\, for galaxies with similar kinematic asymmetry \mk. The contours show where 68\% and 95\% of the data are, and the median uncertainty is shown in the top-right corner of every panel. We color-code our data by the mean azimuthal deviation from the galaxy minor axis (top row), or by the mean distance from the center in units of \re\, (bottom row). Regular rotators (left panel; $\mk \leq 0.04$) show a clear anti-correlation between $h_3$ and \vs. The quasi-regular rotating galaxies ($0.04<\mk\leq 0.08$) and the non-regular rotators ($\mk>0.08$) show a steeper vertical relation in $h_3$. We find that the strongest $h_3$ signal originates from spaxels along the major axis at large radii.}
\label{fig:fig8}
\end{figure*}

\begin{figure*}
\epsscale{0.85}
\plotone{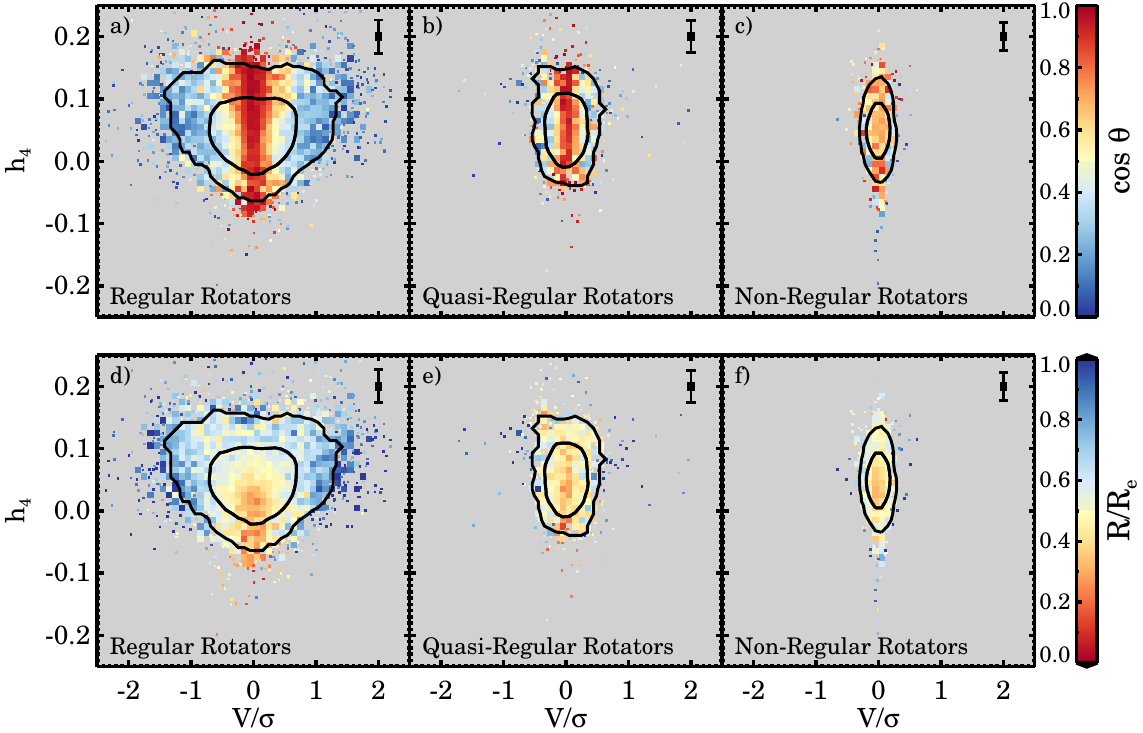}
\caption{Kurtosis $h_4$ versus \vs\, for galaxies with similar kinematic asymmetry \mk. Color-coding and contours are similar to Figure \ref{fig:fig8}. Regular rotators reveal a heart-shaped relation between $h_4$ and \vs. We find a strong correlation of $h_4$ with radius, but not with azimuthal direction. Non-regular and quasi-regular rotators do not show a correlation, but instead show a steep vertical relation in $h_4$ with a small range in \vs.}
\label{fig:fig9}
\end{figure*}


\section{High-Order Stellar Kinematics Features}
\label{sec:hok}

Here, we investigate the relation between the high-order moments ($h_3$, $h_4$) and \vs. \citet{naab2014} provide us with a theoretical framework from hydrodynamical cosmological simulations.  From their mock “IFS observations”, three distinct patterns are identified in $h_3$ versus \vs, that they relate to different assembly histories. 
Specifically, they find that fast-rotating galaxies that formed in gas-rich mergers show a strong  $h_3$-\vs\, anti-correlation, whereas fast-rotators that originated in gas-poor mergers do not.

In our observed data, we anticipate the number of observational and physical parameters that drive the high-order moments to result in more than three $h_3$-\vs\, relations than were seen previously in the \citet{naab2014} simulations. For example, viewing perspective, flattening, rotation versus pressure support (bulge and disk), the presence of bars, and oval distortions are all expected to change the $h_3$ versus \vs\, relation. Some parameters, however, such as flattening and rotation, are expected to be correlated. There are further complications such as dust that can affect these parameters; the observed LOSVD no longer represents the intrinsic LOSVD when the optical depth increases.  Thus, instead of relying solely on the high-order patterns as presented in \citet{naab2014}, here we develop a new method for parameterizing the $h_3$-\vs\, signatures of individual galaxies, which we then apply to our full sample. After this analysis, we then return to the simulations and compare the observational and simulated high-order signatures.

We first follow the approach by \citet{krajnovic2011} where galaxies are selected with similar kinematic asymmetry values and then analyze the stacked $h_3$-\vs\, signatures. Our second approach is to analyze the $h_3$-\vs\, signature for each galaxy individually. We will then try and identify high-order kinematic signatures that occur more often than others and sort them into separate classes.

\subsection{Selecting Galaxies Based on Kinemetry}
\label{sec:kinemetry}

In Section \ref{subsec:kinemetry}, we divided galaxies into three groups: regular rotators ($\mk \leq 0.04$), quasi-regular rotators ($0.04<\mk \leq 0.08$), and non-regular rotating galaxies ($\mk>0.08$). In Figure \ref{fig:fig8}-\ref{fig:fig9}, we show the high-order kinematic signatures of these three groups. The density of spaxels that meet the strict selection criteria Q$_3$ are indicated by the contours drawn at 68\% and 95\%. The data in the top row are color-coded by the mean azimuthal deviation $\cos{\theta}$ from the galaxy's minor axis (top row), such that spaxels along the major axis are shown in blue and spaxels along the minor axis are shown in red. In the bottom row, we color code the data by the mean distance from the center in units of \re.

Different high-order stellar kinematic signatures are clearly visible for regular and non-regular rotating galaxies (Figure \ref{fig:fig8}). Regular rotators ($\mk\leq 0.04$) show a strong anti-correlation between $h_3$ and \vs, indicative of a stellar disk within these galaxies. We find that the strongest $h_3$ signal originates from spaxel along the major axis at large radii. There is a weak anti-correlation, close to being vertical, for quasi-regular rotating galaxies ($0.04<\mk\leq 0.08$). We still detect a correlation between the azimuthal angle and $h_3$, which shows that quasi-regular galaxies still have rotation with a possible small disk. Non-regular rotators ($\mk>0.08$) show a steep vertical relation in $h_3$ versus \vs\, with no relation between $\cos{\theta}$ and $h_3$. As a function of radius, we find that spaxels at larger radii have a stronger $h_3$ signal.

Regular rotators show a distinct, heart-shaped pattern in $h_4$ versus \vs\, (Figure \ref{fig:fig9}). The highest $h_4$ values originate from spaxels at large radii, but not from a specific azimuthal direction, while the lowest $h_4$ spaxels tend to lie in the center along the minor axis. Quasi-regular and non-regular galaxies show no relation in $h_4$ with \vs.

The high-order kinematic signatures that we find with SAMI are similar to the results from \citet{krajnovic2008,krajnovic2011}. Note, however, that they used their \kinemetry\, results to come up with a more elaborate classification scheme, which is beyond the scope of this paper.

\begin{figure*}[!t]
\epsscale{1.15}
\plotone{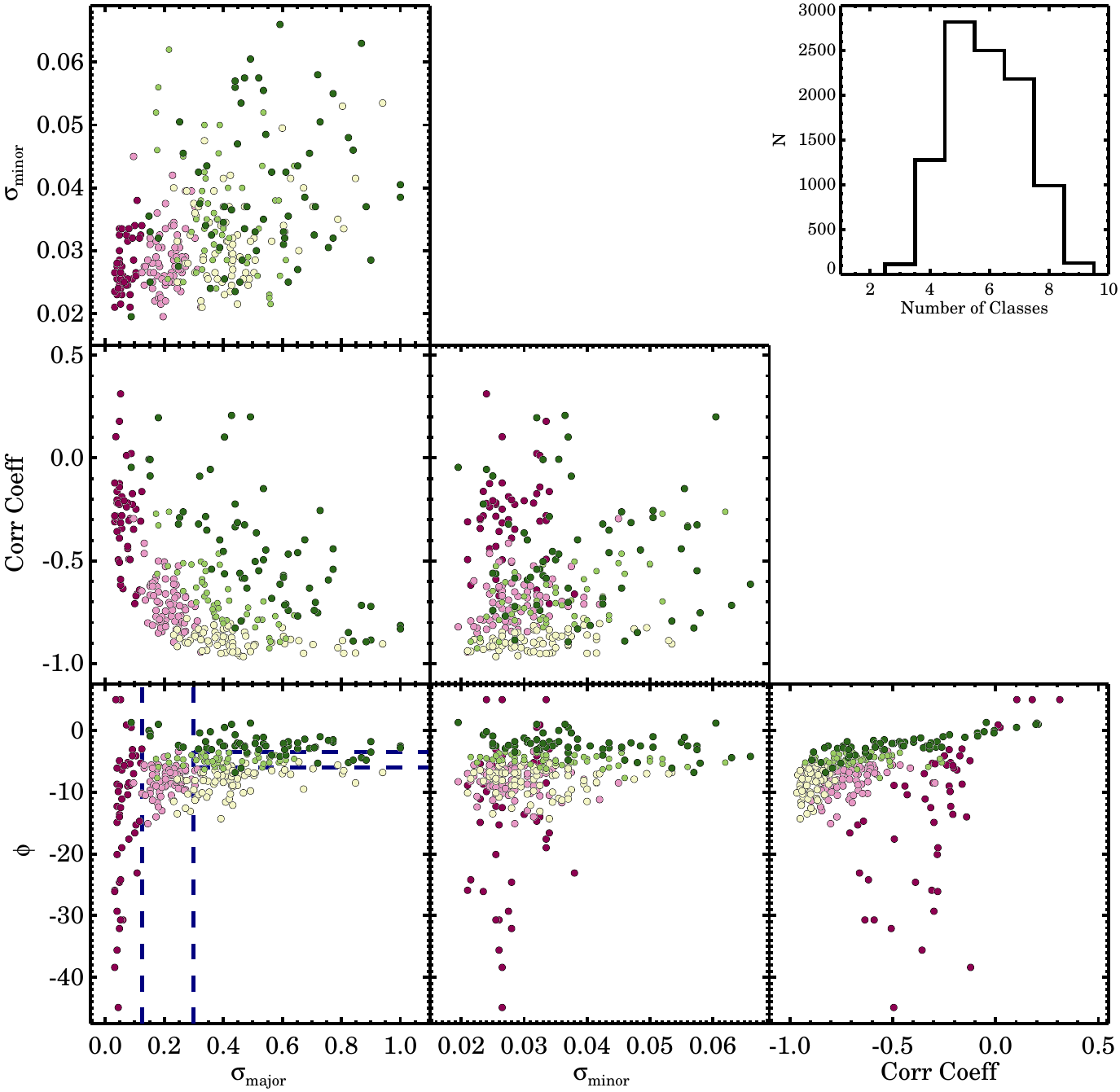}
\caption{Distribution of the best-fit parameters of the high-order kinematic signatures: \smj, \smi, $\phi$, and the Pearson correlation coefficient. Using finite mixture models we identify five classes that are highlighted by the different colors. The blue dashed lines in the bottom left panel show a simplified classification using \smj\, and $\phi$ alone (see also Table \ref{tbl:tbl4}). Top right: distribution of the optimal number of classes from bootstrapping ($N=10000$) our sample.
} 
\label{fig:fig10}
\end{figure*}
%


\subsection{Selecting Galaxies Based on High-order Stellar Kinematic Features}
\label{sec:hok_vs_all}

In this section, we explore classifying {\em individual} galaxies from their high-order signatures alone. We do this by representing the $h_3$ versus \vs\, distribution with a 2D elliptical Gaussian, with dispersion $\sigma_x$, (\smj, along the major axis of the ellipse), $\sigma_y$ (\smi, along the minor axis of the ellipse), and angle $\phi$, centered on the origin:

\begin{equation}
\label{eq:eq5}
f(x,y) = \frac{1}{2\pi\sigma_x\sigma_y}\exp\left[- \left(ax^2 - 2bxy + cy^2 \right)\right],
\end{equation}
with
\begin{equation}
a = \frac{\cos^2\phi}{2\sigma_x^2} + \frac{\sin^2\phi}{2\sigma_y^2}
\label{eq:eq6}
\end{equation}
\begin{equation}
b = -\frac{\sin2\phi}{4\sigma_x^2} + \frac{\sin2\phi}{4\sigma_y^2} 
\label{eq:eq7}
\end{equation}
\begin{equation}
c = \frac{\sin^2\phi}{2\sigma_x^2} + \frac{\cos^2\phi}{2\sigma_y^2}.
\label{eq:eq8}
\end{equation}
A maximum log-likelihood approach is used to determine how well our Gaussian model approximates the $n$ number of data-points. The log-likelihood is defined as:
\begin{equation}
\label{eq:eq9}
\ln {\cal{L}} = \prod_{i=1}^n f(x_i,y_i),
\end{equation}
Here, $f(x_i,y_i)$ is the probability function for a given data-point $i$ at $x_i$ and $y_i$. The log-likelihood is then calculated from the product of all probability functions. For our 2D Gaussian this becomes:
\begin{equation}
\label{eq:eq10}
\ln {\cal{L}} = \sum_{i=1}^n -\ln(2\pi \sigma_x \sigma_y) - ax_i^2 +2bx_iy_i - cy_i^2
\end{equation}
with $a$, $b$, and $c$ defined in Equations \ref{eq:eq6}-\ref{eq:eq8}. We calculate the log-likelihood for a large range of values for $\sigma_x$, $\sigma_y$, and $\phi$, and then derive for which values the maximum log-likelihood is reached. Each galaxy is assigned the corresponding value of $\sigma_x$, $\sigma_y$, and $\phi$. Hereafter we will refer to $\sigma_x$ and $\sigma_y$ as \smj\, and  \smi\, for clarity. In order to get a model independent measurement of the anti-correlation strength, we also calculate the Pearson correlation coefficient for each galaxy. The derived quantities are given in Table \ref{tbl:tbl3}.


\begin{deluxetable}{ c c c c c c}[!th]
\tabletypesize{}
\tabletypesize{\scriptsize} 
\tablecolumns{6} 
\tablewidth{0pt} 
\tablecaption{Log-likehood Estimates and High-Order Classification}
\tablehead{
\colhead{CATID} & \colhead{\smj} & \colhead{\smi} & \colhead{$\phi$} & \colhead{Corr. Coeff.} & \colhead{Class} } \\
\startdata
       15165  &     0.30  &     0.03  &    -8.90  &    -0.82  &      3  \\
       15481  &     0.04  &     0.02  &    -8.70  &    -0.16  &      1  \\
       22582  &     0.15  &     0.03  &    -8.30  &    -0.62  &      2  \\
       22595  &     0.42  &     0.02  &    -8.20  &    -0.93  &      3  \\
       22887  &     0.44  &     0.06  &    -2.50  &    -0.33  &      5  \\
\nodata & \nodata & \nodata & \nodata & \nodata & \nodata \\
\enddata
\tablecomments{This Table will be published in its entirety in the electronic edition of ApJ. A portion is shown here for guidance regarding its form and content.}
\label{tbl:tbl3}
\end{deluxetable}



We show the distribution of the four parameters that quantify each galaxy's $h_3$-\vs\, relation in Figure \ref{fig:fig10}. While each parameter reveals a new insight on the different families of high-order kinematic signatures, no strong groups are immediately apparent in each of the six panels, only small over-densities. Inspired by \citet{milone2015}, we adopt a method based on the Finite Mixture Models by \citet{mclacklan2000}. We use the Mcluster CRAN package \citep{fraley2002,fraley2012} in the statistical software system R, designed for model-based clustering and classification. The package performs a maximum likelihood fit assuming different groups, where the significance of each group is determined from the Bayesian Information Criterion (BIC) given the loglikelihood, the dimension of the data, and number of mixture components in the model.

\begin{deluxetable*}{c r r c c c c c c c c c }[!th]
\tabletypesize{\footnotesize} 
\setlength{\tabcolsep}{2pt}
\tablecolumns{12}
\tablecaption{Stellar kinematic classes and their mean properties}
\tablehead{
\colhead{Class} & \colhead{range $\sigma_{\rm{major}}$} & \colhead{range $\phi$} & \colhead{$N_{\rm{gal}}$} & \colhead{$\left\langle\log_{10} M_{*} /\rm{M}_{\odot}\right\rangle$} & \colhead{$\left\langle g-i\right\rangle$} & \colhead{$\left\langle R_{\rm{e}} \right\rangle$ [kpc]} & \colhead {$\left\langle \epsilon_{\rm{e}} \right\rangle$}  & \colhead{$\left\langle \lambda_{\rm{e}} \right\rangle$}  & \colhead{$\left\langle \overline{k_5 / k_1} \right\rangle$} & \colhead{$\left\langle \rm{FWHM}_{\rm{PSF}}\right\rangle$} & \colhead{$\left\langle R_{\rm{(max,~h3)}} /R_{\rm{e}} \right\rangle$}}
All  &&   &          315 & 10.74 & 1.18 & 3.52 & 0.27 & 0.31 & 0.08 & 2.10 & 0.71 \\ 
\hline
1     &   $0 <\sigma_{\rm{major}} \leqslant 0.125$      & $-45^\circ < \phi < 0^\circ \;\;\;\;\;$   & 52 & 11.03 & 1.23 & 5.57 & 0.15 & 0.08 & 0.41 & 2.14 & 0.60\\
2     &    $0.125 <\sigma_{\rm{major}} \leqslant 0.3\,\;\;\;$ & $-45^\circ < \phi < 0^\circ \;\;\;\;\;$ & 97 & 10.68 & 1.19 & 3.34 & 0.20 & 0.23 & 0.04 & 2.13 & 0.73\\
3     &   $\sigma_{\rm{major}}> 0.3\;\;\;\, $      & $\phi \leqslant -6^\circ \;\;\;$  & 64 & 10.67 & 1.18 & 2.96 & 0.28 & 0.36 & 0.03 & 2.07 & 0.81\\
4     &   $\sigma_{\rm{major}}> 0.3\;\;\;\, $       & $-6^\circ < \phi \leqslant -3.5^\circ$  & 57 & 10.70 & 1.14 & 3.11 & 0.35 & 0.42 & 0.02 & 2.07 & 0.71\\
5     &   $\sigma_{\rm{major}}> 0.3\;\;\;\, $       & $\phi > -3.5^\circ$ & 45 & 10.69 & 1.17 & 2.86 & 0.46 & 0.44 & 0.02 & 2.07 & 0.66\\ 
\enddata
\label{tbl:tbl4}
\end{deluxetable*}


We first run a Gaussian finite mixture model with ellipsoidal, varying volume, shape, and orientation (VVV). Five classes are identified with a best-fit BIC=1378. Using the bootstrap method, i.e., random sampling with replacement, we repeat the fit $10,000$ times to estimate the uncertainty on the number of classes. The distribution of the recovered optimal number of classes is shown in the top-right panel of Figure \ref{fig:fig10}. We find a clear peak at N=5 which confirms the initial classification fit. Next, we fix the number of classes to five and repeat the fit $10,000$ times using bootstrapping to identify how often a galaxy is classified into one of the five groups. Figure \ref{fig:fig10} shows the results of the bootstrap analysis, where each galaxy is assigned the color of the class
it most often resides in.

\begin{figure*}[!t]
\epsscale{1.15}
\plotone{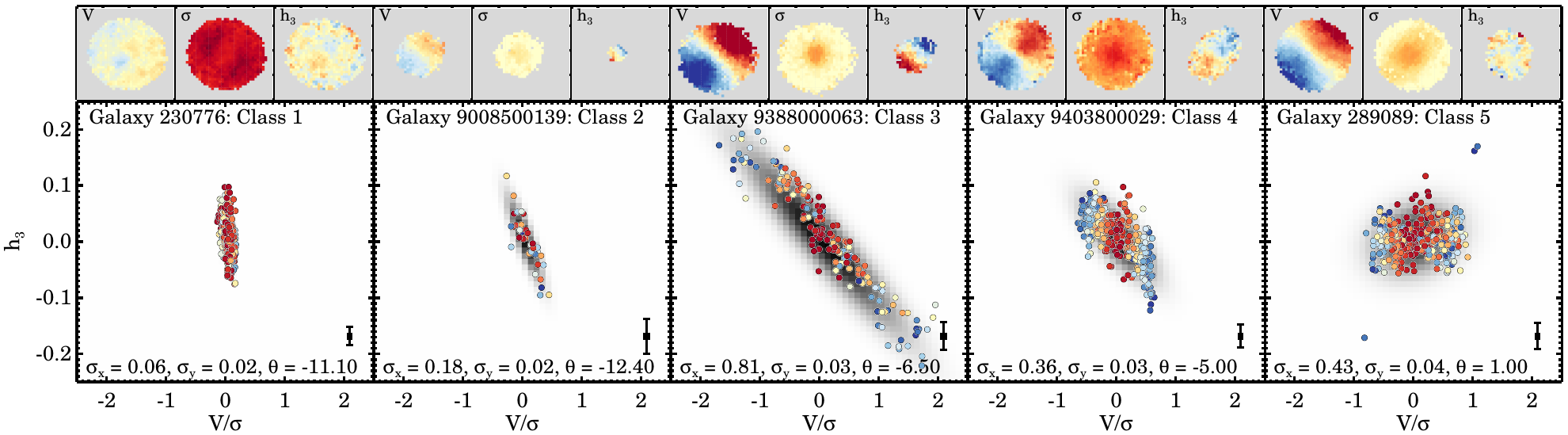}
\caption{Kinematic classes as identified from the high-order stellar kinematic signatures. For each class, the main panel shows the skewness $h_3$ versus \vs\, for the data (color-code by azimuthal deviation from the galaxy's minor axis), and the best-fitting model in grey. Note that the best-fitting model for the Class 1 galaxy is obscured by the observed data. The median uncertainty is shown in the bottom-right corner of every panel. The three panels on top show the stellar velocity ($\pm150$ \kms), velocity dispersion (100-350 \kms), and $h_3$ ($\pm0.15$) kinematic maps. From the kinematic maps and the main panel, it is clear that Class 5 galaxies are classified as fast-rotating galaxies without a strong $h_3$-\vs\, anti-correlation.}
\label{fig:fig11}
\end{figure*}

There is no clear distinct separation of the classes in any of the parameters, instead every distribution shows a gradual transition from one class to another. However, out of all four parameters, $\phi$ and \smj\ are the cleanest set to separate the five classes that we found by using the Gaussian finite mixture models. Given that the five classes are most easily separated in $\phi$ and \smj, we propose a simplified classification based on $\phi$ and \smj\, alone, as indicated by the blue dashed lines in the bottom left panel of Figure \ref{fig:fig10}. The selection criteria are given in Table \ref{tbl:tbl4}. From Class 1 to 3, the mean value in \smj\, increases, which coincides with a steep vertical $h_3$ versus \vs\, relation in Class 1 to a strong anti-correlation in Class 3, respectively. Class 3-5 are selected by similar \smj, but are separated by their angle $\phi$. From Class 3 to 5, the angle $\phi$ goes from a negative angle to zero angle, i.e., the $h_3$ versus \vs\, relation goes from steep to horizontal.

Figure \ref{fig:fig11} shows examples of individual galaxies in Class 1-5, which we will use to describe the five classes in more detail. Class 1 shows an $h_3$ versus \vs\, relation that is steep to vertical with little spread in \smj\, direction. The galaxy has a high average velocity dispersion, and the velocity field shows no sign of rotation, except in the core, where there is evidence for a kinematically decoupled core. The $h_3$ map shows no strong directional anti-alignment, except in the core, where there is an anti-alignment with the kinematic decoupled core. From a visual inspection of the broad-band color images and kinematic maps, we find that Class 1 galaxies are related to non-rotating or slow-rotating elliptical galaxies. 

For Class 2 the $h_3$ versus \vs\, relation is steep, with relatively little spread in the \smj\, direction, but with more spread than Class 1 by definition. Class 2 objects sometimes show a weak vertical boxy signature in $h_3$ versus \vs. From a visual inspection, we find that this class could be further separated into galaxies with boxy-round or weak anti-correlated signatures, but the spatial sampling of the data in combination with the fitting method do not allow for this. The velocity maps show rotation, but the rotation is not as strong as compared to Class 3-5.

Classes 3 and 4 have a strong anti-correlation between $h_3$ and \vs\, which is also clearly evident from the anti-alignment of the velocity field and the $h_3$ map. From the strength of their velocity fields relative to their velocity dispersions, these galaxies would be classified as fast rotators. By definition, the anti-correlation of Class 4 is less steep as compared to Class 3, but Class 4 also has a larger perpendicular spread in the anti-correlation. 

For Class 5 galaxies the relation between $h_3$ and \vs\, is mostly horizontal to slightly inclined, and sometimes show signs of a combined weak anti-correlation and correlation with $h_3$. From the kinematic maps there is evidence for strong rotational support, as is also evident by the large range in \vs. Furthermore, in the kinematic maps we find that the 2D $h_3$ signal shows no directional anti-alignment with the 2D velocity field. All Class 5 galaxies would be classified as fast-rotating galaxies based on their positions in the \lre-$\epsilon_{\rm{e}}$ diagram.

\begin{figure*}
\epsscale{1.2}
\plotone{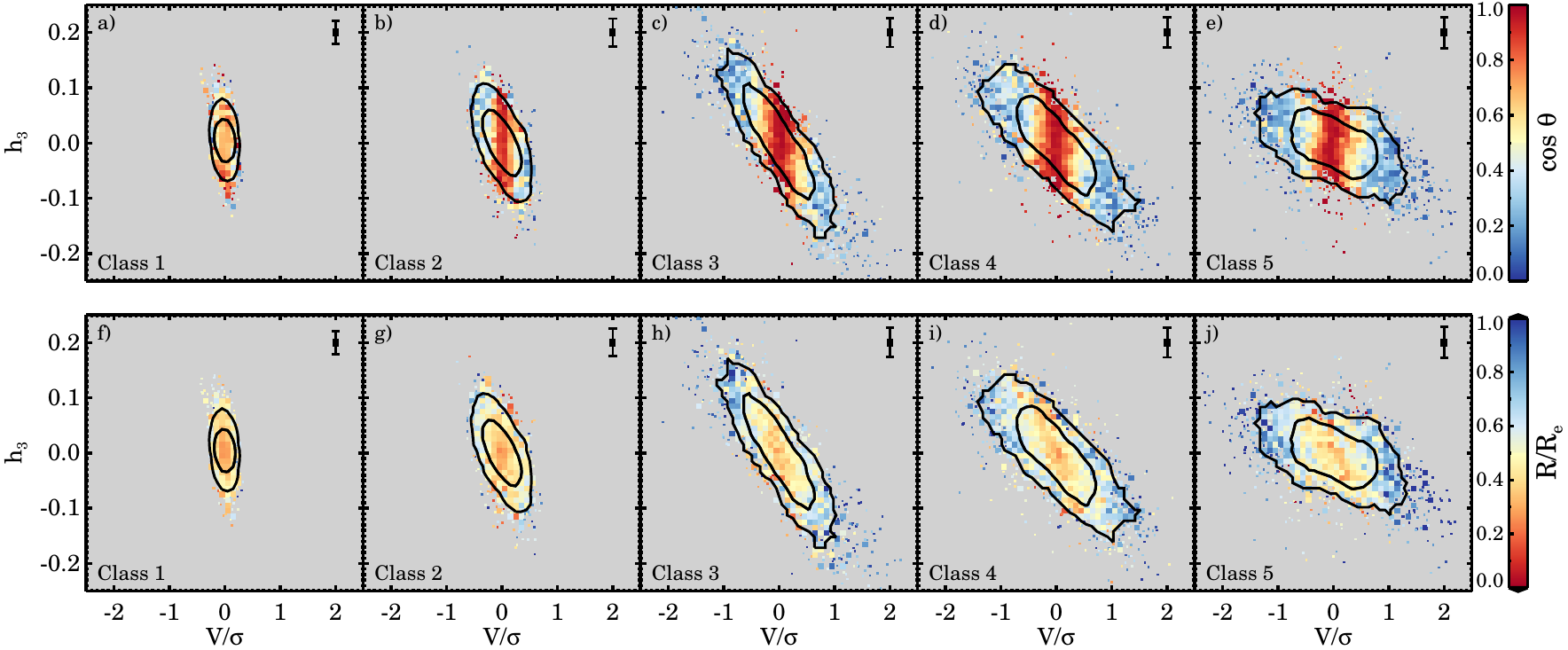}
\caption{Skewness $h_3$ versus \vs\, for the kinematic classes identified from the high-order stellar kinematic signatures. The contours show where 68\% and 95\% of the data are, and the median uncertainty is shown in the top-right corner of every panel. We color-code our data by the mean azimuthal deviation from the galaxy's minor axis (top row), or by the mean distance from the center in units of \re\, (bottom row). We find a vertical relation between $h_3$ versus \vs\, for galaxies in Class 1 but a strong anti-correlation in Class 3. Galaxies in Class 4 show an anti-correlation with a smaller angle $\phi$ and more scatter as compared to Class 3. Galaxies in Class 5 show a weak anti-correlation, and while the spread in \vs\, is similar to Class 3 and 4, the $h_3$ strength of Class 5 is less.
}
\label{fig:fig12}
\end{figure*}

\begin{figure*}
\epsscale{1.2}
\plotone{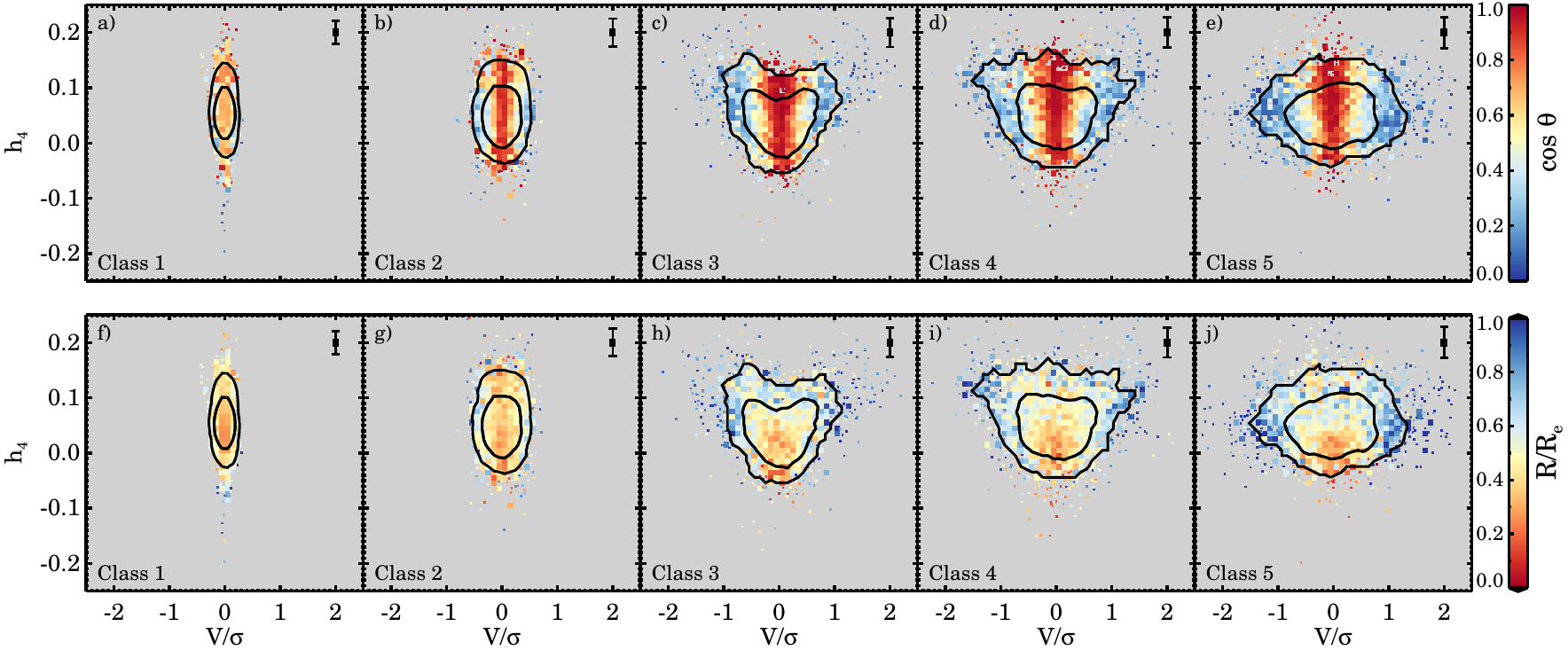}
\caption{Kurtosis $h_4$ versus \vs\, for the kinematic classes identified from the high-order stellar kinematic signatures. Color-coding and contours are similar to Figure \ref{fig:fig8}. For $h_4$ versus \vs, only Classes 3 and 4 show a distinct (heart-shaped) signature. For Class 1 and 2 there is no correlation with radius and $h_4$ strength, whereas for Class 3-5 the lowest $h_4$ values are found in the center.
}
\label{fig:fig13}
\end{figure*}

We show the stacked high-order kinematic signatures of the five classes in Figure \ref{fig:fig12} and \ref{fig:fig13}. The contours indicate 68\% and 95\% of the data, where we applied a boxcar smoothing filter with a width of two. In the top row, we color-coded by the mean azimuthal deviation $\cos{\theta}$ from the galaxy's minor axis, and by the mean distance from the center in units of \re\, in the bottom row. We find five specific $h_3$ versus \vs\, signatures, but with a gradual transition from one class into another. The gradual transition was already indicated by Figure \ref{fig:fig10}, in which we found that our five classes highlight well-defined regions in the four-dimensional parameter space but without strong overdensities. Classes 1-4 show similar $h_3$ versus \vs\, relations as compared to the individual examples in Figure \ref{fig:fig11}, whereas Class 5 now shows a weak anti-correlation that was absent in the example galaxy. Class 3 and 4 show the strongest anti-correlation between $h_3$ and \vs, indicative of a rotating stellar disk within these galaxies.

For galaxies in Class 2-5, the strongest $h_3$ signal originates from spaxel along the major axis at large radii. However, for galaxies in Class 2-4, spaxels that are located along the minor axis (red) show a vertical $h_3$ versus \vs\, relation, whereas minor-axis spaxels in Class 5 galaxies show a slight positive $h_3$ versus \vs\, relation. We detect no relation between $\cos{\theta}$ and $h_3$ for Class 1 galaxies, but spaxels at larger radii do show a stronger $h_3$ signal.

We show $h_4$ versus \vs. in Figure \ref{fig:fig13}. For Class 1 galaxies, the range of \vs\, is very narrow as compared to the range of $h_4$, with no trend, similar to the $h_3$ signatures. Class 2 shows a broader spread in \vs\, as compared to Class 1. For Class 3 and 4 galaxies, we find the heart-shape that was also clear for regular rotators in Figure \ref{fig:fig9}, whereas Class 5 galaxies show a rounder distribution in $h_4$ and \vs. For Class 1 and 2 there is no correlation with radius and $h_4$ strength, whereas for Class 3-5 the lowest $h_4$ values are found in the center. For Class 3 and 4, the highest $h_4$ values originate from spaxels at large radii, but not from a specific azimuthal direction

The main conclusion from Figure \ref{fig:fig12}-\ref{fig:fig13} is that for all five classes we find well-defined signatures in $h_3$ versus \vs\, with a gradual transition from one class into another. In the next section, we investigate how the kinematic signatures relate to integrated global galaxy properties such as stellar mass, color and \lr.
%

\subsection{Galaxy Properties of High-order Stellar Kinematic Classes}
\label{sec:gphoskc}

In the previous section, galaxies were separated into five classes based on their high-order stellar kinematic signatures alone. Here, we will look at the integrated galaxy properties of these classes, and investigate where they lie in known relations between color versus stellar mass, effective radius versus stellar mass, and proxy for the spin parameter versus ellipticity. 

\subsubsection{Global Properties}

\begin{figure*}
\epsscale{1.10}
\plotone{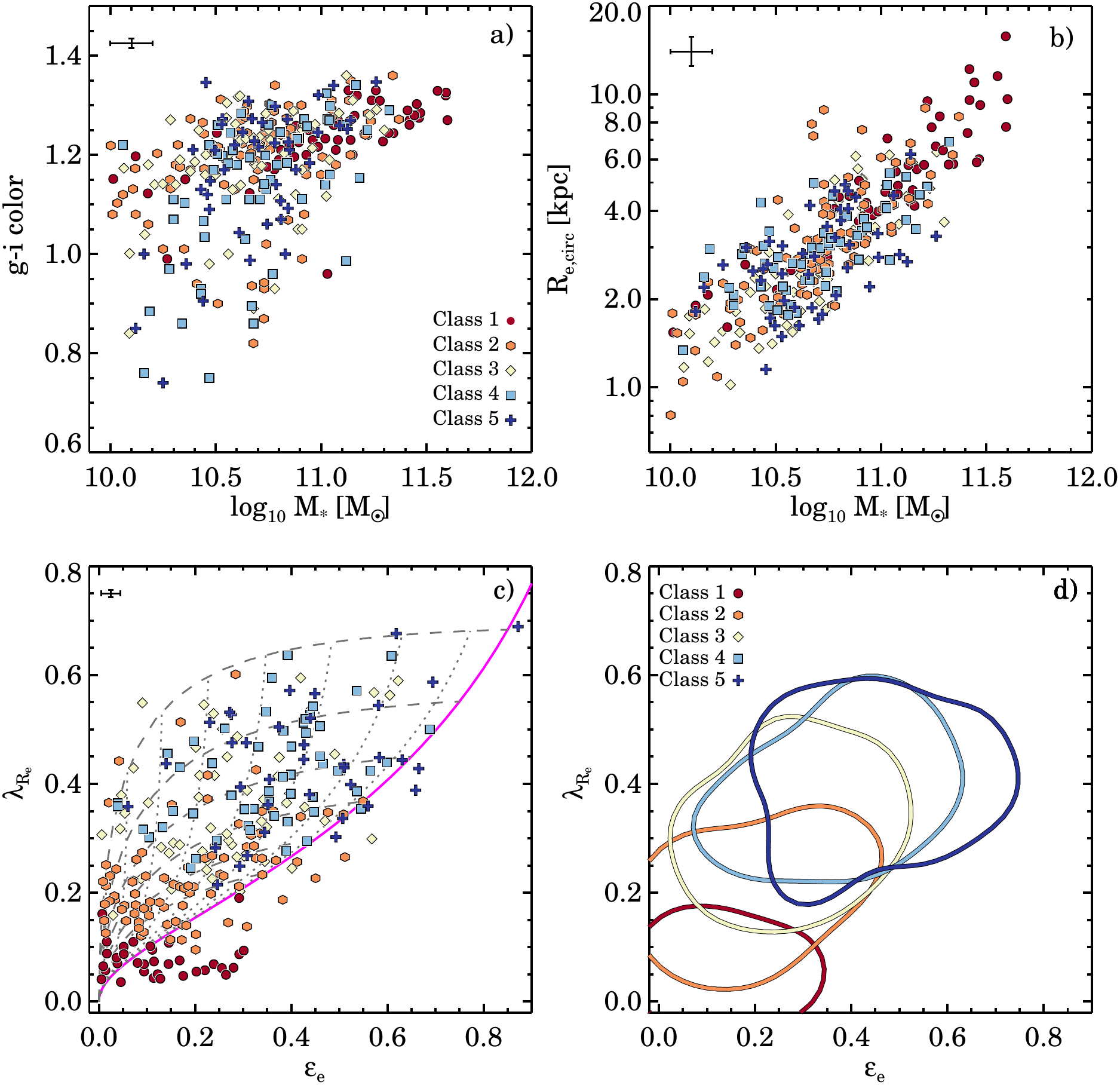}
\caption{Galaxy properties for different stellar kinematic classes. The median uncertainty is shown in the top-left corner of every panel. a) $g-i$ color versus stellar mass. Class 1 galaxies dominate the massive-red end of the distribution, whereas Class 2-5 are evenly distributed. b) Effective radius versus stellar mass. No clear separation between classes is present, except for Class 1 galaxies which are on average larger. 
We note that our sample is biased towards early-type galaxies, and that we are missing galaxies with bluer colors and galaxies with large radii at high stellar mass. c) \& d) Proxy for the spin parameter\lre\, versus ellipticity $\epsilon_{\rm{e}}$. Panel c) shows the data for all galaxies (lines as in Figure \ref{fig:fig7}); in panel d) the contours enclose 68\% of the total probability using kernel density estimates. 
We find that the five classes occupy distinct regions in the \lre-$\epsilon_{\rm{e}}$ diagram, albeit with significant overlap. However, galaxies with similar $\lambda_{R_{\rm{e}}}-\epsilon_{\rm{e}}$ values can show distinctly different $h_3-V/\sigma$ signatures. Thus it is important to realize that the overlapping regions observed here separate more clearly in a higher dimensional space.}
\label{fig:fig14}
\end{figure*}

Our full sample contains 315 galaxies, with 52 galaxies in Class 1, 97 in Class 2, 64 in Class 3, 57 in Class 4 and 45 in Class 5 (see also Table \ref{tbl:tbl4}). In Figure \ref{fig:fig14}a, we show the $g-i$ color versus stellar mass for all galaxies for which a high-order stellar kinematic class could be determined. Class 1 galaxies (red circles) are mostly found on the red-sequence and dominate at the high-mass end ($M_* > 10^{11}$M$_{\odot}$). There are few galaxies at relatively low stellar masses and only two galaxies with blue $g-i$ colors ($< 1.0$). Galaxies in Class 2 (orange hexagons) have lower mean stellar masses than Class 1, but the bulk resides on the red-sequence. Galaxies from Class 3 (beige diamonds), Class 4 (light-blue squares) and Class 5 (blue pluses) have a large range in both color and stellar mass. 


We show the mass-size relation in Figure \ref{fig:fig14}b. Unsurprisingly, Class 1 galaxies are among the largest galaxies in our sample, whereas Classes 3-5 are distributed evenly on the mass-size plane. In Figure \ref{fig:fig14}c and \ref{fig:fig14}d, we show the spin parameter approximation (\lre) versus ellipticity ($\epsilon_{\rm{e}}$). Note that \lr\, within an effective radius could be determined for 269 out of the 315 galaxies for which we derived a high-order stellar kinematic class. The data for all individual galaxies are shown in Figure \ref{fig:fig14}c, while kernel density estimates are used in Figure \ref{fig:fig14}d; the contours show 68\% out of the total probability. For the kernel density estimates we use a Gaussian kernel with a bandwidth of 0.076.

Galaxies in Class 1 populate the region below the magenta line that indicates an edge-on view of axisymmetric model galaxies with $\beta_z = 0.70\times \epsilon_{\rm{intr}}$. From panels a) and b) we already learned that Class 1 galaxies are among the most massive, large, red galaxies in our selected sample, so it comes as no surprise that these galaxies will have complex dynamical structure, and are also classified as slow rotators by the \lre-$\epsilon_{\rm{e}}$ criterion. 

Class 2 galaxies have slightly higher \lre\, and ellipticity values than Class 1. Most Class 2 galaxies reside close to the fast slow separation criteria of \citet{emsellem2011} and \citet{cappellari2016}. A closer inspection of Class 2 galaxies that sit above \lre$>0.35$ reveals that significant number of these outliers have bars (6/10). Classes 3 and 4 are true fast rotators as indicated by their high \lre\, values, but Class 4 galaxies have on average higher \lre\, values than Class 3 galaxies (\lre=0.42, versus \lre=0.36, respectively). For the 36 galaxies in Class 5 with \lre\, measurements, we find on average high ellipticity and high \lre; Class 5 galaxies populate the extreme regions. The 9 galaxies without \lre\ measurements also have high average ellipticity.

In Figure \ref{fig:fig14}d we find that the five classes occupy distinct regions in the \lre-$\epsilon_{\rm{e}}$ diagram, but there is significant overlap between the contours. The key result, however, is that galaxies with similar $\lambda_{R_{\rm{e}}}-\epsilon_{\rm{e}}$ values can show distinctly different $h_3-V/\sigma$ signatures. Thus it is important to realize that the overlapping regions observed here separate more clearly in a higher dimensional space.



\begin{turnpage}
\begin{figure*}[!t]
\epsscale{1.15}
\plotone{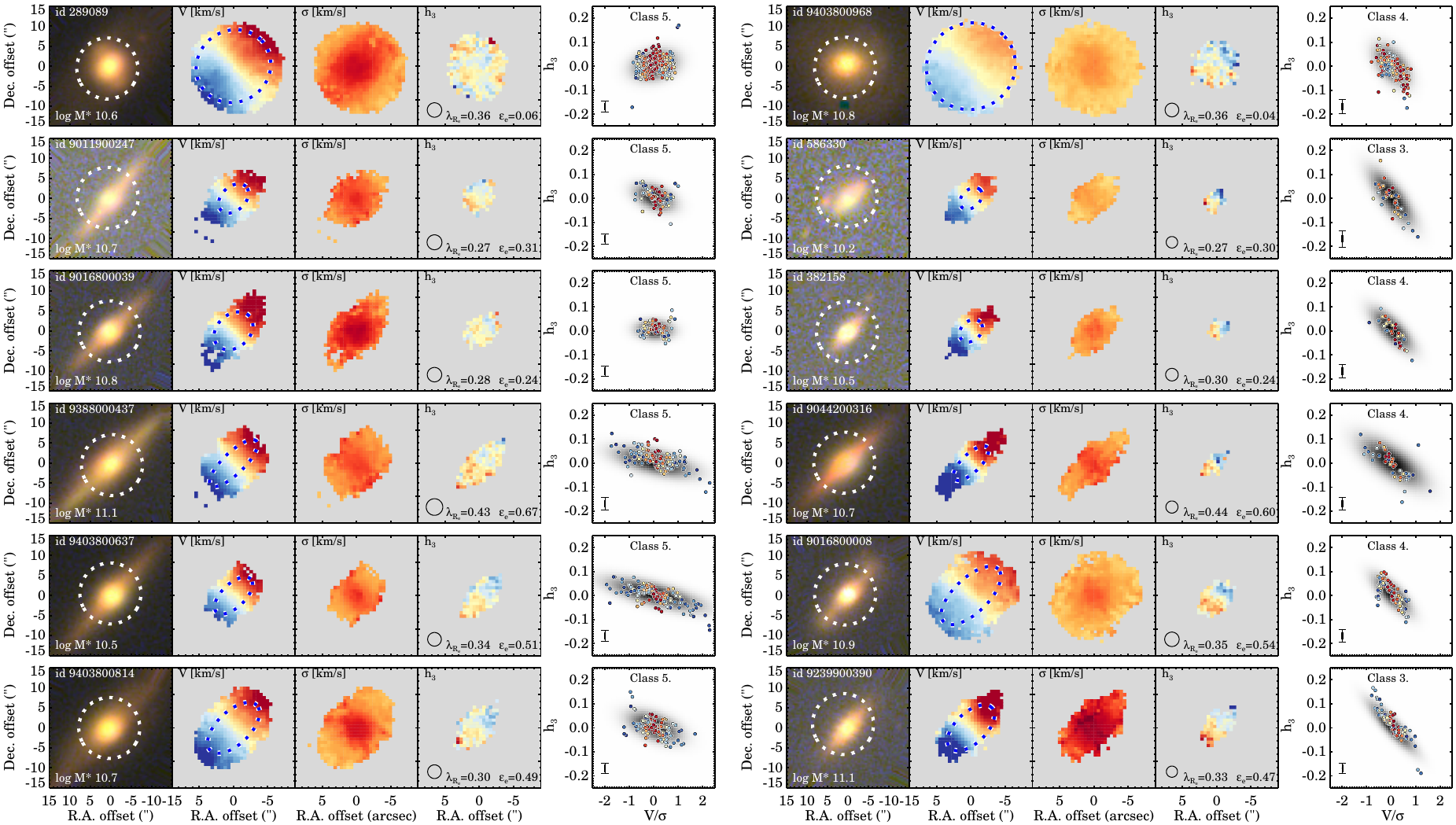}
\caption{Comparing Class 5 objects (left) to galaxies with similar \lre-$\epsilon_{\rm{e}}$ values (right). For each galaxy we show the $ugr$ color image, the stellar velocity ($\pm$150\kms ), velocity dispersion (0-250 \kms), and $h_3$ ($\pm0.15$) kinematic maps, and finally $h_3$ versus \vs\, (the median uncertainty is shown in the bottom-left). The white dashed circle in the $ugr$ color image reflects size of the SAMI hexabundle, whereas the dashed-blue-white ellipse in the velocity map shows the ellipse that contains half the light of the galaxy. Compared to galaxies with similar \lre-$\epsilon_{\rm{e}}$ values, galaxies in Class 5 have similar morphologies, stellar velocity and velocity dispersion maps, yet their $h_3$ maps and $h_3$ versus \vs\, signatures are very different.}
\label{fig:fig15} 
\end{figure*}
\end{turnpage}


\subsubsection{Class 5 Morphologies}

In the previous section, Class 5 galaxies were found to be fast-rotating without an $h_3$-\vs\, anti-correlation. Given the interesting properties of this class, here we will compare the morphologies and stellar kinematic maps of Class 5 to Class 2-4 galaxies. For comparison, galaxies are selected that occupy the same region in the \lre-$\epsilon_{\rm{e}}$ diagram, i.e., for each Class 5 galaxy we select its nearest neighbor from any other class.

Figure \ref{fig:fig15} shows color images and stellar kinematic maps of Class 5 galaxies on the left, and for their nearest \lre-$\epsilon_{\rm{e}}$ neighbors on the right. We find morphologies ranging from spirals to fully edge-on disks, but there are no morphological differences between Class 5 and the selected Class 2, 3 and 4 galaxies. For example, galaxies 9403800814 and 9239900390 are similar in morphology (6th row); both galaxies are edge-on disks with a central bulge.

All galaxies shown in Figure \ref{fig:fig15} are strongly rotating, and some show a dispersion dominated bulge. The $h_3$ maps look different for Class 5 galaxies as compared to Class 2-4 galaxies. For Class 5 galaxies, we find no anti-alignment of the $h_3$ signal with the velocity field, whereas similar \lre-$\epsilon_{\rm{e}}$ galaxies do show this strong anti-alignment. This is also visible from the $h_3$ versus \vs\, panel, where Class 2-4 galaxies show a strong anti-correlation and Class 5 galaxies do not.

Our $h_3$ spatial detection limit (i.e., our minimum requirement of 30 good spaxels) is also not the cause for the discrepancy between the classes. For example, galaxy 586330 (Class 3, 2rd row) and galaxy 382158 (Class 4, 3rd row) have 30-35 spaxels for which $h_3$ could be reliably measured, yet the $h_3$-\vs\, anti-correlation is clearly visible. All Class 5 galaxies are also well above the spatial detection limit. However, if Class 5 on average has lower \rmaxh/\re\, as compared with Class 3-4, then we could be tracing different regions of the galaxies, i.e., bulge versus disk. Class 5 galaxies have lower median \rmaxh/\re\, (0.66) as compared to Class 3 and 4 (0.73; 0.81; 0.71, respectively; see also Table \ref{tbl:tbl4}). From Figure \ref{fig:fig4}b, however, we find that \rmaxh/\re\, ranges from 0.1 to 1.5, so the difference between the median \rmaxh/\re\, of the classes is small. Furthermore, Class 5 galaxies have higher ellipticity as compared to Class 3 and 4. For edge-on galaxies \rmaxh\, could be smaller due to observational effects. If we compare the median \rmaxh/\re\, for galaxies with $\epsilon_{\rm{e}}>0.5$ (0.59) to galaxies with $\epsilon_{\rm{e}}<0.5$ (0.70), a similar trend is detected. This means that any galaxy with high ellipticity would have slightly lower \rmaxh, irrespective of its class. We conclude therefore that the spatial detection limit is not the cause for the different identified classes. 

Finally, we also look into the effects due to seeing, which could be affecting our classification. Because all classes have similar median seeing (see Table \ref{tbl:tbl4}), this is not likely to impact our results. Moreover, the size of the PSF is indicated by the circle on the bottom right in the $h_3$ map (Figure \ref{fig:fig15}), and is always smaller than the $h_3$ detection map.

\section{Discussion}
\label{sec:discussion}

\subsection{Revisiting Kinematic Galaxy Classifications}
\label{subsec:rkgc}

With the introduction of the SAURON instrument \citep{bacon2001} and its survey \citep{dezeeuw2002}, a visual inspection of the stellar kinematics maps of 66 galaxies by \citet{emsellem2004} led to a simple classification of ETGs into two groups. The first group were galaxies with rotating disks seen at different inclinations, whereas the galaxies in the other group were inconsistent with having simple disks. This confirmed earlier results obtained with long-slit spectrographs which revealed that luminous ellipticals are found to rotate slowly \citep{bertola1975,illingworth1977,binney1978,bertola1989}, whereas intrinsically faint ellipticals rotate as rapidly as disk bulges \citep{davies1983}. A more quantitative classification of fast and slow rotators was later proposed that used an approximation for the spin parameter  \lr\, \citep{emsellem2007,cappellari2007}. Early-type galaxies were separated into slow and fast rotators, depending on whether their \lr\, within an effective radius was below or above 0.1, respectively. 

Subsequently, the \at\, survey expanded the sample to 260 galaxies and the classification was further refined as \lre$=0.31\sqrt{\epsilon_{\rm{e}}}$ \citep{emsellem2011}. Galaxies above this limit were classified as fast rotators, galaxies below were defined as slow rotators. This refined classification was motivated by a different classification based on the kinematic asymmetry of the velocity field \citep{krajnovic2011}. Their results showed that galaxies with regular rotation fields can also be classified as fast rotators, whereas non-regular rotators either had (i) no rotation at all, (ii) irregular rotation, (iii) signs of kinematically decoupled cores, or (iv) two counter-rotating disks.

In this paper, we confirm the results from the SAURON and \at\, survey: the majority of early-type galaxies agree with being a family of oblate rotating systems viewed at random orientation (Figure \ref{fig:fig7}). The other group of early-type galaxies show complex dynamical structures, with irregular velocity fields, 2-sigma peaks, or kinematic misalignment, indicating that they are triaxial systems.

\citet{cappellari2016} interprets these results as a kinematic dichotomy: slow and fast rotators are distinct classes that can be separated by a selection in the \lre-$\epsilon_{\rm{e}}$ diagram. Further evidence for a dichotomy is derived from Jeans anisotropic modeling, where the distribution of $\kappa$, the ratio of $V_{\rm{obs}}/V(\sigma_{\phi}=\sigma_{R})$ of the observed velocities and a model with oblate velocity ellipsoid, shows two clear distributions \citep{cappellari2016}.

However, both the \at\, and the \citet{cappellari2016} fast/slow separation in the \lre-$\epsilon_{\rm{e}}$ space are based upon the regular versus non-regular classification by \citet{krajnovic2011}. From Figure \ref{fig:fig6} we find that there is no sharp transition in the \mk\, distribution from regular to non-regular rotation. Additionally, the regular/non-regular selection may depend on data quality: for galaxies in the SAURON survey, regular rotating galaxies were selected as $\mk < 0.02$ \citep{krajnovic2008}, whereas in the \at\ survey a limit of $\mk < 0.04$ was chosen due to lower S/N and higher average \mk\, \citep{krajnovic2008}. Furthermore, both measurement uncertainties and seeing also impact the SAMI \kinemetry\, measurements (see e.g., Appendix \ref{subsec:app_ro} and \ref{subsec:app_sims}. Measurement uncertainties increase the \mk\, parameter, whereas seeing brings it down. With typical 2\farcs0 seeing, both effects on average cancel out, but with added scatter that could be large enough to wash out a sharp transition in the \mk\, distribution.

From Figure \ref{fig:fig7} we find that galaxies with $0.1<$ \lre$<0.2$ have a relatively large range in \mk, with no clear separation. Instead we find a transition zone where galaxies go from having regular and fast rotation to non-regular and slow rotation. For SAMI galaxies, the \lre-$\epsilon_{\rm{e}}$ classification is furthermore sensitive to the data quality (see e.g., Appendix \ref{subsec:app_ro} and \ref{subsec:app_sims}). We find outliers with \lre$\sim0.05-0.1$ in repeat observations when the difference in seeing is large (i.e., from $1\farcs6$ to $2\farcs8$). This is confirmed by SAMI "re-observed" simulated \at\, galaxies, from which we find that both seeing and measurement uncertainties impact the \lre\, measurements. While the effect of seeing is strongest for galaxies with $\lre>0.25$, a sharp transition zone between $\lre=0.1-0.2$ could disappear because of this.

Thus, from the directly observable properties \lre\, and \mk\, we find that a dichotomy is unlikely to be detected due to measurement uncertainties and inclination effects. The application of Jeans Anisotropic Modelling to recover intrinsic properties, however, indicates a dichotomy in the internal velocity moments \citep{cappellari2016}, and does not suffer from these problems. While the division into fast and slow rotators is useful for many studies, e.g., the kinematic-morphology density relation \citep[][Brough et al. in prep; van de Sande et al. in prep]{cappellari2011b,fogarty2014}, we therefore caution against using only the \lre-$\epsilon_{\rm{e}}$ diagram for kinematically classifying galaxies. The addition of \kinemetry, Jeans anisotropic modeling, and/or high-order stellar kinematics should be used to fully understand the stellar kinematic properties of galaxies.


\begin{figure*}[!t]
\epsscale{1.15}
\plotone{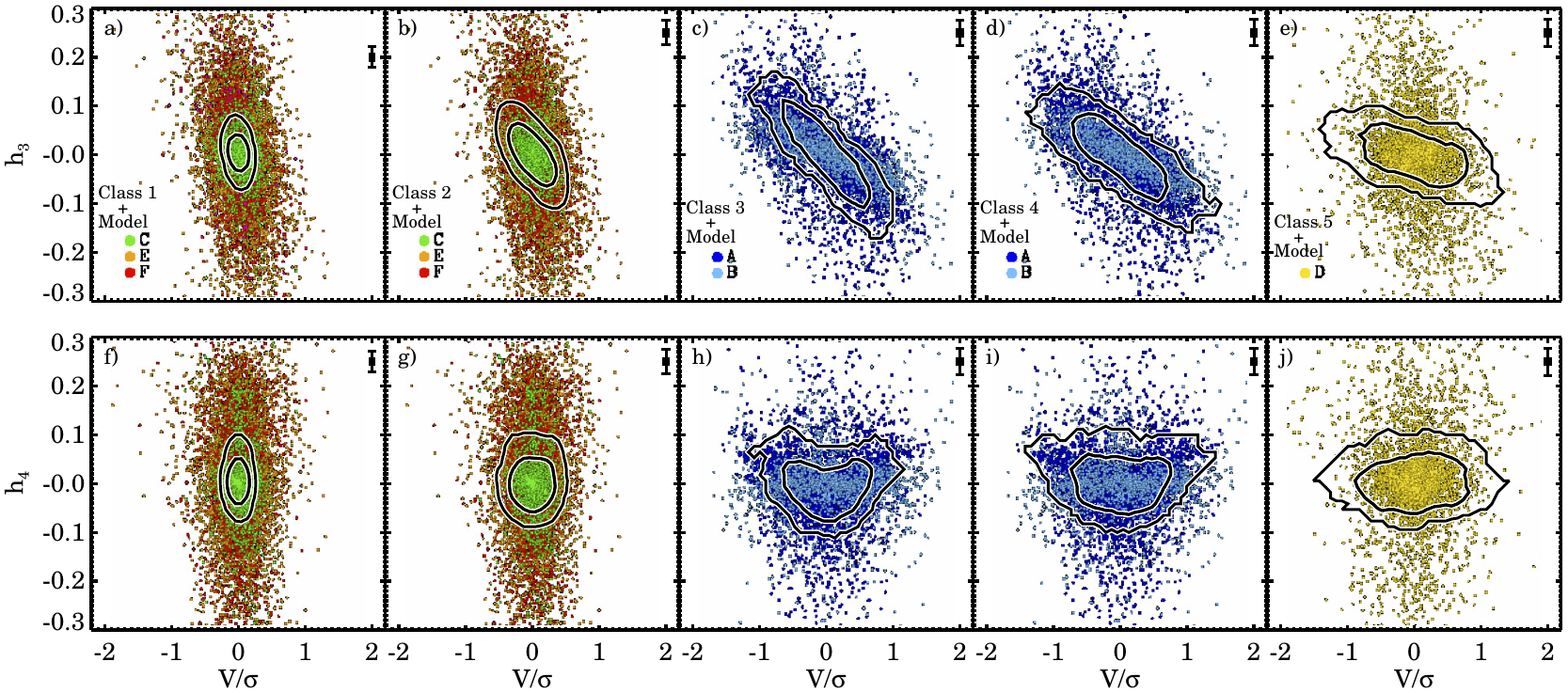}
\caption{Comparing the observed high-order kinematic classes to the model classes from \citet{naab2014}. In the top row we show $h_3$ versus \vs, in the bottom row $h_4$ versus \vs. The contours show where 68\% and 95\% of the observed data are for Class 1-5; the median uncertainty is shown in the top-right corner of every panel. The colored points show the best qualitative model from \citet{naab2014}. Our $h_4$ values are artificially lowered by 0.05 in order to compare the overall observed $h_4$ shape to those of the models.}
\label{fig:fig16} 
\end{figure*}


In this paper, we find that galaxies in similar regions of the \lre-$\epsilon_{\rm{e}}$ diagram can show distinctly different relations between $h_3$ and \vs. While we show that there are \emph{five kinematic classes} that best describe the different high-order kinematic signatures, we emphasize that the four-dimensional space for parameterizing these signatures shows a continuum of properties with five attractors. There are, however, observational and physical motivations for the five classes. Orientation, inclination, flattening, rotation versus pressure support, and the presence of a bar, all impact the high-order stellar kinematic features. Indications for these effects in different kinematic classes are presented in Section \ref{subsec:dgpmfrg} and \ref{subsec:cfg}. Higher spatial and spectral resolution data and/or more detailed simulation are needed, however, to provide a clean separation between all these effects.

\subsection{A Dearth of Gas-Poor Mergers in Fast-Rotating Galaxies?}
\label{subsec:dgpmfrg}

One of the goals of this paper is to compare the high-order stellar kinematic signatures in the SAMI galaxy survey to those as predicted by \citet{naab2014}. They use cosmological hydrodynamical zoom-in simulation to link the assembly history of galaxies to their present day shapes and kinematics. They analyze 44 central galaxies and divide these galaxies into six different model classes. Class A and B are fast rotators (\lre$\sim0.3-0.6$) that experienced gas-rich mergers, and show a strong anti-correlation between $h_3$ and \vs. Class C, E, and F are slow-rotating galaxies that either had late gas-rich mergers or gas-poor minor and/or major mergers. These model classes have a range in ellipticity of $\epsilon_{\rm{e}}\sim0.3-0.5$ with low spin parameter \lre$\sim0.05-0.2$, and show a vertical relation between $h_3$ and \vs. 

Class D consists of galaxies with a late gas-poor merger that either leads to a significant spin-up of the stellar merger remnant, or leaves the rotating properties of galaxy pre-merger intact. These galaxies have low to intermediate spin parameters (\lre$\sim0.1-0.3$); roughly half would be classified as fast rotators as based on the slow-fast selection criteria from \citet{emsellem2011}. Most interestingly, despite the relative fast rotation, Class D galaxies do not show a strong anti-correlation between $h_3$ and \vs. There are also no other signs of embedded disk-like components. We note that this signature has been found before in fast-rotating merger remnants from binary merger simulations \citep{naab2001,naab2006c,naab2006a,jesseit2007}. Due to the expected absence of gas in the late major mergers, there is no dissipative component during the merger. Therefore, no significant disk is able to regrow and the galaxy cannot support stars on tube orbits with high-angular momentum \citep{barnes1996,bendo2000,naab2001,naab2006c,jesseit2007,hoffman2009,hoffman2010,naab2014}. 

The stellar orbits of the model classes are further investigated in \citet{rottgers2014}. They show that the $h_3$-\vs\, anti-correlation in rotating galaxies (model Class A and B) originate from a high-fraction of stars on z-tube orbits. Slow rotators that experienced a recent merger (model Class C, E, and F) are dominated by stars on box and x-tube orbits in the center with an increasing contribution of z-tube orbits beyond one effective radius. For model Class D galaxies they show that the majority of the stars are on box orbits. While stars on prograde z-tubes are present in Class D galaxies, their contribution is too low to generate an LOSVD with a steep leading wing ($h_3$). 

In Figure \ref{fig:fig16} we compare the high-order kinematic signatures our five classes to the six model classes from \citet{naab2014}. The model classes are qualitatively selected to best-match the $h_3$-\vs\, relations of our observed class. In the top row of Figure \ref{fig:fig16} we find that Class 1 and 2 have very similar high-order stellar kinematic signatures as model Classes C, E, and F, but the $h_3$ amplitudes in model Class E and F are more extreme than in the observed data. Class 3 and 4 with the strong anti-correlations are comparable to model Class A and B, and both Class 5 and model Class D galaxies show a weak or no $h_3$-\vs\, anti-correlation. In the bottom row of Figure \ref{fig:fig16} we show $h_4$ versus \vs. We lowered the observed values $h_4$ values by 0.05 in order to compare the $h_4$ shape to those of the models, as we found earlier that the median $h_4$ in the SAMI data is $0.05$ (Figure \ref{fig:fig3}b), whereas the models have mean $h_4=0.0$. We find a good agreement in $h_4$ versus \vs\, between all observed classes and models. The models do show more scatter in $h_4$ as compared to the observational data, in particular for Class F and D. Interestingly, the heart-shaped $h_4$-\vs\, signature of Class 3 and 4 is also seen in model Class A and B, which is opposite to the $h_4$-\vs\, relations from the 3:1 merger simulations by \citep{naab2006c}.

When we compare the observed and model classes with regards to their positions in the \lre-$\epsilon_{\rm{e}}$ diagram, there are some differences however. Observed slow rotators (Class 1) have lower ellipticity than slow rotators from the simulations (model Class C, E, and F), although they agree well in stellar mass. Class 5 galaxies have higher average spin parameter and slightly higher ellipticity ($\epsilon_{\rm{e}}\sim0.46$) than model D, which have on average elipticities of $\epsilon\sim0.35$. Moreover, two of the five model Class D galaxies have \lre\,$< 0.2$, whereas most Class 5 galaxies are well above \lre\,$> 0.2$. Thus, while the high-order stellar kinematic signatures are similar, discrepancies in \lre\, and $\epsilon_{\rm{e}}$ between observed Class 5 and model Class D galaxies shows that the two classes likely have a different formation history.

Given the lack of overlap between Class 5 and Class D galaxies in the \lre-$\epsilon_{\rm{e}}$ diagram, we then wonder if our classification missed other galaxies that do not show an $h_3$-\vs\, anti-correlation. First, the $h_3$-\vs\, relations are investigated for all galaxies with similar \lre\, and $\epsilon_{\rm{e}}$ values as Class D galaxies. We select ellipticities $0.3<\epsilon_{\rm{e}}<0.5$ and spin parameter approximation $0.1 <$ \lre$ < 0.4$. Within this region, we find galaxies from Classes 2-4, but all Class 3 and 4 galaxies show a strong anti-correlation in $h_3$-\vs. Out of 16 Class 2 galaxies in this selected region, five show a possible lack of $h_3$-\vs\, relation. All galaxies show little spread in \vs\, and a square distribution in $h_3$-\vs. For two galaxies we are only tracing the inner part of the bulge, but the three other candidates could be similar to model Class D, although these galaxies suffer from relatively poor spatial $h_3$ sampling. From the available imaging two out of five galaxies show a clear bulge and disk, which leaves three possible matches with model Class D.

We extend our search to all Class 2 galaxies, without the restriction on \lre\, and $\epsilon_{\rm{e}}$. Out of our 85 Class 2 galaxies, for eight additional galaxies we find boxy-round $h_3$-\vs\, profiles. From a visual inspection of the  imaging data, five galaxies show clear signs of an inner bar, while the other three galaxies are round ($\epsilon_{\rm{e}}<0.15$) with low spin parameter (\lre$<0.2$).

In conclusion, the absence of model Class D galaxies in the SAMI galaxy survey data suggests that most fast-rotating galaxies are formed through gas-rich mergers. From the \citet{naab2014} simulations, we would have expected approximately 10\% (5 out of 44) of the SAMI galaxies to be rotating without showing an $h_3$-\vs\, anti-correlation. There are several limitations in the simulations, however, that could skew these predictions. First, only a small number (44) of galaxies were analyzed, but this number could be increased by repeating this analysis using the Illustris \citep{vogelsberger2014,genel2014} and EAGLE \citep{schaye2015,crain2015} simulations. Secondly, the model from \citet{naab2014} disfavors the formation of disks, which could indicate that these scenarios are missing in the cosmological simulation. For example, \citet{naab2006c} present isolated binary mergers where a remnant stellar disk was observed without a correlation in $h_3$-\vs. Thirdly, \citet{sharma2012} show that the orientation of the instrinsic spin of merging halos strongly impacts the orientation and amplitude of the angular momentum in the merger remnant. Therefore, it's vital for binary merger simulations to probe a large ensemble of realistic spin orientations, as observed in large-scale structures. Otherwise, certain merger scenarios might be missed. Fourthly, the feedback prescription in \citet{naab2014} could be improved upon such that the model would reproduce reasonable galaxy population properties, as the current model favors the formation of early-type galaxies. With the latest cosmological simulations such as Illustris and \mbox{EAGLE}, many of these issues can be resolved, which could lead to a better understanding of linking the high-order kinematic signatures in galaxies to their assembly histories.

\subsection{Class 5 galaxies: Fast Rotators Without a Stellar Disk; Merger Remnants or Edge-on Bars}
\label{subsec:cfg}

It then remains unclear, however, why Class 5 galaxies show weak or no $h_3$-\vs\, anti-correlation. When comparing the morphologies, we find that many Class 5 have clear signs of disks, which are absent in the model Class D galaxies from \citet{naab2014}. When we look at the residuals from the photometric Sersic profile fits \citep{kelvin2012}, some galaxies show strong tidal features or large residuals in their center. A recent merger could have destroyed the inner parts of the disks or have altered the orbits of the central stars in such a way that a strong $h_3$ signal is not detected.

Bars in edge-on spirals could also have a strong impact on the $h_3$ signatures \citep{chung2004, bureau2005}. Depending on the orientation and bar strength of a galaxy, \citet{bureau2005} show that the $h_3$ radial profile can show a strong correlation with radius rather than the more common disk-like anti-correlation to no correlation at all. Furthermore, their results show that the radial $h_4$ signal appears to be V-shaped, similar to the heart-shaped patterns we find for fast rotators in Classes 3 and 4. \citet{seidel2015} also study the influence of bars on the kinematics of nearby galaxies. They show interesting relations between $h_3$ and \vs\, at different bar radii. For some of their galaxies, however, the $h_3$ and $h_4$ results are hard to interpret because $\sigma$ falls well below the instrumental resolution and the $h_3$ and $h_4$ drop to zero. Nonetheless, for galaxy NGC4643 for example, they find a strong $h_3$-\vs\, anti-correlation at 0.1R$_{\rm{bar}}$, whereas the anti-correlation disappears at 0.5R$_{\rm{bar}}$ and 1R$_{\rm{bar}}$. 

A complex inner $h_3$-\vs\, structure is also seen in the edge-on S0 galaxy NGC 3115 \citep{guerou2016}. While the outer regions show a strong anti-correlation, the inner regions reveal a zig-zag pattern in $h_3$-\vs. Furthermore, a thin, fast-rotating stellar disc is embedded in the fast-rotating spheroid which leads to another $h_3$-\vs\, inner anti-correlation. Should a similar galaxy be present in our sample, it seems unlikely that the strong anti-correlation from the outer disk would be observed: both the S/N and the velocity dispersion would not meet our selection criterion Q$_3$. However, the inner zig-zag structure in $h_3$-\vs\, could be observed. When we search for this pattern in Class 5 galaxies, we indeed detect three galaxies that might show similar inner high-order signatures. In Figure \ref{fig:fig15} (left panel, row 3-5), galaxy 9016800039, 935880004037 and 9403800637 all show an $h_3$-\vs\, pattern similar to the edge-on S0 NGC 3115, which has tentative evidence for an inner bar \citep{guerou2016}. Thus, it is clear that bars can have a strong impact on $h_3$. While the evidence is far from conclusive for Class 5 galaxies, the connection to edge-on bars and different orientations and inclinations is worth pursuing further with hydrodynamical simulations.

We mention two other studies that also compare their 2D high-order stellar kinematic results to the simulation from \citet{naab2014}. \citet{spiniello2015} study the fast-rotating galaxy NGC4697 using eight VLT-VIMOS pointings, and measure the stellar kinematics out to 0.7\re.  They find a strong anti-correlation between $h_3$ and \vs\, for this system. From a combined stellar kinematic and stellar population analysis, their findings suggest that this system assembled its mass through gas-rich minor mergers. With relatively limited data, \citet{forbes2016} find a large range in high-order kinematic signatures but identify one galaxy (NGC4649\footnote{NGC4649 is classified as a fast-rotating galaxy by \citet{arnold2014} and \citet{forbes2016}. However, they incorrectly quote the \lr\, value for this galaxy to be within one effective radius; the \at\, coverage for this galaxy is only 0.35 $R_{\rm{max}}/R_{\rm{e}}$.})
that does not show a strong $h_3$-\vs\, anti-correlation. However, visual classification of the $h_3$-\vs\, signatures is open to different interpretations, particularly when the spatial sampling and data quality are relatively low. 

\section{Conclusion}
\label{sec:conclusions}

\noindent In this paper, we have used the SAMI Galaxy Survey to study the high-order stellar kinematic signatures of galaxies. We present our method for measuring the stellar kinematics in SAMI data and demonstrate our data quality. Furthermore, Monte Carlo simulations are used to determine the limits for which reliable measurements of the high-order moments can be obtained in SAMI data.

A proxy for the spin parameter (\lr) and ellipticity ($\epsilon_{\rm{e}}$) are used to re-examine the classification of fast and slow rotators. We show the velocity fields in the \lre-$\epsilon_{\rm{e}}$ diagram, and find a transition of slow-rotating galaxies with low values of \lre\, and $\epsilon_{\rm{e}}$, moving towards fast rotators with high values of \lre\, and $\epsilon_{\rm{e}}$. 

We measure the kinematic asymmetry of the velocity fields with \textsc{kinemetry} and find a good agreement in the distribution of regular and non-regular rotating galaxies as compared to the \at\, survey. There is a good correspondence between respectively fast and slow rotators and regular and non-regular rotators. We find that the majority (92\%) of galaxies with regular velocity fields are consistent with being rotating axisymmetric systems with a range in intrinsic ellipticities.

Within the SAMI Galaxy Survey sample, there is no strong evidence for a dichotomy between slow and fast-rotating galaxies, nor between regular and non-regular galaxies. Instead, there is a transition zone where galaxies go from regular fast rotators to slow non-regular rotators. This is not in conflict with the results
by \citet{cappellari2016} who finds evidence for a dichotomy using Jeans Anisotropic Modelling to recover the intrinsic properties of early-type galaxies; a dichotomy is unlikely to be detected from direct observables due to inclination and data quality, in particular with the impact of seeing on our measurements. From SAMI repeat observations and simulations of \at\, galaxies "re-observed" with SAMI, we find that a sharp transition zone in \lre\, could be washed out due to measurement uncertainties.

Using the kinematic asymmetries, we separate galaxies into regular, quasi-regular, and non-regular rotators. Within one effective radius, 71\% of galaxies are classified as regular rotators ($\mk \leq 4$\%) and 29\% are classified as quasi regular or non-regular rotators ($\mk > 0.04$). Regular rotating galaxies show a strong $h_3$ versus \vs\, anti-correlation, which has also been found by \citet{krajnovic2008,krajnovic2011}. This reveals the presence of a stellar disk within regular rotating galaxies. Quasi-regular and non-regular rotators, however, show a more vertical relation in $h_3$ and \vs. 

We develop a new method for kinematically classifying galaxies that is based on a galaxy's $h_3$-\vs\, signatures alone. This assumes that the $h_3$ versus \vs\, relation can be approximated by a two-dimensional Gaussian with parameters \smj, \smi, and angle $\phi$. From the distribution in \smj, \smi, $\phi$, and the Pearson correlation coefficient, we identify five classes with different high-order stellar kinematic signatures. From Class 1 to 5, galaxies show a sharp vertical relation between $h_3$ and \vs\, in Class 1, to a strong anti-correlation in Class 3-4, and finally towards a weak or horizontal relation between $h_3$ and \vs\, in Class 5. Class 1 galaxies have similar properties as slow, non-regular rotating galaxies, whereas Classes 2-5 show stronger rotation fields and are kinematically similar to fast rotators. 

We identify 45 fast-rotating galaxies that do not show an $h_3$ versus \vs\, anti-correlation (Class 5 galaxies). These galaxies occupy the outer regions in \lre-$\epsilon_{\rm{e}}$ space, i.e., they have either high spin parameters and/or high ellipticity. 

Our high-order kinematic classes are compared to recent predictions from \citet{naab2014} who use hydrodynamical cosmological zoom-in simulations. Their simulated galaxies show different $h_3$-\vs\, relations depending on whether the galaxy had experienced predominantly gas-rich or gas-poor mergers in the past. Fast rotators with wet-mergers show a strong $h_3$ versus \vs\, anti-correlation, whereas fast rotators where the last event was a dry-merger do not, because the absence of a dissipative gas component prevented disk formation.

The high-order kinematic signatures of our five classes are well matched by the simulations. However, our Class 5 galaxies occupy a different region in the \lre-$\epsilon_{\rm{e}}$ diagram than the disk-less galaxies formed by gas-poor mergers in the hydrodynamical simulations. From a detailed look at the morphologies of our Class 5 galaxies, we find evidence for large stellar disks. Class 5 objects are therefore more likely to be recently disturbed galaxies or edge-on galaxies with counter-rotating bulges or bars. Thus, we do not find evidence for a significant population of fast-rotating galaxies without a stellar disk. This suggest that gas-poor mergers are unlikely to be a dominant formation path for fast-rotating galaxies. 

In this paper, our novel way of classifying galaxies that is based on their high-order kinematic signatures alone, has focused mainly on the $h_3$ signatures, but several interesting patterns in $h_4$ require further study. Part of our classification is based on the spread in \vs\, and is therefore somewhat similar to previous work that separated galaxies as based on \vse\, or \lr. However, we find that galaxies with similar \lre\, and $\epsilon_{\rm{e}}$ values can have very different high-order signatures. From an observational and physical standpoint we expected to find multiple high-order classes, because orientation, streaming, and rotation versus pressure support all impact the LOSVD. However, higher spatial and spectral resolution data in combination with more detailed simulations are needed to cleanly separate these effects. 

The comparison of the SAMI Galaxy Survey data with hydrodynamical simulations shows great potential for linking the high-order moments to the type of mergers galaxies experienced in their past. Major new integral field spectrographs, such as Hector \citep{blandhawthorn2015,bryant2016}, will have higher resolution ($R\sim4000$) and more hexabundles (50-100) than SAMI, with the aim of observing more than 50,000 galaxies. 
By studying the high-order stellar kinematic signatures in these data, we can start to constrain the assembly histories of galaxies as a function of stellar mass, morphology, environment, and large scale structure.

%

\acknowledgments{
We thank the anonymous referee for the constructive comments which improved the quality of the paper. We also thank Michele Cappellari and Thorsten Naab for useful comments and discussions. The SAMI Galaxy Survey is based on observations made at the Anglo-Australian Telescope. The Sydney-AAO Multi-object Integral-field spectrograph (SAMI) was developed jointly by the University of Sydney and the Australian Astronomical Observatory, and funded by ARC grants FF0776384 (Bland-Hawthorn) and LE130100198. JvdS is funded under Bland-Hawthorn's ARC Laureate Fellowship (FL140100278). JTA acknowledges the award of a SIEF John Stocker Fellowship. M.S.O. acknowledges the funding support from the Australian Research Council through a Future Fellowship Fellowship (FT140100255). SB acknowledges the funding support from the Australian Research Council through a Future Fellowship (FT140101166). SMC acknowledges the support of an Australian Research Council Future Fellowship (FT100100457). SKY acknowledges support from the Korean National Research Foundation (NRF-2014R1A2A1A01003730) and the Yonsei University Future-leading Research Initiative (RMS2 2015-22-0064).

The SAMI input catalog is based on data taken from the Sloan Digital Sky Survey, the GAMA Survey and the VST ATLAS Survey. The SAMI Galaxy Survey is funded by the Australian Research Council Centre of Excellence for All-sky Astrophysics (CAASTRO), through project number CE110001020, and other participating institutions. The SAMI Galaxy Survey website is http://sami-survey.org/.

GAMA is a joint European-Australasian project based around a spectroscopic campaign using the Anglo-Australian Telescope. The GAMA input catalogue is based on data taken from the Sloan Digital Sky Survey and the UKIRT Infrared Deep Sky Survey. Complementary imaging of the GAMA regions is being obtained by a number of independent survey programs including GALEX MIS, VST KiDS, VISTA VIKING, WISE, Herschel-ATLAS, GMRT and ASKAP providing UV to radio coverage. GAMA is funded by the STFC (UK), the ARC (Australia), the AAO, and the participating institutions. The GAMA website is: http://www.gama-survey.org/.

Based on data products (VST/ATLAS) from observations made with ESO Telescopes at the La Silla Paranal Observatory under program ID 177.A-3011(A,B,C).

This research has made use of the NASA/IPAC Extragalactic
Database (NED), which is operated by the Jet
Propulsion Laboratory, California Institute of Technology,
under contract with the National Aeronautics and Space
Administration. 

Funding for SDSS-III has been provided by the Alfred P. Sloan Foundation, the Participating Institutions, the National Science Foundation, and the U.S. Department of Energy Office of Science. The SDSS-III web site is http://www.sdss3.org/.
}


\bibliography{jvds_sami_stellar_kinematics}

%
%

\appendix

%
\begin{figure*}
\epsscale{1.0}
\plotone{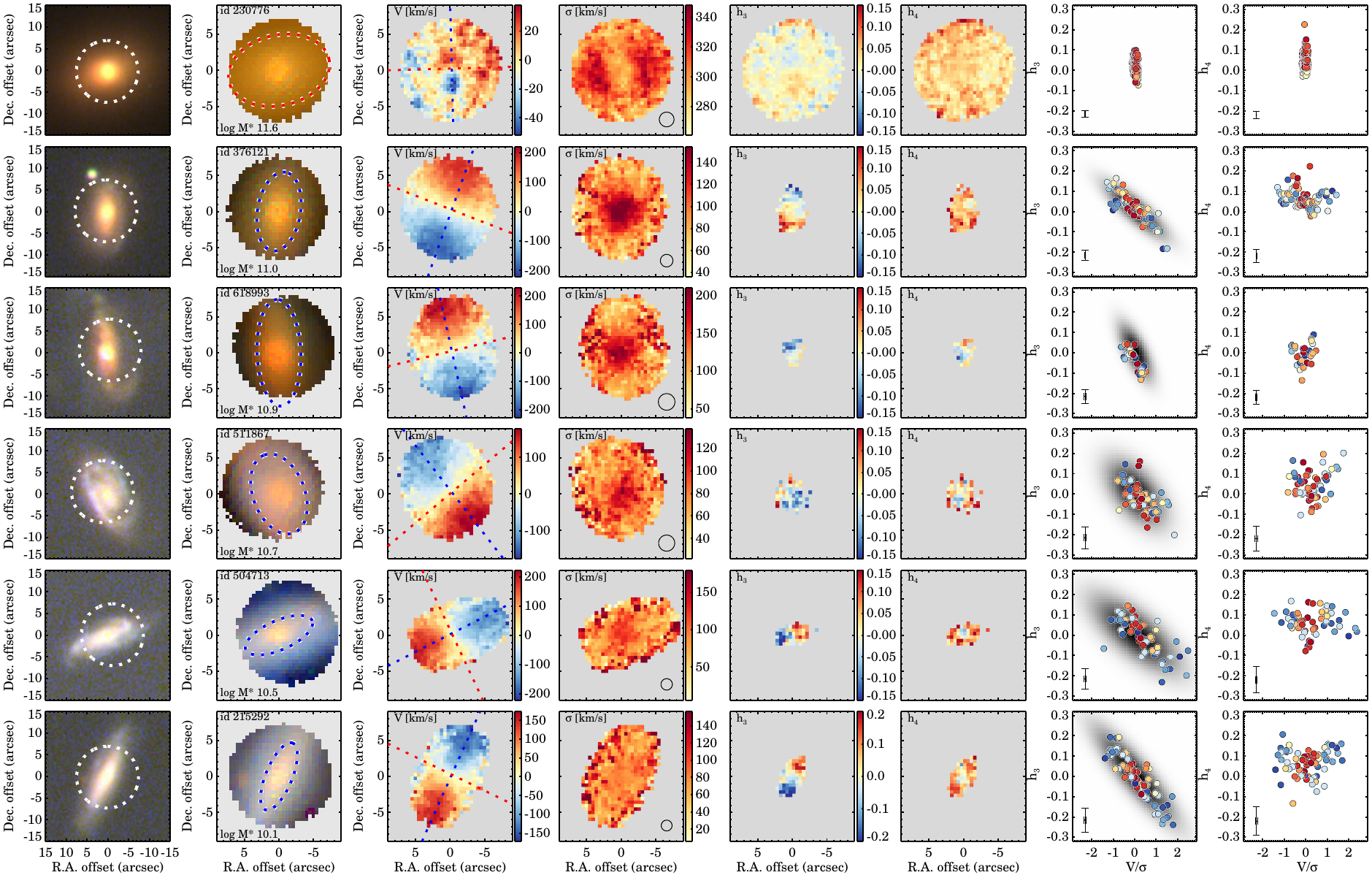}
\caption{Stellar kinematic maps for 6 galaxies in the SAMI test-sample for which the high-order moments could be derived. For each galaxy, from left to right we show: (1) the SDSS $u g r$ color image, where the white circle shows the size of the SAMI hexabundle. (2) Reconstructed color images of the galaxies from the SAMI spectra. The ellipse indicates the semi-major axis equal to one \re\, in blue, or semi-major axis equal to \re/4. in red. (3) The stellar velocity map from the fourth order moment fit (\vmf). Here the blue dashed line shows the kinematic major axis, and the red dashed line shows the kinematic minor axis. (4) The stellar velocity dispersion map from the fourth order moment fit (\smf). The black circle shows the size of the PSF. (5) $h_3$ maps. We only show the spaxels that meet the selection criteria for measuring the high-order moments. (6) $h_4$ maps. (7) $h_3$ versus \vs\, with the best-fitting log likelihood model in grey. (8) $h_4$ versus \vs. $h_3$ and $h_4$ share the same color bar. Galaxy stellar mass increases from bottom to top.
}
\label{fig:fig17}
\end{figure*}
%


\section{Optimising the Recovery of the LOSVD}
\label{sec:app}

Large integral field spectroscopic surveys capable of measuring accurate stellar kinematics, such as \at \citep{cappellari2011a}, CALIFA \citep{sanchez2012}, SAMI \citep{croom2012}, and MaNGA \citep{bundy2015}, are relatively new. All have different instrumental designs, and different target selections. Therefore, in order to derive accurate stellar kinematics, no standard recipes can readily be used, and the stability of the stellar kinematic measurements and assumptions that are made in the process must be tested thoroughly. 

\begin{figure*}
\epsscale{1.1}
\plottwo{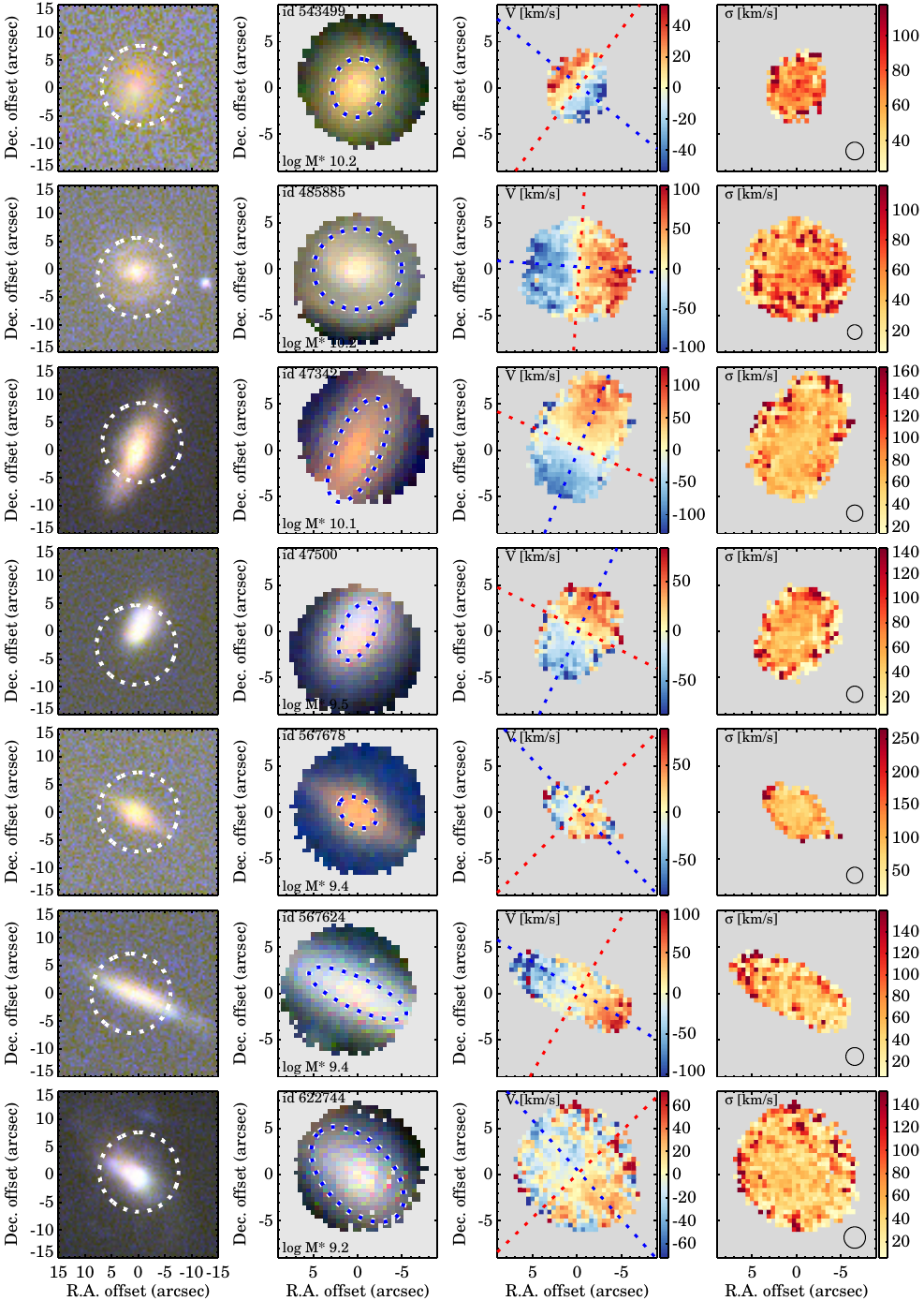}{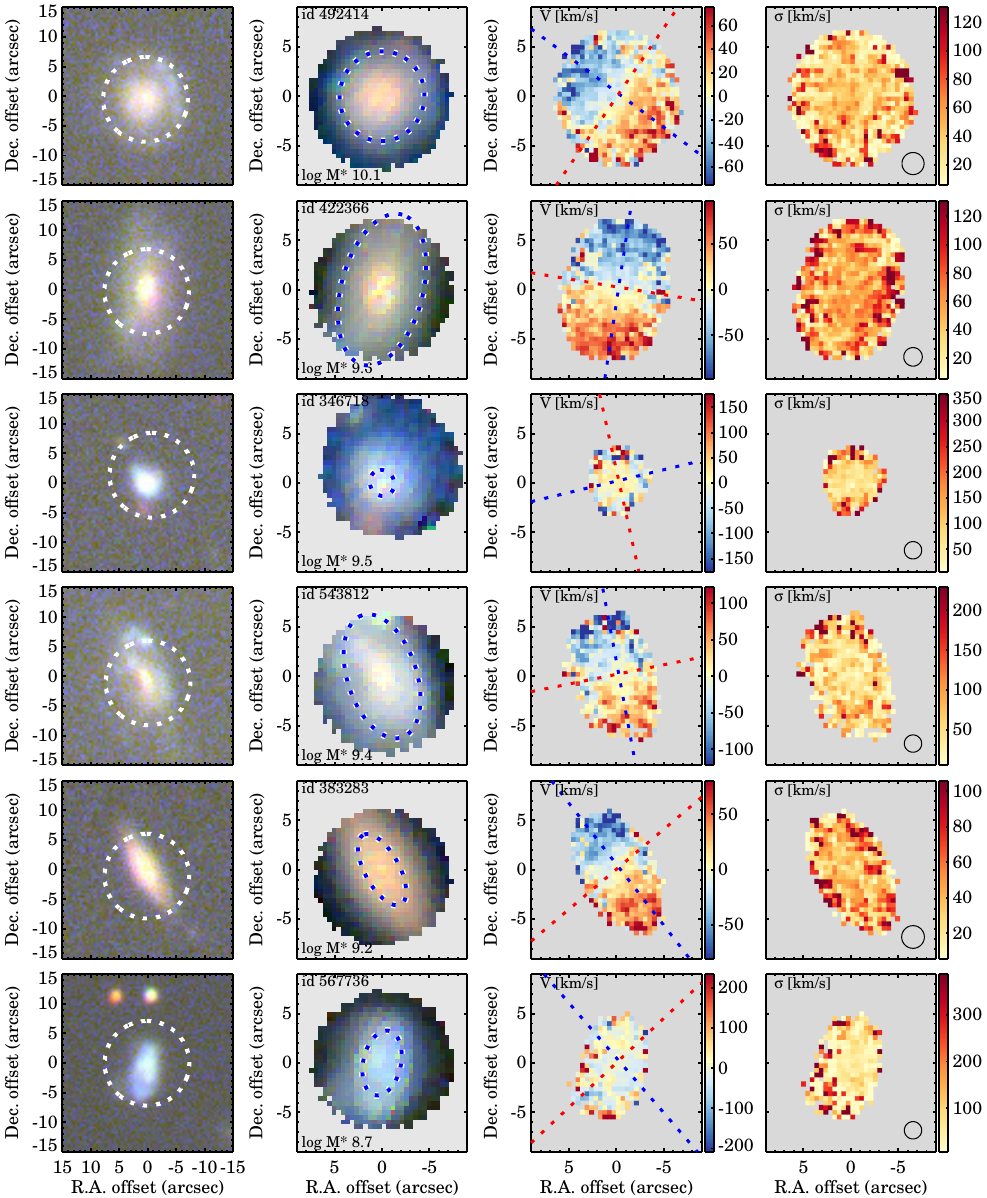}
\caption{Stellar kinematic maps for the remaining 13 galaxies in the SAMI test sample with second order moment stellar kinematic maps. For each galaxy, in similar fashion as for Figure \ref{fig:fig17}, from left to right we show: (1) the SDSS $u g i$ color image. (2) Reconstructed color images of the galaxies from the SAMI spectra. (3) The stellar velocity map from the second order moment fit (\vmt). (4) The stellar velocity dispersion map from the second order moment fit (\smt). Galaxy stellar mass increases from bottom to top.
}
\label{fig:fig18}
\end{figure*}

In this appendix, we test several of the assumptions that were made for measuring the stellar kinematic parameters for SAMI, and explore the parameter space for which the high-order moments can be recovered reliably. Specifically, we look at the impact of the degree of the additive polynomial, penalizing bias, template choice, and the choice of stellar library and stellar populations models. For this purpose, 19 galaxies are selected with a large range in stellar mass, age, and star formation activity. In Figures \ref{fig:fig17} and \ref{fig:fig18}, we show color images and stellar kinematic maps for this test sample. Figure \ref{fig:fig17} shows the six most massive galaxies for which reliable high-order moments could be measured; Figure \ref{fig:fig18} shows the other 13 galaxies with Gaussian LOSVDs only. More details can be found in the figure captions. Finally, we investigate the impact of seeing on our measurements by looking at repeat observations with different atmospheric conditions.


%
%


\subsection{Uncertainty Estimates from Repeat Observations}
\label{subsec:app_ro}

Due to the optimal tiling of the SAMI fields and plate configuration, 24 galaxies were observed twice. These sources are ideal for estimating uncertainties due to weather conditions, seeing, and the use of different hexabundles. We pre-select on quality for comparing between the two observations, i.e., galaxies from the first observation (\textit{obs$_1$}) were observed under better seeing conditions as compared to the second observation (\textit{obs$_{2}$}). 

\begin{figure*}
\epsscale{1.15}
\plotone{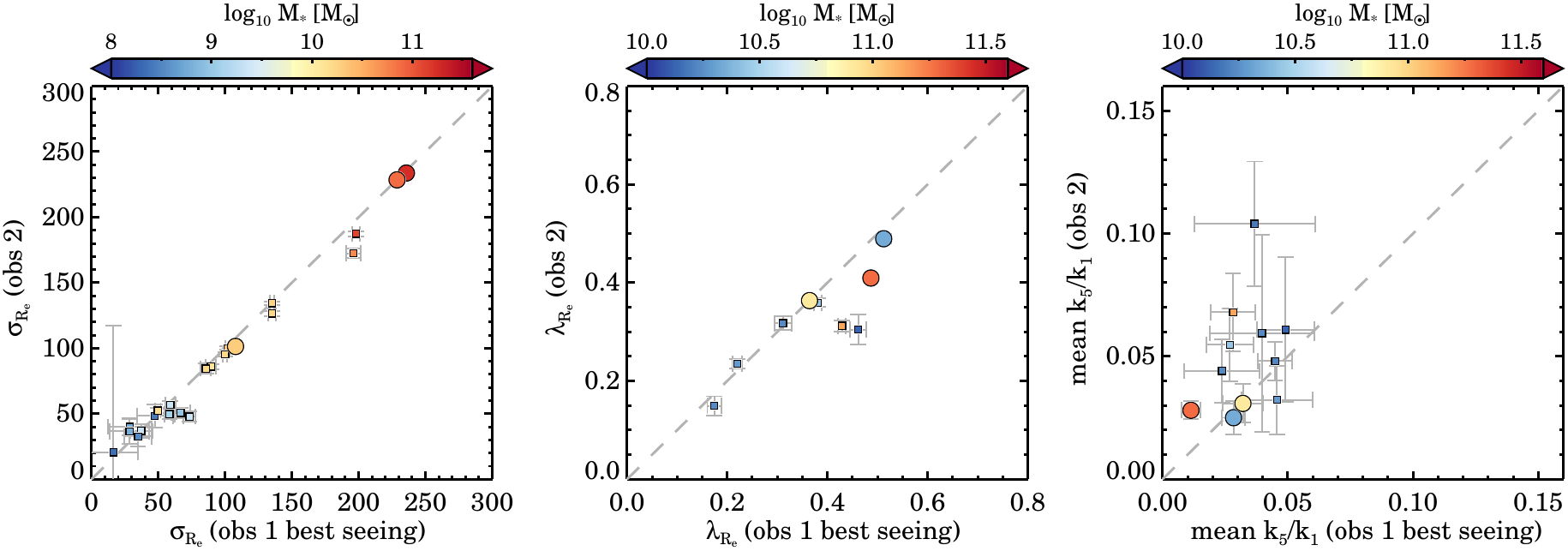}
\caption{Aperture velocity dispersion ($\sigma_{\rm{e}}$), proxy for the spin parameter (\lre), and kinematic asymmetry measurements from galaxies that have been observed twice. We show the best-seeing observations on the horizontal axis. The circles show galaxies that are in the final high-quality sample, squares show the other data with lower quality. All galaxies are color-coded by stellar mass. There is a good agreement for aperture velocity dispersions, whereas for \lre\, we find lower values for three galaxies when the seeing is worse (1\farcs98 versus 2\farcs51). There is also a considerable scatter for \mk, but here the random uncertainties are also larger. For the three galaxies that are in the final sample (circles) we find consistent results. }
\label{fig:fig19}
\end{figure*}

In Figure \ref{fig:fig19} we compare the aperture velocity dispersion, \lre, and kinematic asymmetry measurements for \textit{obs$_1$} and \textit{obs$_1$}. Note that not all galaxies have full \re\, coverage. Galaxies that are in our final high-quality sample are shown as circles, all other galaxies as squares. There is a good agreement between the 
stellar velocity dispersion measurements of \textit{obs$_1$} and \textit{obs$_2$}. For \lre, however, all galaxies have smaller measurement uncertainties than expected from the offset of the one-to-one relation. Moreover, for three, the repeat observations with worse seeing show significantly lower values of \lre\, (0.05-0.1). For the \mk, measurements, there is significant scatter and an offset from the one-to-one relation, but the measurement uncertainty for most outliers is also large. Nonetheless, more galaxies are classified as non-regular or quasi-regular rotators when the seeing is worse. Note, however, that only three galaxies with repeat observations passed quality cut Q$_3$.

\begin{figure*}
\epsscale{1.1}
\plotone{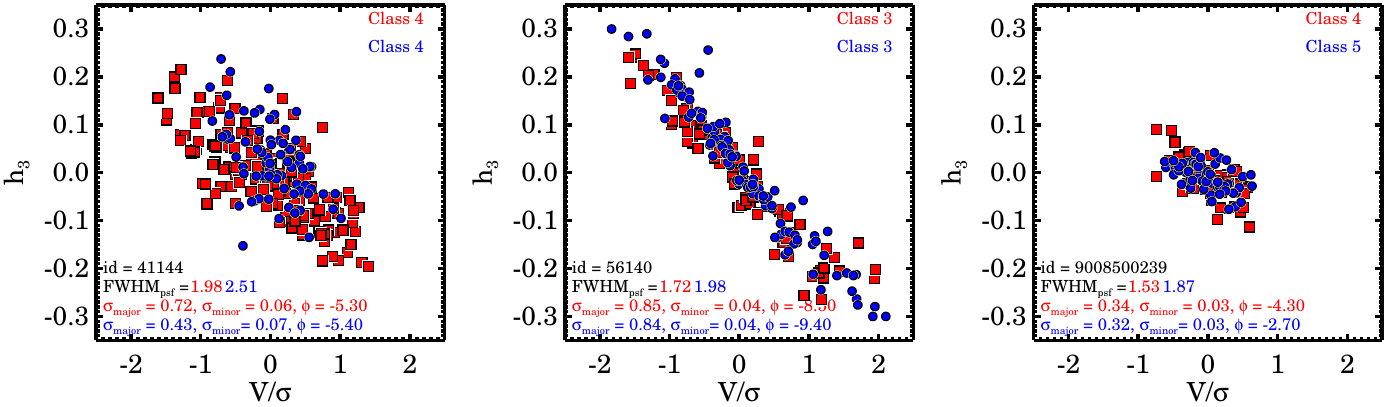}
\caption{$h_3$ versus \vs\, for three galaxies with repeat observations. The observations with the better seeing conditions are shown in red, and the repeat observation with poorer seeing in blue. There is a good agreement in the classification from the first two repeat observations, but the third galaxy changes classification from class 4 to 5 due to a small difference in measured angle $\phi$.}
\label{fig:fig20}
\end{figure*}

In Figure \ref{fig:fig20}, we show the relation between $h_3$ and \vs\, for galaxies from \textit{obs$_1$} (red) and \textit{obs$_2$} (blue). The first repeat observation (galaxy 41144, left panel), shows a similar anti-correlation. With better seeing we find a larger \smj, but the kinematic classification is the same. We find a good agreement between the measurements from the second repeat observation (galaxy 56140), but the anti-correlation is slightly steeper under worse seeing conditions ($\phi=-8.5$ versus $\phi=9.4$). Note, that for this galaxy, we find a significant difference in \lre\, between the two observations: 0.5 versus 0.41. For the third repeat observation (galaxy 9008500239, right panel) we find similar values of \smj\, and \smi\, but a difference in angle $\phi$ that is large enough for the galaxy to change classification.

\begin{figure*}
\epsscale{1.15}
\plotone{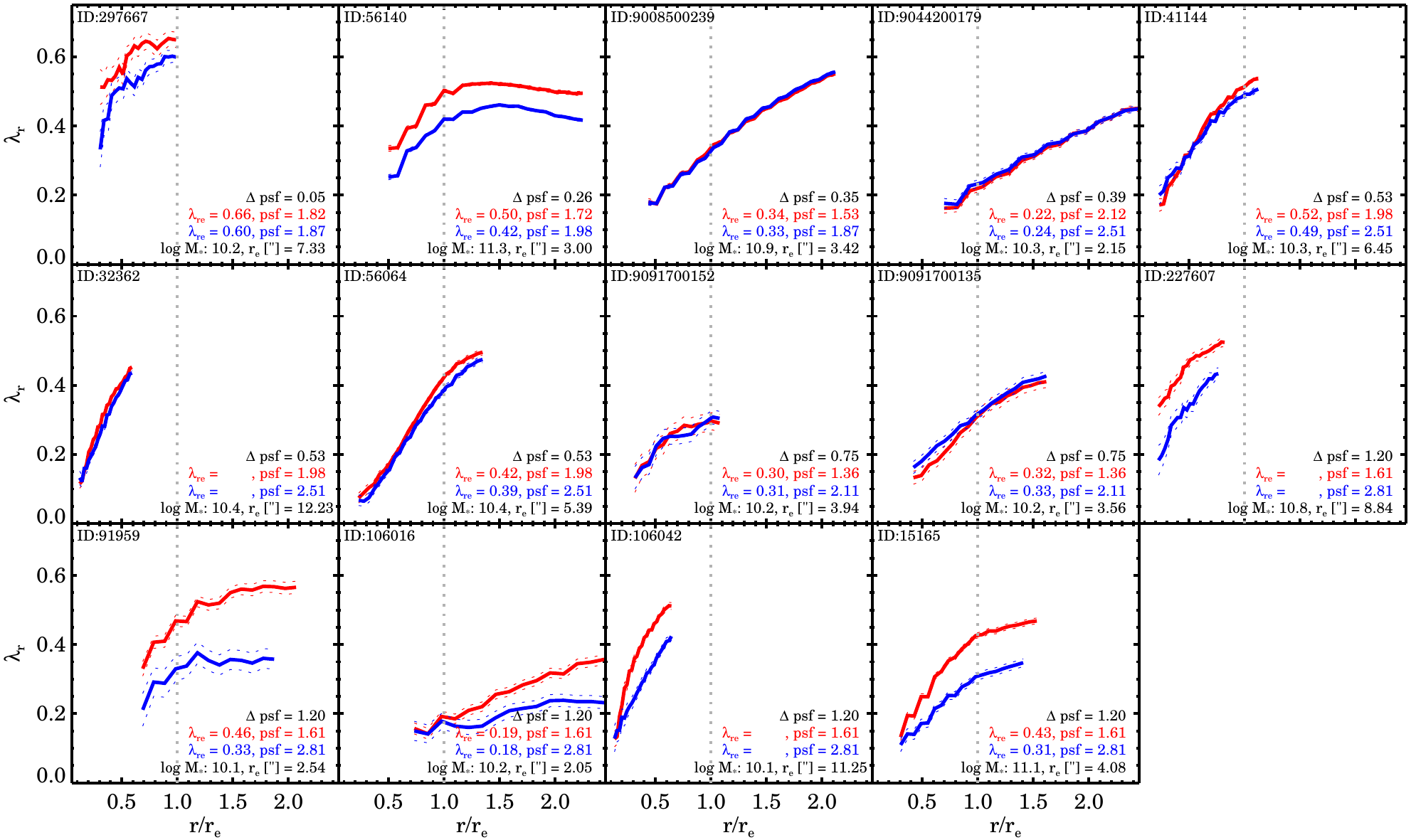}
\caption{Growth curves for the spin parameter approximation for repeat observations. The observations with the better seeing conditions are shown in red, the observation with the worse seeing in blue. For a few galaxies, where the seeing differences are large (1\farcs61 versus 2\farcs81), we find systematically lower values for \lre. For other galaxies with seeing differences ranging from 0\farcs35-0\farcs75, the growth curves from the repeat observations agree well. 
}
\label{fig:fig21}
\end{figure*}

Given the relatively large systematic uncertainty in the \lre\, measurements for the repeat observations, here we further investigate this by looking at the growth curves of the spin parameter proxy. Figure \ref{fig:fig21} shows all repeat observations for which \lr\, could be measured as a function of radius. Note that there are more galaxies with \lr\, measurements compared to Figure \ref{fig:fig19}, because not all galaxies have measurements out to 1\re.

For about half the galaxies, there is a relatively large difference in the growth curves between the repeat observations. In particular, galaxies 15165, 227607, 106016, 106042, and 91959 have been observed under very different seeing conditions (1\farcs6 versus 2\farcs8), and are affected the most. In worse seeing conditions, the radial \lr\, profiles are significantly lower by 0.05-0.1. Note, that all of these galaxies are from the same aperture plate, hence were observed at the same time. Not all galaxies, however, are affected by the seeing to the same extent. For example, galaxies 56064, 32362, and 41144 have a seeing difference of 0\farcs53 (1\farcs98 versus 2\farcs51), but show nearly identical \lr\, profiles. 

In conclusion, from the repeat observations we find that different atmospheric conditions can impact the \lr\, measurements on the order of 0.05-0.1. Furthermore, these results show the importance of repeat observations for large integral field surveys such as SAMI and MaNGA.

%


\subsection{Uncertainty Estimates from Re-Observing Simulated \at Galaxies}
\label{subsec:app_sims}

From repeat observations we found that different atmospheric conditions can impact our \lr\, and \mk\, measurements. Here, we use existing \at\, kinematic measurements to study the effect of seeing and measurement uncertainty on SAMI observations. \at\, measurements are used for this purpose, due to the data's higher spatial sampling as compared to SAMI, and because many of the results presented in this paper are compared to key results from the SAURON and \at\, surveys.

First, we use the publicly available \at\, stellar kinematic data products\footnote{http://www-astro.physics.ox.ac.uk/atlas3d/} and tabulated data from \citet{cappellari2011a}, \citet{emsellem2011}, and \citet{krajnovic2011}. We select galaxies that have full coverage out to at least one effective radius; however, only binned data derived from four or less original spaxels are used in order to avoid step functions in the velocity and dispersion maps. The following 23 galaxies meet these selection criteria: 
\textit{NGC0680, NGC1121, NGC2577, NGC2592, NGC2594, NGC2695, NGC2699, NGC2824, NGC2852, NGC3458, NGC3610, NGC3648, NGC3757, NGC3838, NGC4262, NGC4283, NGC4342, NGC4660, NGC5103, NGC5507, NGC5845, NGC6149, UGC09519}. All selected galaxies are regular rotators \citep{krajnovic2011}, with have a broad range in \lr\, (0.05-0.6) and ellipticity (0.05-0.6) \citep{emsellem2011}.

The flux, velocity, and velocity dispersion data are interpolated onto a regular grid. Outside the maximum measurement radius, we extrapolate the data to avoid step functions because later on the LOSVD is smoothed to mimic seeing. We rebin the data in order to obtain a similar angular size distribution as SAMI galaxies, i.e., to have effective radii between $2^{\prime\prime}$ and $6^{\prime\prime}$ in a 25$^{\prime\prime}\times25^{\prime\prime}$ size box with 0\farcs5 pixel size. Next, we create a three dimensional cube, where for each $(x,y)$ pixel we construct a Gaussian LOSVD in the $z$ coordinate using the flux, velocity and velocity dispersion.

To mimic seeing, we convolve all $x,y$ "images" in the LOSVD cube along the $z$ direction with a 2D Gaussian function with an FWHM ranging from 0.1$^{\prime\prime}$, 0.5$^{\prime\prime}$, 1.0$^{\prime\prime}$, ..., 3.0$^{\prime\prime}$. Note that it is not correct to simply smooth the flux, velocity, and velocity dispersion maps independently, as all three moments are correlated components in an observed LOSVD. For fast-rotating galaxies this is particularly important, as a steep gradient in flux and velocity is present in the center. When convolved with the seeing, the gradient in flux and velocity go down, whereas the velocity dispersion increases.

\subsubsection{Simulated Uncertainty Estimates on \lre}
\label{subsubsec:app_sims_lr}

\begin{figure*}
\epsscale{1.15}
\plotone{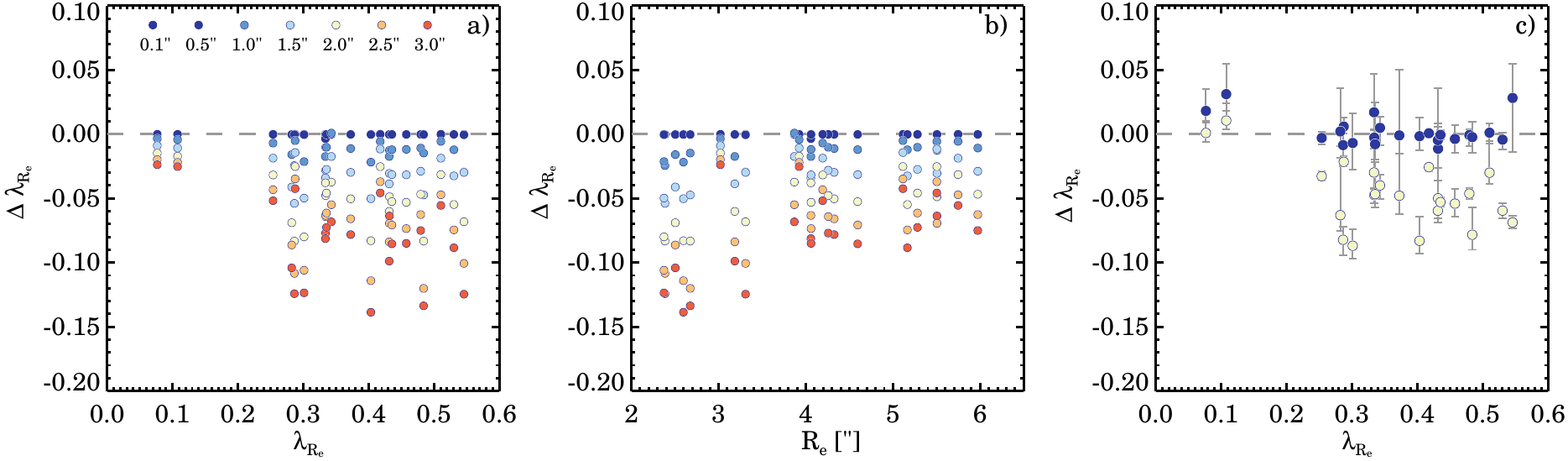}
\caption{$\Delta$\lre\, versus \lre\, for galaxies from the \at\, "re-observed" with SAMI under different simulated seeing conditions. $\Delta$\lre\, is defined as $\lre\, (\rm{FWHM_{PSF}}) - \lre\, (\rm{FWHM_{PSF}~0\farcs1)}$. In Panel a), we find that the median \lre\, decreases by 0.01 when the seeing FWHM$_{\rm{PSF}}$ = 1\farcs0 to $\Delta\lre$=-0.08 when the seeing is 3\farcs0. Seeing effects are stronger for galaxies with higher \lre. We note that the typical seeing for the SAMI galaxy survey is 2\farcs0. In Panel b) we show the effect of seeing as a function the galaxy's \re\, (semi-major axis). Smaller galaxies are on average more strongly impacted by seeing than bigger galaxies. In Panel c) we show the results when including both the effect of seeing and measurement errors. The data points give the median of a 1000 different realizations of the noise, whereas the lower and upper error bars are the 16th and 84th percentiles of the distribution. With FWHM$_{\rm{PSF}}$=2\farcs0 and $\lre<0.2$, the effect of seeing and measurement uncertainties cancel out, whereas for galaxies with $\lre>0.2$ seeing is dominant over measurements uncertainties and causes a median decrease of 0.05 in \lre.
}
\label{fig:fig22}
\end{figure*}

For each simulated galaxy, we measure \lre\, as described in Section \ref{subsec:lambda_r}. Figure \ref{fig:fig22}a-b shows the results for \lre\, under different simulated seeing conditions. We define $\Delta$\lre\, as \lre\, measured under different seeing conditions normalized by \lre\, measured when the seeing is 0\farcs1. At this point we do not include measurement errors, in order to separate the effect of seeing and measurement errors on \lre. Different colors show different realizations of the seeing, from 0\farcs1 in blue to 3\farcs0 in red. We note that typical seeing for the SAMI Galaxy Survey is 2\farcs0, indicated by the beige data.

In Figure \ref{fig:fig22}a we show $\Delta$\lre\, versus \lre. We find that \lre\, decreases with increasing FWHM$_{\rm{PSF}}$. Furthermore, galaxies with higher \lre\, are more strongly impacted by seeing than galaxies with low \lre. For $\lre<0.2$ we find that $\Delta\lre$ is small (-0.025) even when the seeing reaches 3\farcs0. When $\lre>0.2$, the impact of the seeing is stronger, with a median $\Delta$\lre=0.08 when FWHM$_{\rm{PSF}}$ = 3\farcs0. As expected, galaxies with small angular sizes are on average more strongly impacted by seeing than bigger galaxies (Figure \ref{fig:fig22}b).

Next, we look at the impact of measurement errors on the \lre\, measurements. For every galaxy, we add normal random noise to the flux, velocity, velocity dispersion as typically measured for galaxies in the SAMI Galaxy Survey (see Section \ref{subsec:qc}). The noise is weighted by the S/N such that spaxels in the center will have lower velocity and velocity dispersion errors as compared to spaxels in a galaxy's outskirts. For every galaxy, we repeat the process of adding random noise a 1000 times and remeasure \lre. Figure \ref{fig:fig22}c shows the median of the distributions with an FWHM$_{\rm{PSF}}$ of 0\farcs1 (blue) and 2\farcs0 (beige), normalized by the \lre\, 0\farcs1 results without measurement errors from Figure \ref{fig:fig22}a-b. The errors bars show the 16th (lower) and the 84th (higher) percentile of the distribution.

When measurement uncertainties are included, with minimal seeing (FWHM$_{\rm{PSF}}$=0\farcs1), we find a slight increase in \lre\, by $\sim$0.02-0.03 for galaxies with $\lre<0.2$. With 2\farcs0 seeing, we find that the increase due to measurement errors and the decrease due to seeing cancel out for these two galaxies. For galaxies with $\lre>0.2$, seeing is the dominant effect and causes a median decrease in \lre\, of 0.05.

\begin{figure*}
\epsscale{1.15}
\plotone{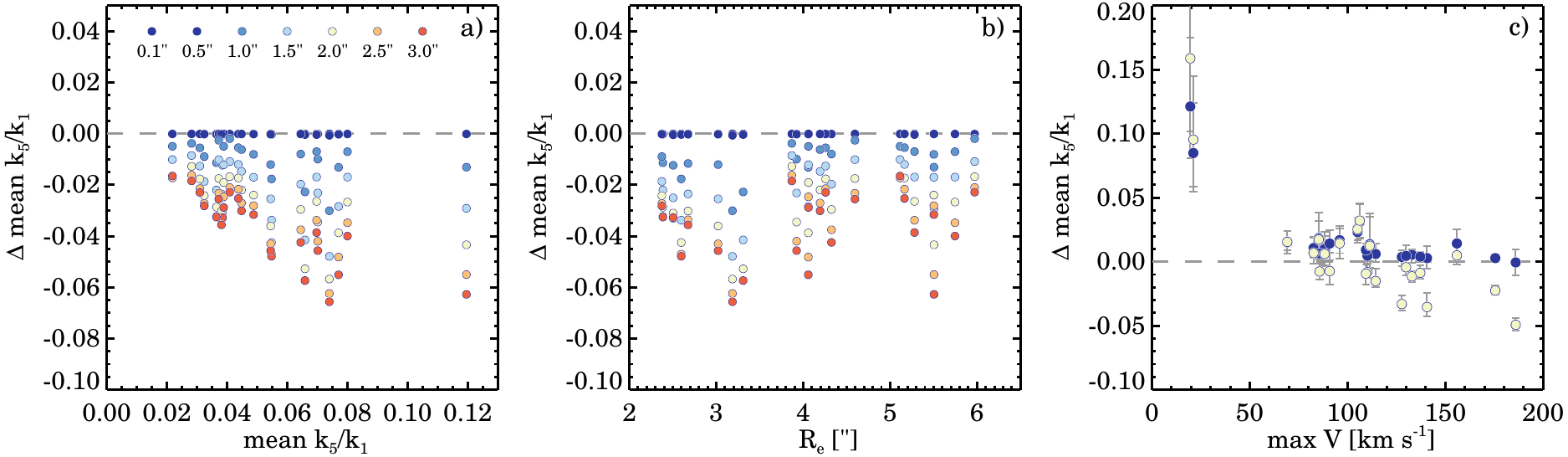}
\caption{$\Delta$\mk\, versus \mk, for galaxies from the \at\, "re-observed" with SAMI under different simulated seeing conditions. $\Delta$\mk\, is defined as $\mk\, (\rm{FWHM_{PSF}}) - \mk\, (\rm{FWHM_{PSF}~0\farcs1)}$. We find that seeing lowers \mk\, i.e., the rotation fields become more regular with increasing seeing (Panel a). Furthermore, the stronger the kinematic disturbance, the stronger the seeing will impact the measurement. In Panel b), we do not find a correlation with the angular size of the galaxy and the impact of seeing. We include the effect of measurement errors in Panel c), where we show $\Delta$\mk\, versus the maximum of the observable velocity field. The data points show the median of a 1000 different realizations of the noise, whereas the lower and upper error bars are the 16th and 84th percentiles of the distribution. We find that measurement uncertainties raise \mk, making the velocity field less regular, in particular when the galaxy is slowly rotating. For fast-rotating galaxies (maximum $V>50$\kms), observed with 2\farcs0 seeing, the effect of measurement uncertainty and seeing cancel out for most galaxies (median $\Delta$\mk=0.01).
}
\label{fig:fig23}
\end{figure*}

\subsubsection{Simulated Uncertainty Estimates on Kinemetry}
\label{subsubsec:app_sims_mk51}

The selected \at\, galaxies for studying the impact of seeing and measurement errors on \lre\, and \mk\, are all regular rotators with $\mk < 0.04$ \citep{krajnovic2011}, whereas the full observed range in \mk\, is 0.0-0.25 (see Figure \ref{fig:fig6}). If we were to use these galaxies for testing the impact of seeing on \kinemetry\, measurements the results would be trivial: regular velocity fields smoothed by the seeing will become more regular. Instead, it would be more interesting to test which kinematic features or disturbances disappear due to the effect of seeing.

Therefore, we add an artificial kinematically decoupled core to every velocity field in order to create a range in observable \mk\, parameters. We mimic a kinematically decoupled core by adding to the velocity field: two 2D Gaussians with opposite sign, at a random orientation, and peak velocity strength of 75\% of $V_{\rm{maximum}}$. The positive and negative peak of the decoupled core are placed at 3/4 of the semi-minor axis to ensure that the decoupled core is always observable.

The results without measurements uncertainties are shown in Figure \ref{fig:fig23}a-b. We find that $\Delta$\mk\, decreases when the seeing FWHM$_{\rm{PSF}}$ increases. The effect of seeing is also stronger for higher \mk. We find that the median \mk\, decreases by 0.01 when the seeing FWHM$_{\rm{PSF}}$ = 1\farcs0, to $\Delta$\mk=-0.03 when the seeing is 3\farcs0. Opposite to \lre\, we don't find a strong correlation between the impact of seeing on \mk\, and the angular size of the galaxy. This could be partially due to the fact that we artificially inserted kinematically decoupled cores, which might not be a fully realistic description of observed kinematic disturbances.

Next, we include noise in the same way as we did for the \lre\, simulations. Figure \ref{fig:fig23}c shows the impact of both measurement uncertainties and seeing on \mk\, versus the maximum of the velocity field. With minimal seeing (FWHM$_{\rm{PSF}}$=0\farcs1), we find that \mk\, increases. In particular for galaxies with little to mild rotation, noise on the velocity measurement is misinterpreted as a kinematic disturbance by \kinemetry. With 2\farcs0 seeing, typical for the SAMI Galaxy Survey, we find that the decrease in \mk\, caused by seeing is counteracted by the increase due to measurement uncertainties. The median decrease in \mk\, is 0.01 when the maximum $V>50$ \kms, but for the two galaxies with little rotation, the measurement errors are dominant over seeing effects ($\Delta$\mk=0.10-0.16).

\begin{figure*}
\epsscale{1.15}
\plotone{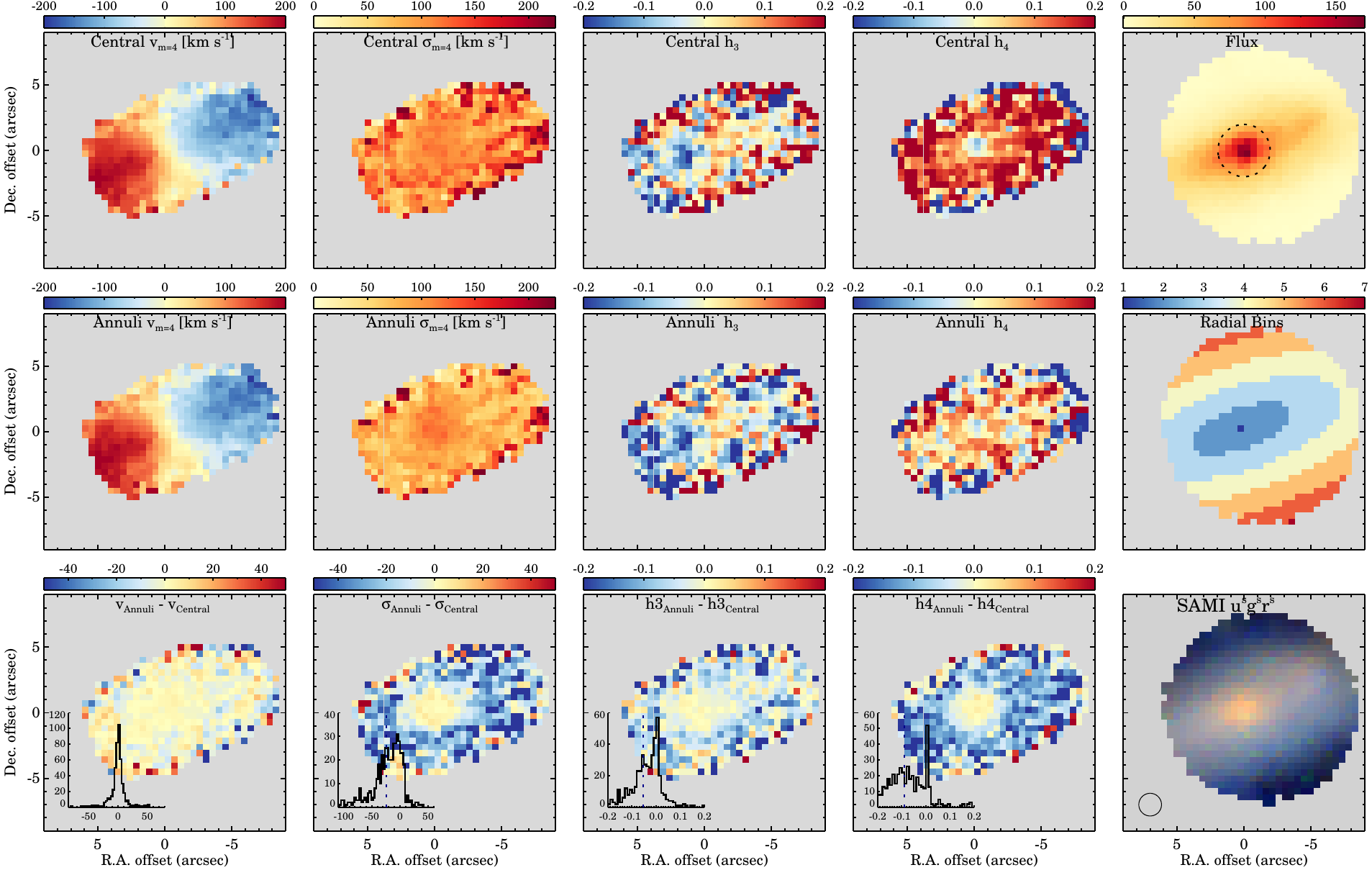}
\caption{Effect of the optimal template on the measured stellar kinematics for galaxy 504713. First four panels, top row: stellar kinematic parameters ($v, \sigma, h_3$, and $h_4$) measured using a single optimal template constructed from the mean spectrum from the $2^{\prime\prime}$ central region. Middle row: stellar kinematic parameters when using optimal templates constructed from annuli as described in Section \ref{subsec:app_tc}. Bottom row: difference between the central and annuli derived stellar kinematics. We also show the SAMI flux map (top-right) with the $2^{\prime\prime}$ aperture, the annular bins (middle-right), and the color image of the galaxy reconstructed from the SAMI spectra. Next to the color image we show the size of the PSF as the black circle. The stellar kinematic maps derived with the central optimal template overestimate $h_4$ and $\sigma$ in the disk. In the velocity dispersion map, the bulge and disk are not distinguishable for a single template, whereas they are clearly present in the velocity dispersion map for the stellar kinematics derived with annuli templates.}
\label{fig:fig24}
\end{figure*}

\subsection{Dependence of the Recovered LOSVD on the Template Choice}
\label{subsec:app_tc}

Template mismatch can significantly impact the measured stellar kinematics \citep[][]{vandesande2013}, and the high-order moments \citep[][]{gerhard1993}. In this section, we investigate the impact of the optimal template on the measured SAMI stellar kinematics. Two effects are analyzed: the spatial region in which the optimal template has been determined, and the choice of template library. Figure \ref{fig:fig24} shows the stellar kinematic maps for galaxy 504713, where the optimal template was derived using two different methods. For the first method, the central $2^{\prime\prime}$ spectrum is used for deriving a single optimal template, and every spaxel in the galaxy is then fit with this optimal template. For the second method, we use binned annuli spectra for constructing an optimal template. The second method is the SAMI default, and is described in more detail in Section \ref{subsec:otc}. Note that we do not derive optimal templates from individual spaxels, which generally do not meet our S/N requirement of 25 that is needed to derive a reliable optimal template.

The first method shows larger values of $h_4$ and $\sigma$ in the disk of the galaxy as compared to method number two. In the centre of the galaxy, the templates gives reasonable results, but in the outskirts $h_4$ reaches relatively high values ($\sim0.2$), which is often associated with template mismatch \citep{gerhard1993}.

On the bottom right, the reconstructed color image from the SAMI spectra are displayed. Galaxy 504713 shows a clear red bulge and a blue disk. The template from method one was derived from the central bulge, whereas method two has optimal templates that vary as a function of radius. It is therefore not too surprising that the template from method one does not provide an adequate match to the stellar population in the disk. 

We conclude that using annular bins for constructing optimal templates provides better results as compared to using centrally derived optimal template.


\begin{figure*}
\epsscale{1.15}
\plotone{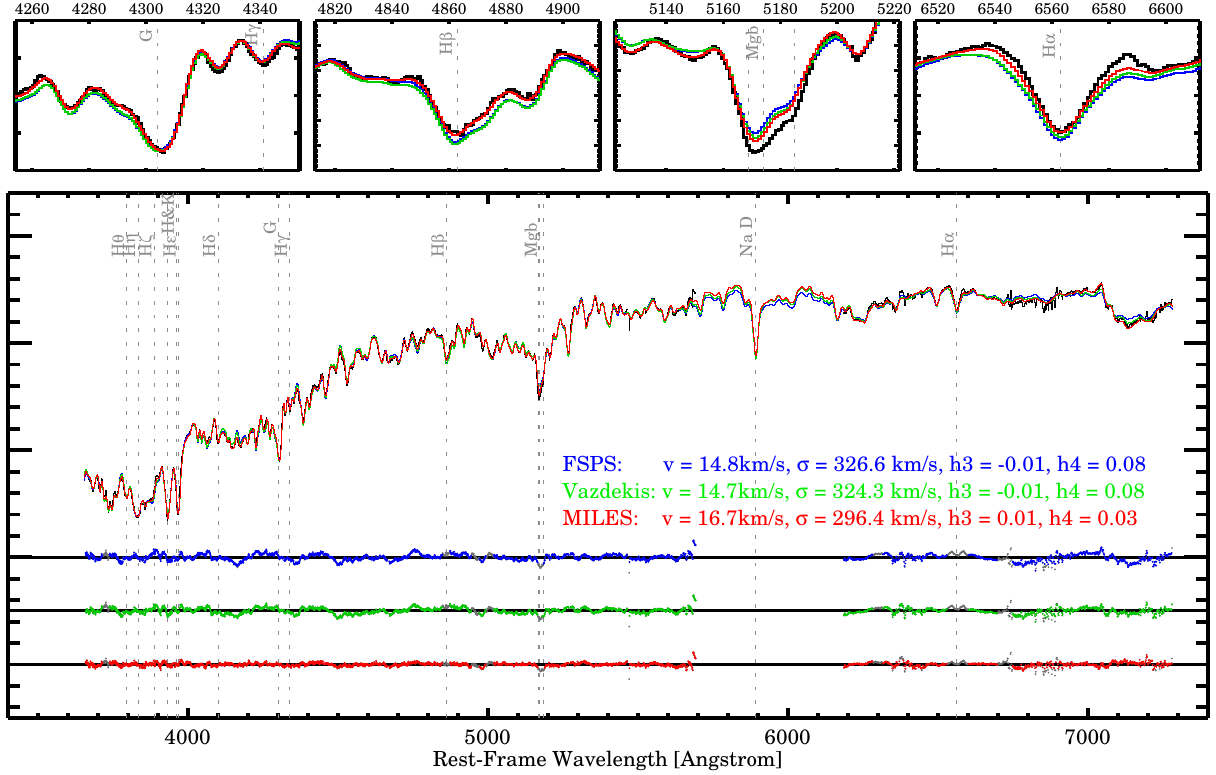}
\caption{Central $2^{\prime\prime}$ aperture spectrum for galaxy 230776 (black) and its best-fit templates from different stellar (population) libraries. Note that the observed spectrum in black is hard to distinguish from the best-fit templates due to the high S/N of this spectrum. From top to bottom, we show the residuals of the galaxy minus the best-fit for the FSPS models in blue, Vazdekis models in red, and the MILES stellar library in red. Wavelength regions and pixels that are masked are shown in grey. Smallest residuals are obtained with the MILES stellar library. The velocity dispersion is also systematically lower when the MILES stellar library is used as compared to the stellar population models.}
\label{fig:fig25}
\end{figure*}

Next, we test the impact of using different stellar libraries for measuring stellar kinematics. The default MILES stellar library is compared to the Flexible Stellar Population Synthesis models (FSPS; v2.5) of \citet{conroy2009} and \citet{conroy2010}, and to the models by \citet[v8.0;][]{vazdekis2010,vazdekis2015}. One advantage of using SPS models over a stellar library with single stars is reducing the degrees of freedom in deriving an optimal template, which is particularly useful when the S/N is low. Because SPS models are already pre-made optimal templates with only age and metallicity as a degree of freedom, using SPS can reduce the uncertainty on the stellar kinematic parameters \citep[e.g.,][]{vandesande2013}. Another advantage of using SPS models is an increased fitting speed. Stellar libraries typically contain a thousand stars, whereas SPS models are distilled to a few hundred templates. A disadvantage of SPS models is the lack of exotic templates. If the stellar populations in a galaxy are highly mixed due to multiple epochs of star formation and merger events, a combination of SPS templates may no longer describe the integrated light adequately.

For both SPS models, we picked the version that use the stellar MILES library as their input. All galaxies from the test sample are fit with the three models, and the optimal templates are constructed from the binned annular spectra. In general, no systematic offsets in the kinematic maps are found (\vmf, \smf, $h_3$, $h_4$) when we use different libraries. The scatter is consistent with the random uncertainties that are expected from the Monte-Carlo simulations. 

There is one exception: galaxy 230776, a massive, red ($\log_{10}M_*/$\msun$ = 11.6$), slow-rotating elliptical galaxy with a counter-rotating core. For the measured velocity there is a good agreement between different models, but for the velocity dispersion we find a systematic difference of ~30\kms. This difference is much larger than expected, given the high S/N of the galaxy. Also, there is a systematically higher value of $h_4$ (0.05) when we use the FSPS and Vazdekis models as compared to when using the MILES models. Figure \ref{fig:fig25} shows the central $2^{\prime\prime}$ spectrum of galaxy 230776, with the best-fitting templates when using the MILES stellar library (red), the FSPS models (blue), and the Vazdekis models (green). Both the FSPS and Vazdekis models show an obvious residual, whereas the best-fit template with the MILES library stars shows very little residual.

SAMI galaxies are selected to have a large range in mass and star formation activity, for which we expect mixed stellar populations. In light of the results shown here, we therefore decided to use the MILES stellar library for deriving optimal templates as opposed to using SPS models. 


\begin{figure*}
\epsscale{1.15}
\plotone{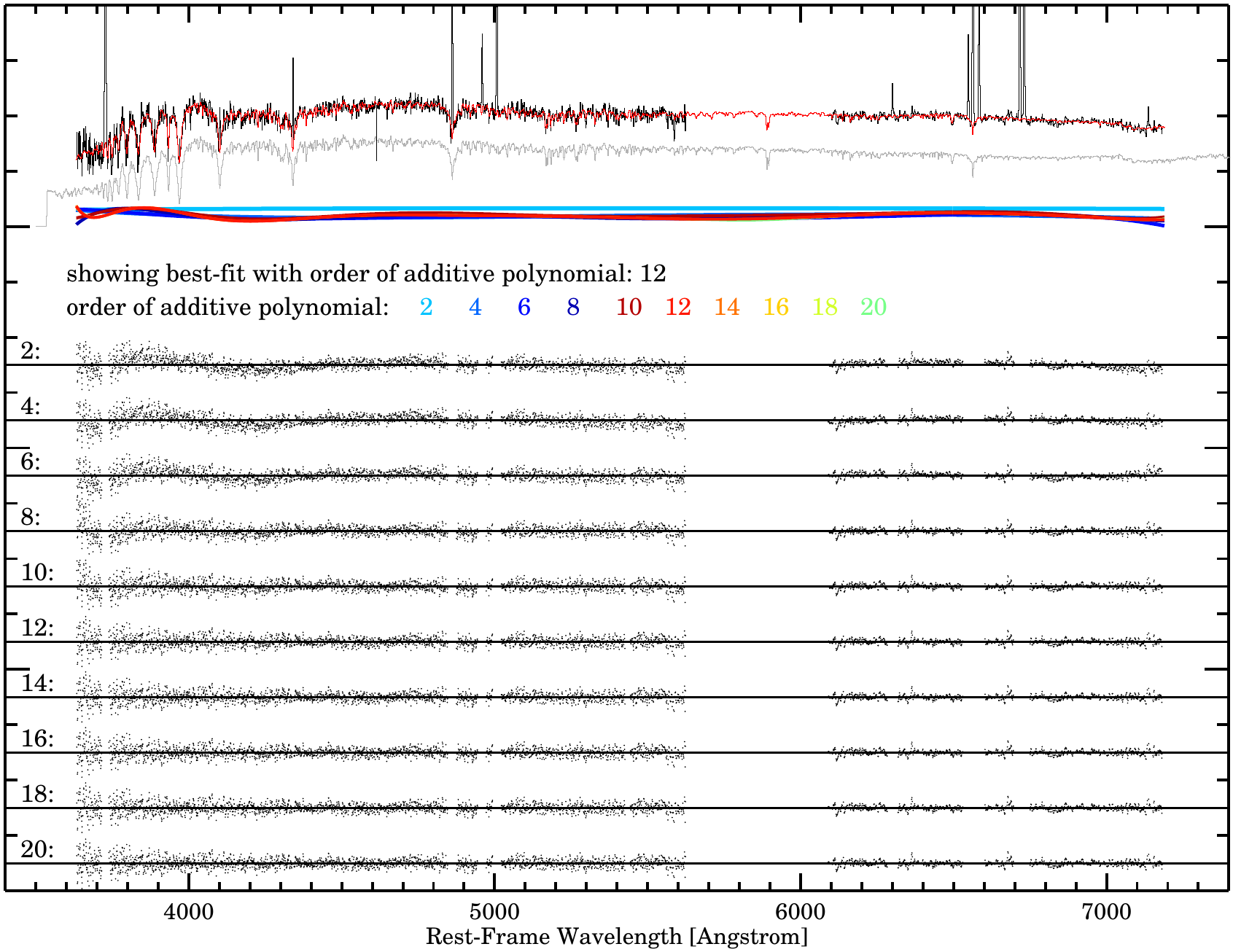}
\caption{Central spectrum for galaxy 47500 (black) and its best-fit template (red), optimal template (grey). From top to bottom, we show the residuals of the galaxy minus the best-fit template with different additive order polynomials. For polynomials with order less than 10 there are clear residuals that vary as a function of wavelength, which disappear for polynomials with order 12 or higher.}
\label{fig:fig26}
\end{figure*}
%



\subsection{Dependence of the Recovered LOSVD on the Order of the Additive Polynomial}
\label{subsec:app_ap}

Small errors in the flux calibration can create possible mismatches between the stellar continuum emission from the observed galaxy spectrum and the stellar template, that could impact the estimated kinematic parameters. Here, we test which additive order Legendre polynomial is needed for the SAMI data in order to account for flux calibration errors. Multiplicative polynomials are not used here, because these could change the depth of the line as a function of wavelength, which would impact the measurement of kurtosis ($h_4$) of the absorption lines. For the \at\, data, a fourth order additive polynomial was used over a wavelength range of 113\AA. With the SAMI wavelength range of $\sim3650$\AA, this would imply a 32nd order polynomial. Every added polynomial order makes the fit more time consuming, which is why we aim to use the lowest order possible while still correcting for possible flux calibration errors.

Using the test sample, for each galaxy the central circular $2^{\prime\prime}$ spectrum is fit with a range in additive polynomials of 1-20. We then attempt to find the additive Legendre polynomial degree that gives consistent and stable results, as based on the residuals, velocity, velocity dispersion and reduced $\chi^{2}$. Figure \ref{fig:fig26} shows the observed flux of galaxy 47500 in the central circular $2^{\prime\prime}$ in black, the optimal template in grey, and the best-fit template with a 12th order polynomial in red. Below the galaxy spectrum, the additive polynomials are shown as derived by \textsc{pPXF} for the optimal fit. Different colors indicate different polynomial degrees, and below the polynomials we show the fit residual (galaxy - best-fit template). A clear non-constant residual, that varies with wavelength, appears when a second order polynomial is used. This non-constant residual disappears when a 12th order polynomial or higher is used.

\begin{figure*}
\epsscale{1.15}
\plotone{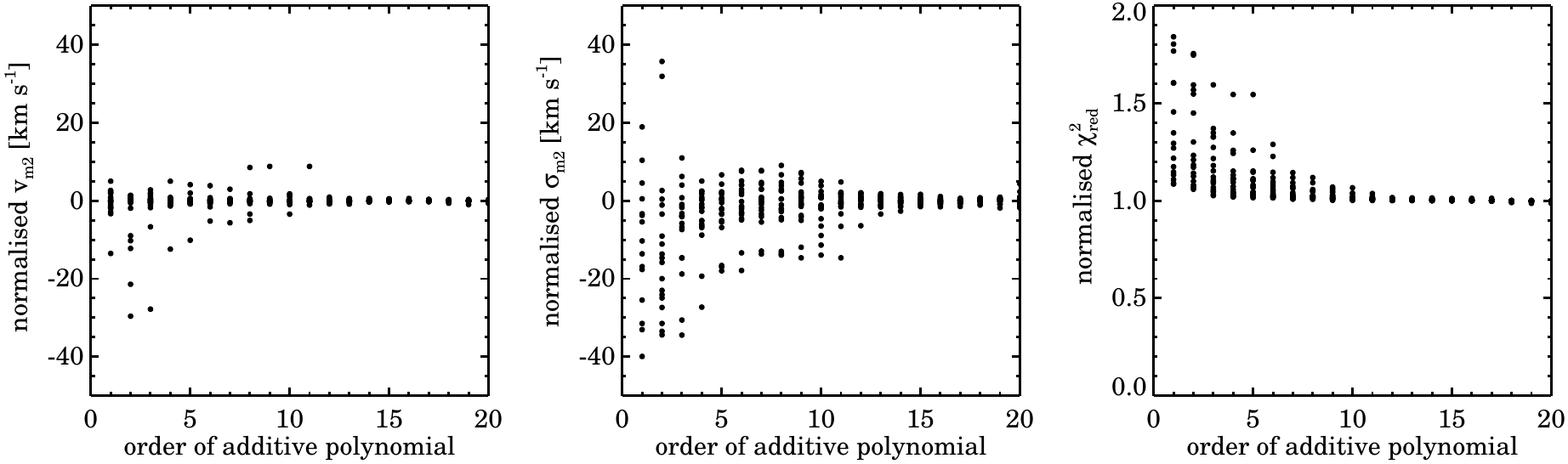}
\caption{Normalized velocity, velocity dispersion, and reduced $\chi^2$ measured with different order additive polynomials. The scatter in the measured velocities and velocity dispersions decreases as a function of polynomial up to order twelve. The reduced $\chi^2$ also decreases up to order twelve, but stabilizes thereafter.}
\label{fig:fig27}
\end{figure*}

In Figure \ref{fig:fig27}, we show the stellar velocity, velocity dispersion, and reduced $\chi^{2}$ versus the additive order of the polynomials, where the data are from the stellar kinematic fits on all galaxies in the test sample. In each panel, and for each galaxy individually, the data are normalized to the mean of the five fits with the highest order polynomials. For \vmt\, and \smt, the scatter decreases as a function of polynomial order. The least amount of scatter is reached when we use a polynomial with order $>12$. The reduced $\chi^2$ also decreases as a function of polynomial order up to order 12, after which the $\chi^2$ no longer varies. We conclude that an additive Legendre polynomial with order 12 suffices to account for possible mismatches between the stellar continuum emission from the observed galaxy spectrum and the template.


%


\subsection{Optimising the Penalizing Bias for SAMI data}
\label{subsec:app_pb}

Fitting high-order moments in spectra can be problematic when the S/N is low or if the velocity dispersion is close to, or lower than, the instrumental resolution \citep[see e.g., ][]{cappellari2011a}. \textsc{pPXF} was designed to employ a maximum penalized likelihood, i.e., forcing a solution to a Gaussian if the high-order moments are unconstrained by the data \citep{cappellari2004}. Whereas \textsc{pPXF} can derive an automatic penalizing bias value based on the $\chi^2$, but this is often too strict. A penalizing bias value that is too high can systematically offset the recovered velocity and velocity dispersion so should be avoided if possible. Therefore, we have derived the optimal penalizing bias value as a function of S/N for SAMI spectra from Monte Carlo simulations. 

From the MILES-based optimal template for galaxy 215292, we create a representative template galaxy spectrum. This spectrum is rebinned onto a logarithmic wavelength scale with constant velocity spacing, using the code \textsc{log\_rebin} provided with the \textsc{pPXF} package. A large ensemble of mock galaxy spectra is created by convolving the template spectrum with a LOSVD. For the LOSVD, we use $h_3$ = 0.1, $h_4$ = 0.1, and 2500 random velocity dispersions between 0 and 200 \kms, with random velocities between -50 and 50 \kms. Random Gaussian noise is added to the spectra to obtain a full sample of mock galaxy spectra with S/N of 5-100 \AA$^{-1}$.

\textsc{pPXF} is used to measure the LOSVD from the simulated galaxy spectra. The program is limited to only one template, because we intend to study the impact of the penalizing bias and S/N, not template mismatch. A fourth order additive Legendre polynomial removes possible continuum mismatching between the templates and the galaxy.

\begin{figure*}
\epsscale{1.15}
\plotone{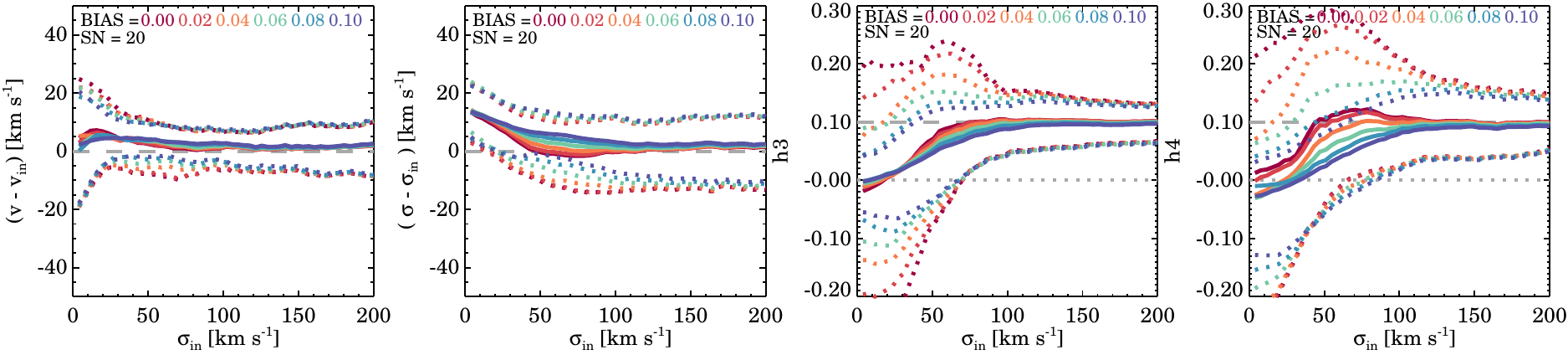}
\caption{Impact of the penalizing bias on the recovered LOSVD from MC-simulations of SAMI spectra with LOSVD parameters $h_3$=0.1, $h_4$=0.1, $5<\sigma [\rm{km} \rm{s}^{-1}]<200$, and S/N = 20 \AA$^{-1}$. The difference between the recovered and the input LOSVD parameters as measured with \textsc{pPXF} are shown for the velocity a), and velocity dispersion b). In the right two panels we show recovered $h_3$ c), and $h_4$ d) as compared to their input values ($h_3$=0.1, $h_4$=0.1). The solid lines show the median, the dotted lines show the 16th and 84th percentile, and the different colors represent different values for the penalizing bias. For an S/N = 20 \AA$^{-1}$, The optimal bias setting for SAMI data with S/N = 20 \AA$^{-1}$ is between 0.04 and 0.06.}
\label{fig:fig28}
\end{figure*}

\begin{figure*}
\epsscale{1.15}
\plotone{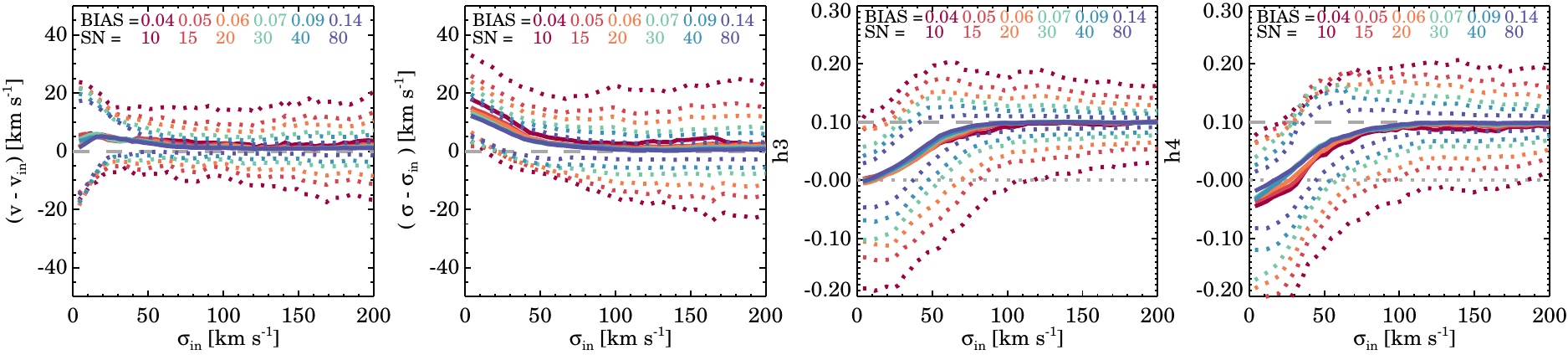}
\caption{Impact of the penalizing bias and S/N on the recovered LOSVD from MC-simulations of SAMI spectra with LOSVD parameters $h_3$=0.1, $h_4$=0.1, $5<\sigma<200$, and S/N = 20 \AA$^{-1}$. We show the difference between the recovered and the input LOSVD parameters for different input S/N values (10-80 \AA$^{-1}$) and their optimal bias setting in a similar fashion as Figure \ref{fig:fig28}. For SAMI spectra with S/N$>20$ \AA$^{-1}$, we can reliably recover $h_3$ and $h_4$ when $\sigma>70$\kms, i.e., the uncertainties on $h_3$ and $h_4$ are $< 0.1$.
}
\label{fig:fig29}
\end{figure*}

In Figure \ref{fig:fig28}, we demonstrate the impact of the penalizing bias value on the recovered LOSVD for an S/N = 20 \AA$^{-1}$, in Figure \ref{fig:fig29} the impact of the S/N on the recovered LOSVD with the optimal penalizing bias for SAMI applied is shown. The solid colored lines are the median of Monte Carlo simulations, whereas the dotted lines show the 16th and the 84th percentile ($1\sigma$). In the two left panels, we present the differences between the measured values and the input values of the velocity dispersion $\sigma_{in}$. In the panels on the right, the recovered values of $h_3$ and $h_4$ are shown than the input values of 0.1 (dashed line). Note that the curves in the following figures have been smoothed with a box-car filter (10\kms), in order to wash out numerical noise due to a limiting number of MC simulations.

%

Figure \ref{fig:fig28}a-d shows the stellar kinematic parameters and recovered LOSVD for six different penalizing bias values. In Figure \ref{fig:fig28}d, we see that when the penalizing bias is higher, the fit is more penalized towards $h_3=0$ and $h_4=0$. A penalizing bias value that is too high will also systematically offset the recovered velocity and velocity dispersion (Figure \ref{fig:fig28}b around $\sigma=50$\kms). An optimal penalizing bias value, however, will reduce the scatter in the velocity dispersion, $h_3$ and $h_4$, without creating a systematic offset in the velocity and velocity dispersion. For an S/N = 20 \AA$^{-1}$, we find that the optimal bias value is between 0.04 and 0.06.

For thirteen different S/N values between 0 and 100 \AA$^{-1}$, we estimated the optimal penalizing bias value (Figure \ref{fig:fig29}a-d). The different colored lines show the recovered LOSVD for six S/N values and their optimal bias-value. For spectra with S/N $<$ 20 \AA$^{-1}$\, the S/N is too low for an accurate measurement or the high-order moments $h_3$ and $h_4$. With an S/N = 20 \AA$^{-1}$, at $\sigma=200$\kms\, the typical uncertainties are $v_{\rm{err}} = 8$\kms, $\sigma_{\rm{err}} = 12.5$\kms, $h_{3,\rm{error}} = 0.04$, $h_{4,\rm{error}} = 0.05$. Below $\sigma=70$\kms\, the systematic offset in $h_3$ and $h_4$ becomes too large ($> 0.02$) for reliable estimates.

\begin{figure*}
\epsscale{0.45}
\plotone{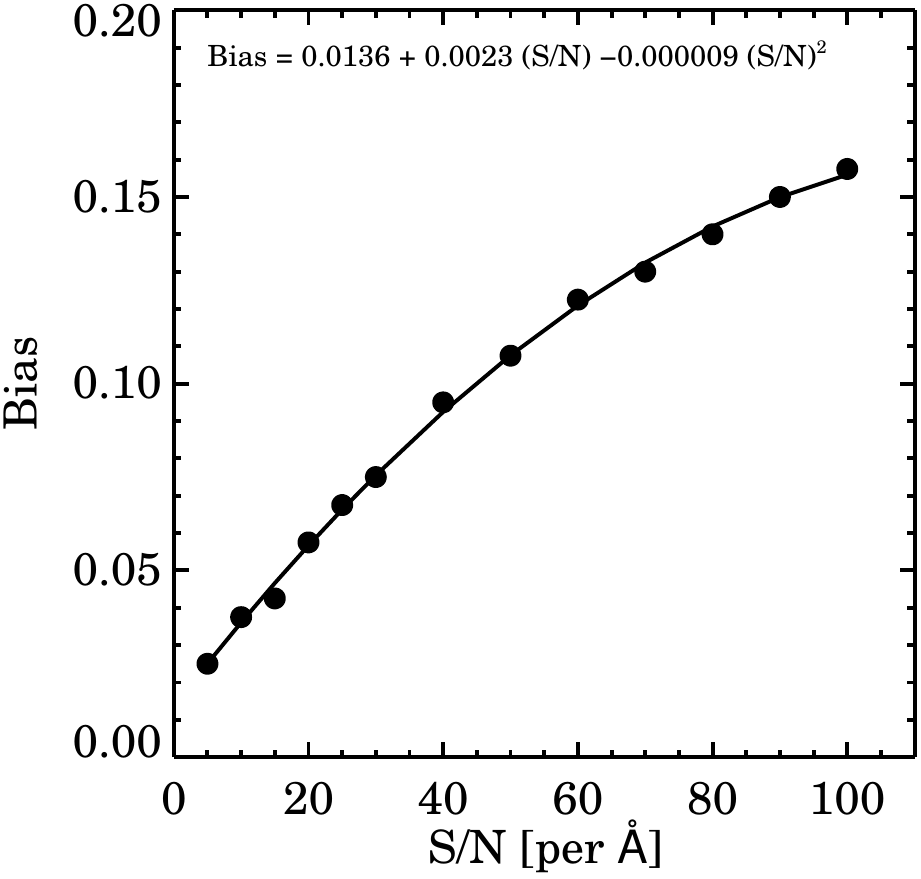}
\caption{Optimal penalizing bias versus S/N for SAMI spectra derived from MC-simulations. The solid line is the best-fit to the data, with Bias = 0.0136 + 0.0023 (S/N)-0.000009 (S/N)$^{2}$.
}
\label{fig:fig30}
\end{figure*}

We use the results from the Monte-Carlo simulation to derive a relation between the optimal penalizing bias value
and the S/N. Figure \ref{fig:fig30} shows the optimal bias values as a function of S/N, together with a second order polynomial fit. From the best-fit relation, we obtain a simple analytic expression for the ideal penalizing bias as a function of S/N:
\begin{equation}
\label{eq:eq11}
Bias = 0.0136 + 0.0023 (S/N) - 0.000009 (S/N)^2.
\end{equation}
Note, that our penalizing bias values are significantly lower as compared to the \at\, values \citet{cappellari2011a}. This can be explained, however, by the larger wavelength range and higher instrumental resolution of the SAMI instrument.

%

%

%

\clearpage

\end{document}